\documentclass[submission,copyright,creativecommons]{eptcs}
\providecommand{\publicationstatus}{} 

\usepackage{iftex}
\usepackage{amsmath}
\usepackage{amssymb}
\usepackage{amsthm}
\usepackage{xcolor}
\usepackage{MnSymbol}
\usepackage{stmaryrd}
\usepackage{xspace}
\usepackage{listings}
\usepackage[scr]{rsfso}
\usepackage{mathpartir} 
\usepackage{prftree}
\usepackage{graphics}
\usepackage[inline]{enumitem}

\usepackage{hyperref}
\usepackage{tikz-cd}
\usepackage{array}

\newcommand{\limplies}{\rightarrow}
\newcommand{\lequiv}{\leftrightarrow}

\newcommand{\inh}[1]{\ensuremath{\top_{\!\!#1}}}

\newcommand{\cln}{{\kern0.2ex:\kern0.2ex}}
\newcommand{\ld}{\mathord{.\;}}

\newcommand\sbullet[1][.5]{\mathbin{\vcenter{\hbox{\scalebox{#1}{$\bullet$}}}}}

\newcommand{\EV}{\mathit{EV}}
\newcommand{\SV}{\mathit{SV}}

\newcommand{\pset}[1]{\mathbb{P}(#1)}
\newcommand{\abs}[1]{\left|#1\right|}

\newcommand{\floor}[1]{\left\lfloor#1\right\rfloor}
\newcommand{\prule}[1]{\scalebox{.95}[1]{\textnormal{{(#1)}}}}
\newcommand{\pto}{\rightharpoonup}

\newcommand{\ev}[2]{\left|#1\right|_{#2}}

\newcommand{\evMrho}[2]{\left|#1\right|_{#2,\rho}}

\newcommand{\apd}{\mathbin{\sbullet}}
\newcommand{\uapdu}{\_{\apd}\_}
\newcommand{\ceil}[1]{\lceil#1\rceil}
\newcommand{\pr}[1]{\left\langle #1 \right\rangle}
\newcommand{\smbceil}{\mathsf{def}}

\newcommand{\nil}{\mathit{nil}}

\newcommand{\cons}{\mathit{cons}}
\newcommand{\pcons}{\mathit{pcons}}

\newcommand{\hd}{\mathit{hd}}
\newcommand{\tl}{\mathit{tl}}
\newcommand{\out}{\mathit{out}}
\newcommand{\nxt}{\mathit{nxt}}

\newcommand{\List}{\mathit{List}}
\newcommand{\Lst}{\mathit{Lst}}

\newcommand{\Sort}{\mathit{Sort}}

\newcommand{\Sorts}{\inh{\mathit{Sort}}}

\newcommand{\Id}{\mathit{ID}}
\newcommand{\Mo}{\mathcal{M}}

\newcommand{\DEF}{\mathsf{EQUALITY}}

\newcommand{\INH}{\mathsf{SORT}}

\newcommand{\PAIR}{\mathsf{PAIR}}

\newcommand{\Smbs}{{\color{blue}\small\textsf{Symbols}:}}
\newcommand{\Axms}{{\color{blue}\small\textsf{Axioms}:}}
\newcommand{\Sugar}{{\color{blue}\small\textsf{Notations}:}}

\newcommand{\Impts}{{\color{blue}\small\textsf{Imports}:}}
\newcommand{\axname}[1]{\textnormal{(\textsc{#1})}}

\lstdefinelanguage{matchinglogic}
{morekeywords={spec, endspec, theory, endtheory, model, endmodel, of},
sensitive=false,
morecomment=[l][\textit]{//},
morecomment=[s]{/*}{*/},
mathescape=true,
escapeinside={(*}{*)},
frame=none,
xleftmargin=0ex,
xrightmargin=0ex,
aboveskip=0ex,
belowskip=0ex,
basicstyle=\ttfamily\small,        
keywordstyle=\color{blue},       
}

\theoremstyle{plain}
\newtheorem{thm}{Theorem}[section]
\newtheorem{lem}[thm]{Lemma}
\newtheorem{prop}[thm]{Proposition}
\newtheorem{cor}[thm]{Corollary}
\theoremstyle{definition}
\newtheorem{defn}[thm]{Definition}
\newtheorem{exmp}[thm]{Example}

\theoremstyle{remark}
\newtheorem{rem}[thm]{Remark}

\newcommand{\catC}{\mathbb C}
\newcommand{\objZero}{\mathbf 0}
\newcommand{\objOne}{\mathbf 1}
\newcommand{\initMorph}{\raisebox{2pt}{\hbox{!`}}\xspace}
\newcommand{\initMorphIt}{\mathit{iniMor}\xspace}
\newcommand{\finMorph}{\hbox{!}\xspace}
\newcommand{\finMorphIt}{\mathit{finMor}\xspace}

\ifpdf
  \usepackage{underscore}         
  \usepackage[T1]{fontenc}        
\else
  \usepackage{breakurl}           
\fi

\title{Matching-Logic-Based Understanding of Polynomial Functors and their Initial/Final Models Revisited}
\author{Dorel Lucanu
\institute{Alexandru Ioan Cuza University of Iași, Romania}
\email{dorel.lucanu@gmail.com}
}
\def\titlerunning{Matching-Logic-Based Understanding of Polynomial Functors Revisited}
\def\authorrunning{D. Lucanu}

\begin{document}
\maketitle

\begin{abstract}
Many data types used in programming are initial algebras or final coalgebras of polynomial functors. We show how to specify them uniformly in matching logic, using the representation of polynomial functors as unary container functors $X\mapsto\sum_{a:A}X^{B[a]}$. For each container functor, we give a matching logic theory of its initial algebra and one of its final coalgebra. Besides the no-junk and no-confusion axioms and their duals, these theories include axioms about the elements that represent functions, for the initial algebra, and a bisimilarity axiom and a realisation axiom, for the final coalgebra; the existence axioms use free set variables to range over all subsets of a model. We prove that, in every model, the algebra extracted from the model is an initial algebra (a W-type) and the extracted coalgebra is a final coalgebra (an M-type) of the container extracted from the model, i.e., the one given by the model's interpretation of the shapes and positions, and that such models exist. So the data types are captured relative to the interpretation of the sorts of shapes and positions, and absolutely whenever these sorts are fixed by auxiliary specifications, which we show for finite sets, products, dependent sums, and (co)inductive sorts built from them. We derive the induction, coinduction, and bisimulation principles as theorems, and illustrate the approach on lists and on Moore machines.
\end{abstract}

\paragraph{Note on this version}
This paper is a revised version of~\cite{Lucanu23} (arXiv:\href{https://arxiv.org/abs/2309.13798v1}{2309.13798v1}), which had a WIP status at that time. Several axioms and proof principles were corrected, the adequacy and satisfiability results were proved, and the case studies were completed. Appendix~\ref{ap:changes} lists the changes.

\section{Introduction}
\label{sec:intro}

It is known that many data types used in programming are defined as initial algebra or final coalgebra  for an appropriate functor $F:\catC\to \catC$, where $\catC$ is a category of data types. In this paper we assume that $\catC$ is the category of sets (see, e.g.,~\cite{Adamek03}).
If $F$ is bicontinuous, i.e., it preserves the colimits of $\omega$-sequences and the limits of $\omega^\mathit{op}$-sequences, then the initial algebra (model) is obtained via colimit of the $\omega$-sequence
\[
\objZero \xrightarrow{\initMorph} F\;\objZero \xrightarrow{F\;\initMorph} F\;F\;\objZero=F^2\;\objZero \xrightarrow{F^2\;\initMorph} F^3\;\objZero \xrightarrow{F^3\;\initMorph} \cdots \tag{\sc Ini}\label{eq:ini}
\]
where $\objZero$ is the initial object in $\catC$, and $\objZero \xrightarrow{\initMorph} X$ is the unique arrow from the initial object~\cite{Adamek03}, and the final coalgebra (model) is the limit of the $\omega^\mathit{op}$-sequence
\[
\objOne \xleftarrow{\finMorph} F\;\objOne \xleftarrow{F\;\finMorph} F\;F\;\objOne=F^2\;\objOne \xleftarrow{F^2\;\finMorph} F^3\;\objOne \xleftarrow{F^3\;\finMorph} \cdots \tag{\sc Fin}\label{eq:fin}
\]
where $\objOne$ is the final object in $\catC$, and $\objOne \xleftarrow{\finMorph}  X$ is the unique arrow to the final object~\cite{Adamek03}.

This is a nice abstract framework, but, as we know, the evil is hidden in details.  How the elements of the initial and final models look like for various concrete functors? How could they be handled in practice?

A possible answer can be obtained by capturing these objects in Matching Logic (ML), the logical foundation of the K Framework, where the program languages and the properties of their programs can be specified in a uniform way (see, e.g., ~\cite{Ros17,CR19,Rosu020,CLR21a,LinCTWR23}). First steps are done in~\cite{CLR20a}, where the initial algebra semantics is captured in ML (see~\cite{ChenLR26} for the journal version, which also derives induction and iteration principles), and in~\cite{CLR21a}, where it is shown how examples of inductive/coinductive data types are fully specified in ML.
Data types are usually parametric: lists depend on a set of elements, and, more generally, the data types considered in this paper depend on a set of shapes and a family of sets of positions (Section~\ref{sec:pf}). We write $\mathit{DT}(P)$ for a data type with parameter sets $P$, which are represented in ML by parameter sorts.
We say that ML captures a (inductive/coinductive) data type $\mathit{DT}(P)$ if there is an ML theory $\mathsf{Th}^{\sf ML}(\mathit{DT})$ such that:
\begin{itemize}
\item from each $\mathsf{Th}^{\sf ML}(\mathit{DT})$-model $M$ we may extract a structure $\alpha(M)$ that is isomorphic to $\mathit{DT}(P_M)$, where $P_M$ is the interpretation of the parameter sorts in $M$, and
\item the deduction principles considered here for $\mathit{DT}$, namely induction for initial algebras, coinduction for final coalgebras, and the bisimulation principle for equality, can be expressed as theorems within ML using its proof system.
\end{itemize}
Other principles, such as iteration and primitive recursion, are outside the scope of this paper (see the Conclusion). In this paper, the first condition is established by the adequacy theorems, which are semantic statements about all models, and the second by derivations in the ML proof system; Section~\ref{sec:summary} classifies all the results in this way.
If, moreover, the theory captures the parameters themselves, i.e., $P_M\cong P$ in every model (e.g., for finite or inductively defined sets), then $\alpha(M)$ is isomorphic to $\mathit{DT}(P)$ in every model. Without such axioms, a model may interpret the parameter sorts differently: for a constant functor with one shape and no positions, a model with two shapes yields a two-element initial algebra.

In this paper, we investigate how data types specified as initial $F$-algebras or as final $F$-coalgebras, where $F$ is a polynomial functor, can be captured in ML. 
The term ``polynomial functor'' is used in the literature for two different classes of functors:
\begin{enumerate}
\item the \emph{classical polynomial functors} (see, e.g., \cite{rutten00}): the smallest class of functors that includes the constant and the identity functors and is closed under sum, product, and exponentiation by a constant object;
\item the \emph{(unary) container functors} (see, e.g., \cite{abbottAG05,altenkirchGHMM15}): the functors of the form $X\mapsto \sum_{a:A}X^{B[a]}$, where $a:A\vdash B[a]$ is an $A$-indexed family.
\end{enumerate}
Every classical polynomial functor is naturally isomorphic to a container functor (Section~\ref{sec:poly-as-cont}).
In this paper, \emph{polynomial functor} means polynomial \emph{in the container sense}, i.e., a unary container functor. In particular, everything we say applies to the classical polynomial functors.

The constant and the identity functors, together with their initial and final models, can be easily captured in ML. Moreover, if we exclude the exponent functor, then the initial algebra can be captured using the approach similar to that from~\cite{CLR20a}. The exponent functor complicates things.
A possible approach for the classical definition is as follows: supposing that we have captured $F_1\;X$ and $F_2\;X$ by the ML theories (specifications) $\mathrm{SPEC}(F_1\;X)$ and resp. $\mathrm{SPEC}(F_2\;X)$, then use these specifications to build $\mathrm{SPEC}(F_1\;X~\mathit{op}~F_2\;X)$, i.e., to obtain something like 
\[
\mathrm{SPEC}(F_1\;X~\mathit{op}~F_2\;X) = \mathrm{SPEC}(F_1\;X)~\overline{\mathit{op}}~\mathrm{SPEC}(F_2\;X)
\]
where $\overline{\mathit{op}}$ reflects $\mathit{op}$ to the level of ML specifications, and is yet to be defined.
In order to accomplish that, we need a ``uniform standard'' definition for the specifications $\mathrm{SPEC}(F\;X)$.

The container functors already have a uniform standard definition, and therefore they are more tempting for our investigation.
The results of this paper show that:
\begin{itemize}
\item it is possible to specify unary container functors, together with their initial algebras and final coalgebras, as well-formed ML theories (Definition~\ref{def:gamma}), which are satisfiable (Theorem~\ref{thm:standard-model}), for arbitrary unary containers, including the boundary cases of empty shape and position sets, empty exponent domains, infinite position sets, and empty initial algebras with non-empty final coalgebras (Section~\ref{sec:boundary});
\item the specification of the initial algebra captures it (a semantic result): for every model $M$, the algebra $\alpha_\mu(M)$ extracted from $M$ is an initial algebra, isomorphic to the W-type of the shapes and positions of $M$ (Theorem~\ref{thm:adequacy-ini}); such models exist (Proposition~\ref{prop:soundness-ini});
\item the specification of the final coalgebra, which includes an axiom identifying bisimilar elements and an axiom ensuring that every behaviour is realised, captures it (a semantic result): for every model $M$, the coalgebra $\alpha_\nu(M)$ extracted from $M$ is a final coalgebra, isomorphic to the M-type of the shapes and positions of $M$ (Theorem~\ref{thm:adequacy-fin}); such models exist (Proposition~\ref{prop:soundness-fin});
\item these results depend on how a model interprets the sorts of shapes and positions; when these sorts are given by finite sets, products, dependent sums, or (co)inductive sorts built from them, the isomorphisms hold with the intended shapes and positions in every model (Theorem~\ref{thm:captured}); for parameter sets that are meant to be arbitrary, the results are relative to their interpretation;
\item the induction principle, the coinduction principle, and the bisimulation principle are derivable in the ML proof system, from the Knaster--Tarski rule and the axioms;
\item the ML reasoning can be used to understand better the intimate structure of the initial algebra and the final coalgebra.
\end{itemize}
The approach is instantiated on the lists example, in order to see the relationship with the classical approach of these data types, and on that of Moore machines, in order to see the (constant) exponential functor at work.

The paper is structured as follows. Section~\ref{sec:pf} recalls the classical polynomial functors and the unary container functors, and shows that the former class is included in the latter.
Section~\ref{sec:ml} briefly recalls the main elements of matching logic, together with the theories of the equality and of the sorts.
Section~\ref{sec:fc-ml} includes the main contribution, showing how the unary container functors and their related concepts can be specified in matching logic. The instantiation of the general approach on the examples of lists and Moore machines are included in Section~\ref{sec:lists} and Section~\ref{sec:Moore}, respectively. The paper ends with some concluding remarks.

\section{Polynomial Functors and Container Functors}
\label{sec:pf}

This section briefly recalls (co)algebras of functors, the classical polynomial functors, and the container functors, and compares the two classes of functors.
We consider only the particular case of the functors defined over the category of sets $\catC$.

\begin{defn}\cite{jacobsRutten2012Introd}
Given a functor $F: \catC\to \catC$, an \emph{$F$-algebra} consists of an object $X$ in $\catC$ and an arrow $\alpha: F\;X\to X$.
An algebra morphism $(X,\alpha)\to (X',\alpha')$ is an arrow $h:X\to X'$ in $\catC$ such that $h\circ \alpha=\alpha'\circ F\;h$.
An \emph{initial algebra} for the functor $F$ is an initial object in the category of $F$-algebras and $F$-algebra
morphisms.
\end{defn}

\begin{exmp}[Lists]\label{ex:lists}
The lists over a set of elements $E$ can be defined as an $\Lst$-algebra $\alpha: \Lst\; X\to X$ for the functor $\Lst:\catC\to \catC$ given by $\Lst\; X = \objOne + E \times X$.
Usually, such an algebra is defined by a constant $\nil:\objOne\to X$ and a binary operation $\cons: E\times X\to X$. The initial $\Lst$-algebra is isomorphic to the finite lists inductively defined~\cite{sangiorgi-bisim-coind} by the grammar
\begin{align}
\List ::= nil \mid \cons(E, \List) \label{bnf:lists}\tag{\textsc{Lst}}
\end{align}
We use $\mu X\ld \Lst\;X$ or $\mu \Lst$ to denote the initial model of the functor $\Lst$. 
\end{exmp}

\begin{defn}\cite{jacobsRutten2012Introd}
Given a functor $F: \catC\to \catC$, an \emph{$F$-coalgebra} consists of an object $X$ in $\catC$ and an arrow $\gamma: X\to F\;X$.
A coalgebra morphism $(X,\gamma)\to (X',\gamma')$ is an arrow $h:X\to X'$ in $\catC$ such that $F\;h\circ \gamma=\gamma'\circ h$.
A \emph{final coalgebra} for the functor $F$ is a final object in the category of $F$-coalgebras and $F$-coalgebra
morphisms.
\end{defn}

\begin{exmp}[Colists]
The colists over a set of elements $E$ can be defined as an $\Lst$-coalgebra $\gamma: X\to \Lst\; X$, where $\Lst$ is the functor used for lists (see Example~\ref{ex:lists}).
Usually, such a coalgebra is defined by
 \begin{itemize}
\item  a total operation ${\_}\ld\!\! ?:X\to \{1,2\}$ such that $x\ld\!\! ? = \mathsf{if}~\gamma(x)=\iota_1(\star)~\mathsf{then}~1~\mathsf{else}~2~\mathsf{fi}$, where $\star$ is the unique element of $\objOne$, and 
\item two partial operations:  ${\_}\ld\hd: X\pto E$, and ${\_}\ld\tl: X\pto X$ such that $x\ld\hd=e$ and $x\ld\tl=x'$  iff $x\ld\!\! ?=2$ and $\gamma(x)=\iota_2(\pr{e,x'})$. 
\end{itemize}

The final coalgebra is isomorphic to the possible infinite lists coinductively defined by the grammar~\eqref{bnf:lists}~\cite{sangiorgi-bisim-coind}.
We use $\nu X\ld \Lst\;X$ or $\nu \Lst$ to denote the final model of the functor $\Lst$.
\end{exmp}

\begin{exmp}[Moore machines]\label{ex:moore}
A Moore machine is an $\Mo$-coalgebra $\alpha: X\to \Mo\;X$, where $\Mo$ is the functor $\Mo: X \mapsto O \times X ^I$, where $I$ is the set of inputs and $O$ is the set of outputs. Usually, a Moore machine is defined by an \emph{output} function $\mathit{out}: X\to O$ and a \emph{transition}  function $\mathit{tr}:X\to X^I$~\cite{rutten00}.
The final coalgebra $\nu X\ld \Mo\;X$ is isomorphic to $(\overline{\mathit{out}},\overline{\mathit{tr}}): O^{I^*}\to O\times (O^{I^*})^I$, $\overline{\mathit{out}}(f)=f(\varepsilon)$, $\overline{\mathit{tr}}(f)=g$ with $g(i)(w)=f(\langle i\rangle\cdot w)$ for $i\in I$ and $w\in I^*$, where $f$ is a function $f: I^*\to O$, $g$ a function $g:I \to O^{I^*}$, $\varepsilon$ the empty sequence, and $\cdot$ the concatenation of sequences.
\end{exmp}

\begin{defn}\cite{rutten00}\label{def:classical-poly}
The class of \emph{classical polynomial functors} is inductively defined as follows: 
\begin{itemize}
\item the constant functor $A$ (where $A$ is an object in $\catC$) is a classical polynomial functor; 
\item the identity functor $\Id$  is a classical polynomial functor; 
\item the sum $F_1+F_2$ of  two classical polynomial functors $F_1$ and $F_2$  is a classical polynomial functor; 
\item the product $F_1 \times F_2$ of  two classical polynomial functors $F_1$ and $F_2$  is a classical polynomial functor;  and 
\item the exponent $F^C$ of a classical polynomial functor $F$ by an object $C$, given by $F^C\,X=(F\,X)^C$, is a classical polynomial functor.
\end{itemize}
\end{defn}

\begin{rem}
Rutten~\cite{rutten00} defines the class as all functors that can be built from the constant functors, the identity functor, sum, product, and the function space functor $X\mapsto X^C$; building here includes composition. Definition~\ref{def:classical-poly} gives the same class: $X\mapsto X^C$ is $\Id^C$, and the class of Definition~\ref{def:classical-poly} is closed under composition, by induction on the outer functor $H$, since $A\circ F=A$, $\Id\circ F=F$, $(H_1+H_2)\circ F=H_1\circ F+H_2\circ F$, $(H_1\times H_2)\circ F=H_1\circ F\times H_2\circ F$, and $H^C\circ F=(H\circ F)^C$. Rutten also notes that including the function space functor is somewhat non-standard.
\end{rem}

The (unary) container functors \cite{abbottAG05,altenkirchM09,altenkirchGHMM15} form a second class of functors, which includes the classical polynomial functors.

\begin{defn}
An \emph{$A$-indexed family $a:A \vdash B[a]$} is a family of objects of $\catC$ indexed by elements of $A$. Categorically, it is an object $B$ of $\catC/A$ and $B[a]$ denotes the elements of $B$ mapped to $a$. 
\end{defn}

\begin{defn}
Given an $A$-indexed family $a:A \vdash B[a]$, the \emph{dependent product} $\prod_{a\cln A}B[a]$ is the object of the \emph{dependent functions}, which maps an $a\cln A$ into a $b\cln B[a]$. Set theoretically, we have
\[
\prod_{a\cln A}B[a]=\left\{f\in (\bigcup_{a\cln A}B[a])^A \middle\mid \forall a\cln A. f(a)\in B[a]\right\}  
\]
The \emph{dependent sum} $\sum_{a\cln A}B[a]$ is the dual of the dependent product and it consists of the pairs $\pr{a,b}$ with $a\cln A$ and $b\cln B[a]$.
\end{defn}

\begin{defn}
A functor $F : \catC \to \catC$ is a \emph{ (unary) container functor} iff it is naturally isomorphic
to a functor of the form $F\; X =  \sum_{a\cln A} X^{B[a]}$, for some objects $A$ in $\catC$ and an $A$-indexed family $a\cln A \vdash B[a]$.
\end{defn}

\begin{rem}
An element of $\sum_{a\cln A} X^{B[a]}$ is a pair $\pr{a, f}$, where $a \cln A$ is the \emph{shape} and $f: B[a] \to X$ is the function that labels the \emph{positions} $B[a]$ with elements from $X$. 
Another way to define $B$ is as an object in $\catC/A$.
\end{rem}

\subsection{Classical Polynomial Functors are Container Functors}
\label{sec:poly-as-cont}

Container functors are closed under sum, product, and exponentiation by a constant object~\cite{abbottAG05}. We recall these constructions below. Together with the container representations of the constant and the identity functors, they show, by induction on Definition~\ref{def:classical-poly}, that every classical polynomial functor is naturally isomorphic to a container functor.

\subsubsection*{Constant Functor} 
The main idea is to identify the constant value with the shapes $A$.
Consider $B$ as being $a\cln A\vdash \objZero$ (no positions of shape $a$),  where $\objZero$ denotes the initial object of $\catC$. 
We get  $\sum_{a\cln A} X^{B[a]} \approxeq \sum_{a\cln A} X^{\objZero} \approxeq A$.
\begin{rem}
The elements of $ \sum_{a\cln A} X^{\objZero}$ are pairs $\pr{a, f: \objZero\to X}$, where $a\in A$. Since $f: \objZero\to X$ is unique, we obtain $\pr{a, f: \objZero\to X}\approxeq a$.
\end{rem}

\subsubsection*{Identity Functor}
Consider $A$ as being $\objOne$ (just one shape) and $B$ as being $\star\cln \objOne\vdash\objOne$ (just one position), where $\star$ is the unique element in $\objOne$. It follows that
$\sum_{a\cln A} X^{B[a]} \approxeq \sum_\star X^\objOne\approxeq X$.
\begin{rem}
The elements of $\sum_\star X^\objOne$ are pairs $\pr{\star, f: \objOne\to X}$. Since $f$ selects just one element $x$ in $X$, it follows that $\pr{\star, f: \objOne\to X}\approxeq x$.
\end{rem}

\subsubsection*{Sum}
Assume that $F\;X = \sum_{a\cln A} X^{B[a]}$ and $F'\;X = \sum_{a'\cln A'} X^{B'[a']}$. Then $(F+F')\;X$ is  $\sum_{a''\cln A+A'} X^{[B,B'][a'']}$, where $[B,B'][a'']=B[a]$ if $a''=\iota_1\;a$, and $[B,B'][a'']=B'[a']$ if $a''=\iota_2\;a'$.
\begin{rem}
 The following commutative diagram may help to understand the definition of $a''\cln A+A'\vdash [B,B'][a'']$:
\begin{center}
\begin{tikzcd}
 & A+A'  & \\
A \arrow[ur,"\iota_1"] && A'  \arrow[ul, "\iota_2"']\\
B \arrow[u]  \arrow[r] & \hbox{$B + B'$} \arrow[uu]  & B' \arrow[u] \arrow[l]
 \end{tikzcd}
\end{center}
The arrow $B+B'\to A+A'$ is equivalently written as the $A+A'$-indexed set $a''\cln A+A'\vdash [B,B'][a'']$.
Set theoretically,  $\sum_{a''\cln A+A'} X^{[B,B'][a'']}$ is the set of pairs $\pr{a'', [f, f']: [B,B'][a''] \to X}$, where 
\begin{itemize}
\item either $a''=\iota_1\,a$, $\pr{a, f:B[a]\to X}$ in $F\;X$, and $\forall x\cln B[a]\ld [f, f']\; x= f\;x$, or
\item   $a''=\iota_2\,a'$, $\pr{a',f':B'[a']\to X}$ in $F'\;X$, and $\forall x'\cln B'[a']\ld [f, f']\; x'= f'\;x'$.
\end{itemize}
 In other words, the shapes of the sum are the disjoint union of the component shapes, and a shape selects one component: the positions and the labelling function of an element of the sum are those of the selected component, so $[f,f']$ is $f$ or $f'$. The object $B+B'$ in the diagram is the total object of positions over $A+A'$, not the position set of a single shape; position sets of a single shape are combined by disjoint union only in the product construction below. 
\end{rem}

\subsubsection*{Product}
With $F\;X$ and $F'\;X$ as for the sum, we have $(F\times F')\;X \approxeq  \sum_{\pr{a,a'}\cln A\times A'} X^{\langlebar B,B'\ranglebar[\pr{a,a'}]}$, where $\langlebar B, B'\ranglebar[\pr{a,a'}]=B[a] + B'[a']$ (the positions of the product are the disjoint union of the component positions); in contrast with the sum, where a shape selects the position set of one component, a shape $\pr{a,a'}$ of the product has the disjoint union of the position sets of both components. 
\begin{rem}
We write $\langlebar B, B'\ranglebar[a,a']$ for $\langlebar B, B'\ranglebar[\pr{a,a'}]$.
Set theoretically,  $\sum_{\pr{a,a'}\cln A\times A'} X^{\langlebar B, B'\ranglebar[a,a']}$ is the set of pairs $\pr{\pr{a,a'}, [f, f']: B[a]+B'[a'] \to X}$, where $f:B[a]\to X$ and $f':B'[a']\to X$. In other words, the shape of the product is the product of the component shapes and a labelling function for the product is the sum of two corresponding component labelling functions. 
\end{rem}

\subsubsection*{Exponentiation}
Assuming $F\;X = \sum_{a\cln A} X^{B[a]}$, the (constant) exponent functor $(F\;X)^C$ is 
$\sum_{g:C\to A} X^{\sum_{c\cln C}B[g\;c]}$.
\begin{rem}
An element of $\sum_{g:C\to A} X^{\sum_{c\cln C}B[g\;c]}$ is a pair $\pr{g, f}$ consisting of a function $g:C\to A$ assigning  shapes to $C$ and a $C$-indexed function $c\cln C\vdash f_c: B[g\;c]\to X$ labelling the positions $B[g\;c]$ for each $c$ in $C$.
\end{rem}

\section{Matching Logic (ML)}
\label{sec:ml}

 Matching logic \cite{Ros17,CR19,CLR21a} provides a unifying framework for defining semantics of programming languages. 
A programming language is defined in 
matching logic as a \emph{logical theory}, i.e., a set of axioms.
The key concept in matching logic is that of \emph{patterns},
which can be \emph{matched} by certain elements. 
By building complex patterns, we can match elements that have
complex structures or certain properties, or both. 
The presentation of matching logic in this review section 
follows \cite{CLR21a}. 

\begin{defn}
\label{def:AML-syntax}
Let us fix two sets $\EV$ and $\SV$.
The set $\EV$ includes \emph{element variables} $x,y,\dots$.
The set $\SV$ includes \emph{set variables} $X,Y,\dots$.
A \emph{matching logic signature} $\Sigma$ is a set of \emph{(constant) symbols}, denoted $\sigma,\sigma_1,\sigma_2,\dots$.
Let us fix a signature $\Sigma$.
The set of \emph{($\Sigma$-)patterns} is inductively defined as follows:
\begin{align*}\label{eq:MLsyntax}
\varphi ::= 
& \ x 
\mid X 
\mid \sigma
\mid \varphi_1 \  \varphi_2
\mid \bot
\mid \varphi_1 \limplies \varphi_2
\mid \exists x \ld \varphi
\mid \mu X \ld \varphi 
\end{align*}
where in $\mu X \ld \varphi$, called a least-fixpoint pattern,
we require that $\varphi$ is positive in $X$,
i.e., every free occurrence of $X$ in $\varphi$ lies within an even number of left-hand sides of implications $\varphi_1 \limplies \varphi_2$. 
\end{defn}

\begin{defn}
\label{def:AML-models}
A \emph{(matching logic) $\Sigma$-model} $M$ consists of
\begin{enumerate}
\item a nonempty \emph{carrier set}, which we also denote $M$; 
\item an \emph{application function} $\uapdu \colon M \times M \to 
\pset{M}$,   where $\pset{M}$ is the powerset of $M$; and
\item a \emph{symbol interpretation} $\sigma_M \subseteq M$ as a subset
     for $\sigma \in \Sigma$.
\end{enumerate}
\end{defn}

\begin{defn}
\label{def:AML-semantics}
Given $M$ and 
a \emph{variable valuation} $\rho \colon (\EV \cup \SV) \to  M 
\cup \pset{M}$ such that
$\rho(x) \in M$ for $x\in \EV$ 
and $\rho(X) \subseteq M$ for $X \in \SV$,
we inductively define \emph{pattern valuation} $\evMrho{\varphi}{M}$
as follows:
\begin{enumerate}
	\item $\evMrho{x}{M} = \{ \rho(x) \}$ for  $x \in \EV$ 
	\item $\evMrho{X}{M} = \rho(X)$ for  $X\in \SV$
	\item $\evMrho{\sigma}{M} = \sigma_M$ for $\sigma \in \Sigma$
	\item $\evMrho{\varphi_1 \, \varphi_2}{M} = 
	\bigcup_{a_i \in \evMrho{\varphi_i}{M},i \in \{1,2\}} a_1 \apd a_2$;
	note that $a_1 \apd a_2$ is a subset of $M$.
	\item $\evMrho{\bot}{M} = \emptyset$
	\item $\evMrho{\varphi_1 \limplies \varphi_2}{M} = M \setminus 
	(\evMrho{\varphi_1}{M} \setminus \evMrho{\varphi_2}{M})$
	\item $\evMrho{\exists x \ld \varphi}{M} = \bigcup_{a \in M} 
  \ev{\varphi}{M,\rho[a/x]}$
  \item $\evMrho{\mu X \ld \varphi}{M} = \mathbf{lfp} (A \mapsto \ev{\varphi}{M, \rho[A/X]})$
\end{enumerate}
where
$\rho[a/x]$ (resp. $\rho[A/X]$) is the valuation $\rho'$
with $\rho'(x) = a$ (resp. $\rho'(X) = A$) and
agreeing with $\rho$ on all the other variables, i.e., on $\EV \cup 
\SV \setminus \{x\}$ (resp. $\EV \cup \SV \setminus \{X\}$).
We use $\mathbf{lfp} (A \mapsto \ev{\varphi}{M, \rho[A/X]})$ to denote the 
smallest set $A$ such that $A = \ev{\varphi}{M, \rho[A/X]}$.
The existence of such an $A$ is guaranteed by the requirement
that $\varphi$ is positive in $X$.
We abbreviate $\ev{\varphi}{M,\rho}$ as $\ev{\varphi}{\rho}$
and further as $\ev{\varphi}{}$ if $\varphi$ is closed.
\end{defn}

\begin{defn}
\label{def:valid}
We say that $\varphi$ \emph{holds} in $M$, written
$M \vDash \varphi$,
if $\ev{\varphi}{M,\rho} = M$ for all $\rho$.
For a pattern set $\Gamma$,  
we write $M \vDash \Gamma$, if $M \vDash \psi$ for all $\psi \in \Gamma$.
We write $\Gamma \vDash \varphi$, if $M \vDash \Gamma$ implies
$M \vDash \varphi$  for all $M$. 
\end{defn}

The following common constructs can be defined from the basic pattern syntax
as syntactic sugar in the usual way:
$$
\begin{aligned}
 \neg \varphi &\equiv \varphi \limplies \bot \qquad
 &\varphi_1 \vee \varphi_2 &\equiv \neg \varphi_1 \limplies \varphi_2 \qquad
 &\varphi_1 \wedge \varphi_2 &\equiv \neg(\neg \varphi_1 \vee \neg \varphi_2) \\
 \top &\equiv \neg \bot 
 &\forall x \ld \varphi &\equiv \neg \exists x \ld \neg \varphi
 &\nu X \ld \varphi &\equiv \neg \mu X \ld \neg \varphi[\neg X / X]
\end{aligned}
$$
We assume the standard precedence between the above constructs.

\subsubsection*{Equality}
The equality can be defined as a derived construct (see~\cite{Ros17,CLR21a}).
The pattern $\varphi_1 = \varphi_2$ is equivalent to $\top$ if $\varphi_1$ and $\varphi_2$ are matched by the same elements, and to $\bot$ otherwise. To express that in ML,
a new symbol $\mathit{def} \in \Sigma$ is introduced, called the \emph{definedness} symbol,
and specify it with the axiom $\prule{Definedness}$. The resulted theory can be described as follows:
\begin{lstlisting}[language=matchinglogic, basicstyle=\linespread{1.2}\ttfamily\small]
spec $\DEF$
  (*\Smbs*)  $\smbceil$
  (*\Sugar*) $\ceil{\varphi} \equiv \smbceil \  \varphi$
  (*\Axms*)   $\prule{Definedness} \quad \forall x \ld \ceil{x} $
  (*\Sugar*) 
    $\floor{\varphi} \equiv \neg \ceil{\neg \varphi}
    $                 (*// totality*)
    $\varphi_1 = \varphi_2 \equiv \floor{\varphi_1 \lequiv \varphi_2}
    $                 (*// equality*)
    $\varphi_1 \subseteq \varphi_2 \equiv \floor{\varphi_1 \limplies \varphi_2}
    $                 (*// set inclusion*)
    $x \in \varphi \equiv x \subseteq \varphi
    $                 (*// membership*)
endspec
\end{lstlisting}
Membership is defined for an element variable $x$. We also write $\varphi\in\psi$ when $\varphi$ is a \emph{functional} pattern, i.e., one matched by exactly one element, such as a pair $\pr{b,f\;b}$ of defined components; then $\varphi\in\psi$ abbreviates $\exists z\ld \varphi=z\land z\in\psi$, where $z$ is fresh, which is equivalent to $\varphi\subseteq\psi$.

\subsubsection*{Sorts}
Matching logic has no builtin support for sorts.
Instead, we define a \emph{theory of sorts}
to support arbitrary sort structures
following the ``sort-as-predicate'' paradigm.
A \emph{sort} has a name and 
is associated with a set of its \emph{inhabitants}. 
In matching logic, we use a symbol $s \in \Sigma$ to represent the sort name
and use $(\mathsf{inh} \ s)$ to represent all its inhabitants,
where $\mathsf{inh} \in \Sigma$ is an ordinary symbol.
For better readability, we define
the notation $\inh{s} \equiv \mathsf{inh} \ s$. 
\begin{lstlisting}[language=matchinglogic, basicstyle=\linespread{1.1}\ttfamily\small]
spec $\INH$  (*\Impts*) $\DEF$
  (*\Smbs*) $\mathit{inh},\Sort$
  (*\Sugar*)
   $\inh{s} \equiv \mathit{inh} \  s
   $                              (*// inhabitants of sort $s$*) 
   $s_1\le s_2\equiv \inh{s_1} \subseteq \inh{s_2}
   $                              (*// subsort relation*)
   $\neg_s \varphi \equiv (\neg \varphi) \wedge \inh{s}
   $                              (*// negation within sort $s$*) 
   $\forall x \cln s \ld \varphi \equiv \forall x \ld x \in \inh{s} \limplies 
 \varphi
   $                              (*// $\forall$ within sort $s$*) 
   $\exists x \cln s \ld \varphi \equiv \exists x \ld x \in \inh{s} 
 \land 
 \varphi
   $                              (*// $\exists$ within sort $s$*) 
   $\mu X \cln s \ld \varphi \equiv \mu X \ld \inh{s} \land \varphi
   $                              (*// $\mu$ within sort $s$*)
   $\nu X \cln s \ld \varphi \equiv \nu X \ld \inh{s} \land \varphi
   $                              (*// $\nu$ within sort $s$*)
   $\varphi \cln s \equiv \exists z \cln s \ld \varphi = z
   $                              (*// \textrm{``}typing\textrm{''}*) 
   $f\cln s_1\otimes\cdots\otimes s_n\to s\equiv
 \forall x_1\cln s_1\ldots\forall x_n\cln s_n\ld \exists y\cln s\ld f\, x_1\ldots x_n = y$
                               (*// functional*) 
   $f\cln s_1\otimes\cdots\otimes s_n\pto s\equiv
 \forall x_1\cln s_1\ldots\forall x_n\cln s_n\ld (f\, x_1\ldots x_n = \bot) \lor \exists y\cln s\ld f\, x_1\ldots x_n = y$
                               (*// partially functional*)
  (*\Axms*) 
   $\exists x\ld \Sort = x$
   $\Sort\in\Sorts$
endspec
\end{lstlisting}
In $\mu X\cln s\ld\varphi$ and $\nu X\cln s\ld\varphi$, we require that $\varphi$ is positive in $X$ and that $X$ does not occur free in $s$. Note that $X\subseteq\inh{s}$ cannot be used as a conjunct of the body, since $X$ would occur negatively in it; intersecting the body with $\inh{s}$ keeps it positive in $X$, and all fixpoints are then included in $\inh{s}$.
The alternative $f\,x_1\ldots x_n=\bot$ in the notation for partially functional symbols makes it meaningful also when $s$ is empty; requiring $\exists y\cln s\ld f\,x_1\ldots x_n\subseteq y$ instead would be unsatisfiable in that case, e.g., for a partial destructor $f\cln s_1\pto s_2$ with $s_1$ non-empty and $s_2$ empty, such as the head of a list over an empty sort of elements (Section~\ref{sec:lists}).

The ML specifications for the sum, product, and function sorts are given in \cite{CLR21a}. For reader convenience, we recall that of the product sorts in Appendix~\ref{ap:sorts}; sum sorts and function sorts are not needed (Definition~\ref{def:gamma}). 

\section{Specifying Initial Algebra and Final Coalgebra for Container Functors in ML}
\label{sec:fc-ml}

In this section we show how to specify a unary container functor in ML and how to extract the structures for their initial algebra and the final coalgebra.
The theories used in the adequacy results are fixed by the following definition, which refers to the specifications given in this section and in Appendix~\ref{ap:sorts}. The remarks after it also refer forward to the parameter specifications of Definition~\ref{def:captured} (Section~\ref{sec:params}) and to the models constructed in Appendix~\ref{ap:model}.

\begin{defn}\label{def:gamma}
Let $A$ be a sort symbol, the sort of \emph{shapes}, and $B$ a family symbol, the family of \emph{positions}, and let $X\mapsto F\;X=\sum_{a\cln A}X^{B[a]}$ be the intended container functor; in a model $M$, the functor determined by the theories is the container functor $F_M$ extracted from $M$, defined in Section~\ref{sec:ini-ml} (Notation for models). The \emph{base theory} $\Gamma_0$ consists of $\DEF$ and $\INH$ (Section~\ref{sec:ml}), $\PAIR$ (Appendix~\ref{ap:sorts}, with \prule{Pair Total}), the specification of the initial morphism $\initMorph$ (Section~\ref{sec:elements}), and the axioms \axname{Shape Sort} and \axname{Position Sort} of the indexed family $a\cln A\vdash B[a]$ (Section~\ref{sec:families}). Then:
\begin{itemize}
\item $\Gamma^\mu_F$ is $\Gamma_0$ together with the axioms \axname{Mu Sort}, \axname{Functional}, \axname{No Confusion}, \axname{No Junk}, \axname{Ext}, and \axname{Tab} of Section~\ref{sec:ini-ml};
\item $\Gamma^\nu_F$ is $\Gamma_0$ together with the dependent sum $P\equiv\sum_{a\cln A}B[a]$ (Section~\ref{sec:depsum}), the axioms of $\mathit{List}(P)$, and the axioms \axname{Nu Sort}, \axname{Functional}, \axname{No Cojunk}, \axname{No Coconfusion.1}, \axname{No Coconfusion.2}, \axname{Bisimilarity}, and \axname{Realisation} of Section~\ref{sec:fin-ml};
\item $\Gamma_F=\Gamma^\mu_F\cup\Gamma^\nu_F$.
\end{itemize}
\end{defn}

The following remarks make the definition precise.
\begin{itemize}
\item \emph{Derived axioms.} \axname{No Confusion} and \axname{No Coconfusion.3} hold in every model of $\Gamma^\nu_F$ and are derivable from its axioms (Lemma~\ref{lem:derived}); they are not listed among the axioms of $\Gamma^\nu_F$, but we use them freely.
\item \emph{Signature.} The signature of $\Gamma_F$ consists of $\smbceil$, $\mathit{inh}$, and $\mathit{Sort}$ (of $\DEF$ and $\INH$); $\mathit{Pair}$, $\mathit{pair}$, $\mathit{fst}$, and $\mathit{snd}$ (of $\PAIR$); $\initMorphIt$; $A$, $B$, and $\mathit{depOf}$; $\mathit{DepSum}$, $\mathit{List}$, $\mathit{pnil}$, and $\pcons$; and $\cons$, $\mu F$, $\nu F$, $\out$, and $\nxt$. Despite the notation, $\mu F$ and $\nu F$ are constant symbols, the names of two sorts, and not fixpoint patterns; their inhabitants are given by the fixpoint patterns of \axname{No Junk} and \axname{No Cojunk}.
\item \emph{Instances.} $\PAIR$ is one set of axioms, quantified over all sorts; the specification of dependent sums is a scheme, instantiated in $\Gamma^\nu_F$ once, for $P$.
\item \emph{Free set variables.} The axioms \axname{Tab} and \axname{Realisation} contain free set variables, $R$ and $T$. Since an axiom holds in a model when it holds for all valuations, they speak about all subsets of the carrier. This is why the adequacy results are semantic (Section~\ref{sec:summary}).
\item \emph{Parameters.} The theories constrain the shapes and positions only by the axioms of the indexed family. For a concrete container, the theory used is $\Gamma_F\cup\Gamma_{\mathit{par}}$, where $\Gamma_{\mathit{par}}$ specifies $A$ and $B$, e.g., by the standard parameter specifications of Definition~\ref{def:captured}; the results of this section are stated for $\Gamma_F$, so they hold for every such extension. The specifications of $\objZero$, $\star$, $\objOne$, and $\finMorph$ in Section~\ref{sec:elements} are not part of $\Gamma_F$: $\objZero$, $\star$, and $\objOne$ occur in parameter specifications, e.g., for lists (Section~\ref{sec:lists}), and $\finMorph$ only in the categorical constructions of Sections~\ref{sec:intro} and~\ref{sec:pf}.
\item \emph{Sums and functions.} The sum sorts and function sorts of~\cite{CLR21a} are not part of these theories, and dependent products (Section~\ref{sec:pf}) are used only at the level of sets, not specified as sorts. A finite sum of sorts, such as the shapes $\objOne+E$ of lists, is represented as a dependent sum over a finite sort of tags, $\sum_{i\cln\mathit{Tag}}C[i]$, by the rules (Fin), (Fam), and (Sum) of Definition~\ref{def:captured}. This representation covers sums with any finite number of summands, the positions are then defined by cases on the tags, and its properties follow from \prule{Pair Inj}, which $\Gamma_0$ already contains (Section~\ref{sec:lists}). So the sum sorts $s_1\oplus s_2$ of~\cite{CLR21a}, with their injections and partial ejections, are not needed, and we do not recall their specification. Functions on positions are represented by the notation $X^{B[a]}$ (Section~\ref{sec:cont-ml}), whose codomain is a set variable and whose domain may be empty, with the axioms \axname{Ext} and \axname{Tab}; so the function sorts of~\cite{CLR21a} are not needed either.
\end{itemize}

By Theorem~\ref{thm:standard-model} in Appendix~\ref{ap:model}, $\Gamma_F$ has a model, so $\Gamma_F$, $\Gamma^\mu_F$, and $\Gamma^\nu_F$ are consistent.

\paragraph{Typing of the symbols}
The axioms of these theories give each symbol the following typing; in each case, the application is defined (matched by exactly one element) on the indicated arguments.
\begin{center}\small
\begin{tabular}{lll}
\hline
Symbol & Typing & Guaranteed by\\
\hline
$\mathit{depOf}$ & $B$, $a\cln A$ $\mapsto$ a sort $B[a]$ & indexed family\\
$\mathit{DepSum}$ & $A$, $B$ $\mapsto$ the sort $P$ & dependent sum\\
$\mathit{pair}$ & $x$, $y$ $\mapsto$ $\pr{x,y}$; $s_1\otimes s_2$ on sorted arguments & \prule{Pair Total}, $\PAIR$\\
$f\;b$ & $f\in X^{B[a]}$, $b\cln B[a]$ $\mapsto$ an element of $X$ & the notation $X^{B[a]}$\\
 & ($\bot$ for $b$ outside $B[a]$) & \\
$\cons$ & $\pr{a,f}$, $a\cln A$, $f\in\top^{B[a]}$ $\mapsto$ an element; in $\inh{\mu F}$ & \axname{Functional};\\
 & (resp.\ $\inh{\nu F}$) when the values of $f$ are in & \axname{No Junk} (resp.\ \axname{No Cojunk})\\
 & $\inh{\mu F}$ (resp.\ $\inh{\nu F}$) & \\
$\out$, $\nxt$ & $x\cln\nu F\mapsto$ $x.\out\cln A$ and $x.\nxt\in\top^{B[x.\out]}$ & \axname{No Coconfusion.1--2}, Lemma~\ref{lem:derived}\\
$\mathit{List}$ & $P$ $\mapsto$ the sort $\mathit{List}(P)$ & \axname{List Sort}\\
$\mathit{pnil}$, $\pcons$ & constructors of $\mathit{List}(P)$ & axioms of $\mathit{List}(P)$\\
$\initMorph$ & $x\mapsto\bot$; the unique function on an empty domain & Section~\ref{sec:elements}\\
\hline
\end{tabular}
\end{center}
The destructors are axiomatized only on $\nu F$; the theory $\Gamma^\mu_F$ does not use them.

\subsection{Capturing Elements from the Category $\catC$}
\label{sec:elements}

Of the specifications below, only that of $\initMorph$ is part of the theories of Definition~\ref{def:gamma}; $\objZero$, $\star$, and $\objOne$ are used in parameter specifications, where they are the instances $n=0$ and $n=1$ of (Fin) in Definition~\ref{def:captured}, and $\finMorph$ only in the categorical constructions of Sections~\ref{sec:intro} and~\ref{sec:pf}.
\begin{itemize}
\item the empty set $\objZero$ is specified by
\begin{itemize}
\item a sort $\objZero$; and
\item the axiom\\
$~~~~\begin{aligned}
&\objZero\cln\Sort\land\inh{\objZero}=\bot && \axname{captures~$\objZero$}
\end{aligned}$
\end{itemize}
so $\objZero$ is a sort without inhabitants; the symbol $\objZero$ denotes the sort name, not $\bot$, which allows equations between sorts such as $B[a]=\objZero$ (Definition~\ref{def:captured});
\item $\objOne=\{\star\}$ is specified by
\begin{itemize}
\item  a symbol $\mathit{star}\in \Sigma$; 
\item a notation: $\star \equiv \mathit{star}$; 
\item a sort $\objOne$; and
\item the axioms\\
$~~~~\begin{aligned}
&\exists y\ld \star = y  && \axname{singleton}\\
&\objOne\cln\Sort\land\inh{\objOne}=\star && \axname{captures~$\objOne$}
\end{aligned}$
\end{itemize}
\item $\objZero\xrightarrow{\initMorph} X$ is specified by
\begin{itemize}
\item  a symbol $\initMorphIt\in \Sigma$; 
\item a notation: $\initMorph \equiv \initMorphIt$; and
\item two axioms\\
$~~~~\begin{aligned}
&\exists y\ld \initMorph = y && \axname{Ini Functional}\\
&\forall x \ld \initMorph\; x = \bot  && \axname{\textrm{captures~}$\objZero\xrightarrow{\initMorph} X$}\\
\end{aligned}$
\end{itemize}
\item $\objOne\xleftarrow{\finMorph} X$ is specified by
\begin{itemize}
\item  a symbol $\finMorphIt\in \Sigma$; 
\item a notation: $\finMorph \equiv \finMorphIt$; and
\item an axiom\\
$~~~~\begin{aligned}
&\forall x \ld \finMorph\; x = \star  && \axname{\textrm{captures~}$\objOne\xleftarrow{\finMorph} X$}\\
\end{aligned}$
\end{itemize}
\end{itemize}

\subsection{Expressing Indexed Families in ML}
\label{sec:families}

An $A$-indexed family $a\cln A \vdash B[a]$ is specified by a constant symbol $\mathit{depOf}\in\Sigma$, a notation $B[a]\equiv \mathit{depOf}\;B\;a$, and two axioms of the same kind:\\
$~~~~\begin{aligned}
&A\cln \Sort && \axname{Shape Sort}\\
&\forall a \cln A\ld B[a]\cln\Sort && \axname{Position Sort}
\end{aligned}$\\
Both use the typing notation $\varphi\cln\Sort\equiv\exists z\cln\Sort\ld\varphi=z$ of Section~\ref{sec:ml}: the first states that $A$ denotes one element, which is a sort, and the second that, for each $a\cln A$, $B[a]$ denotes one element, which is a sort.

\subsection{Expressing Dependent Sums in ML}
\label{sec:depsum}

A dependent sum $\sum_{a\cln A}B[a]$ is specified by a constant $\mathit{DepSum}\in \Sigma$, a notation $\sum_{a\cln A}B[a] \equiv \mathit{DepSum}\;A\;B$ ($a$ plays a local role and its name can be changed\footnote{Actually, $\Sigma$ should be captured as a binder~\cite{ChenR20}, but this is not needed for the purpose of this paper.}), and by adding the axioms:
\begin{itemize}
\item $\sum_{a\cln A}B[a]$ is a sort: $\sum_{a\cln A}B[a]\cln\Sort$ 
\item $\inh{\sum_{a\cln A}B[a]}$ is the set of dependent pairs: \[\inh{\sum_{a\cln A}B[a]}=\exists a\cln A\ld \exists b\cln B[a]\ld \pr{a,b}\]
\end{itemize}

\subsection{Expressing Unary Container Functors in ML}
\label{sec:cont-ml}

Let $X\mapsto F\;X= \sum_{a\cln A}X^{B[a]}$ be a container functor.
Recall that $F\;X$ is the set of pairs $\pr{a,f}$ with $a\cln A$ and $f\cln X^{B[a]}$ (or, equivalently, $f\cln B[a]\to X$).
We specify $X^{B[a]}$ explicitly by the notation
\[
\begin{aligned}
X^{B[a]} \equiv \exists f\ld f \land \big(&((\inh{B[a]}\not=\bot)\land(\forall b\cln B[a]\ld \exists x\ld f\;b = x \land x\in X)\land(\forall b'\ld\neg(b'\in\inh{B[a]})\limplies f\;b'=\bot))\\
&\lor ((\inh{B[a]}=\bot)\land f=\initMorph)\big)
\end{aligned}
\]
and use it directly in the specification of $\sum_{a\cln A}X^{B[a]}$:
\[
\exists a\cln A\ld \exists f\ld \pr{a, f} \land f\in X^{B[a]}
\]
which is equivalent to
\[
\begin{aligned}
\exists a\cln A\ld \exists f\ld \pr{a, f} \land \big(&((\inh{B[a]}\not=\bot)\land (\forall b\cln B[a]\ld \exists x\ld f\;b = x \land x\in X)\land(\forall b'\ld\neg(b'\in\inh{B[a]})\limplies f\;b'=\bot))\\
&\lor ((\inh{B[a]}=\bot)\land f=\initMorph)\big)
\end{aligned}
\]
The notation $X^{B[a]}$ is a pattern, not a sort: it is matched by the elements that behave as strict functions from $\inh{B[a]}$ to $X$, where $X$ may be any pattern, in particular a set variable, and $\inh{B[a]}$ may be empty. This is what the fixpoint axioms of Sections~\ref{sec:ini-ml} and~\ref{sec:fin-ml} require, so we do not use the function sorts of~\cite{CLR21a}, which relate two sorts.
Recall that if $\inh{B[a]}=\bot$ ($B[a]\approxeq\objZero$) then there is just one function $\objZero\xrightarrow{\initMorph} X$ specified by $\initMorph \equiv \initMorphIt$.
The last conjunct of the first case makes the representation \emph{strict}: $f$ is undefined outside $B[a]$. Without it, if $B[a]\subsetneq B[a']$ in some model, an element representing a function on $B[a']$ would also represent its restriction to $B[a]$, and the axiom \axname{Ext} of Section~\ref{sec:ini-ml} would identify elements that represent different functions on $B[a']$.

\subsection{Specifying Initial Algebra}
\label{sec:ini-ml}

Let $X\mapsto F\;X= \sum_{a\cln A}X^{B[a]}$ be a container functor.
Lemma~\ref{lem:continuity} below and the description of $\inh{\mu F}$ by approximations assume that $F$ is \emph{finitary}, i.e., every position set $B[a]$ is finite; the adequacy result (Theorem~\ref{thm:adequacy-ini}) does not.
We write $\top^{B[a]}$ for the pattern $X^{B[a]}[\top/X]$, which is matched by the elements that behave as functions from $B[a]$ to arbitrary elements.
The initial $F$-algebra is specified by using the characterization given by the ``no junk and no confusion'' properties for the constructors~\cite{CLR21a}:
\begin{itemize}
\item a constructor $\cons\in\Sigma$ specified by the following axioms:\\
$\begin{aligned}
&\forall a\cln A\ld \forall f\ld f\in \top^{B[a]}\limplies \exists x\ld \cons\;\pr{a, f} = x  && \axname{Functional}\\
&\forall a,a'\cln A\ld \forall f, f'\ld f\in \top^{B[a]}\land f'\in \top^{B[a']}\land{}\\
&\qquad\cons\;\pr{a, f} = \cons\;\pr{a',f'} \limplies a = a' \land f = f'&& \axname{No Confusion}
\end{aligned}$
\item a sort $\mu F$ with initial semantics:\\
$\begin{aligned}
&\mu F\cln\Sort && \axname{Mu Sort}\\
&\inh{\mu F} = \mu X\ld \exists a\cln A\ld  \cons\;\pr{a,X^{B[a]}}&& \axname{No Junk}
\end{aligned}$
\end{itemize}

\paragraph{Notation for models}
Fix a model $M$ of $\Gamma^\mu_F$ (or of $\Gamma^\nu_F$, in Section~\ref{sec:fin-ml}) and a valuation $\rho$; we write $x\apd y$ for application in $M$.
Let $A_M=\ev{\inh{A}}{M}$ and, for $a\in A_M$, let $B_M[a]$ be the interpretation of $\inh{B[a]}$ when the variable $a$ is valued to $a$. The \emph{container extracted from $M$} is the set-theoretic container $a\cln A_M\vdash B_M[a]$, i.e., the container given by the interpretation in $M$ of the shapes and positions, and the \emph{container functor extracted from $M$} is its functor
\[
F_M\,X=\sum_{a\in A_M}X^{B_M[a]} .
\]
It may differ from the intended functor $F$, since the theories do not fix the interpretation of $A$ and $B$ (Section~\ref{sec:params}). We say that $F$ is \emph{finitary in $M$} if every $B_M[a]$ is finite.
For $S\subseteq M$ and $a\in A_M$, let $\mathit{Fun}_a(S)$ be the interpretation of $X^{B[a]}$ when $X$ is valued to $S$ and $a$ to $a$.
If $B_M[a]=\emptyset$, then $\mathit{Fun}_a(S)=\{\initMorph_M\}$, where $\initMorph_M$ is the unique element matching $\initMorph$.
Otherwise, $f\in\mathit{Fun}_a(S)$ iff for every $b\in B_M[a]$ there is $s_b\in S$ with $f\apd b=\{s_b\}$, and $f\apd b=\emptyset$ for every $b\in M\setminus B_M[a]$; we then write $\abs{f}$ for the function $B_M[a]\to S$, $b\mapsto s_b$, and call $f$ a \emph{code} of $\abs{f}$. For $B_M[a]=\emptyset$, the only element of $\mathit{Fun}_a(S)$ is $f=\initMorph_M$, and $\abs{f}$ is the empty function.
Note that $f\in\mathit{Fun}_a(S)$ iff $f\in\mathit{Fun}_a(M)$ and all values of $\abs{f}$ are in $S$.
By \axname{Functional}, for $a\in A_M$ and $f\in\mathit{Fun}_a(M)$ the pattern $\cons\;\pr{a,f}$ is matched by exactly one element, which we denote $c(a,f)$; by \axname{No Confusion}, for $f\in\mathit{Fun}_a(M)$ and $f'\in\mathit{Fun}_{a'}(M)$, $c(a,f)=c(a',f')$ implies $a=a'$ and $f=f'$.

Table~\ref{tab:notation} summarizes this notation. Note the difference between $\inh{B[a]}$ (subscript), the positions, and $\top^{B[a]}$ (superscript), the elements representing functions on the positions.
\begin{table}[t]
\centering
\small
\begin{tabular}{>{\raggedright}p{0.17\textwidth}>{\raggedright}p{0.12\textwidth}>{\raggedright\arraybackslash}p{0.62\textwidth}}
\hline
Notation & Kind & Meaning\\
\hline
$B[a]$ & sort & the sort of positions of the shape $a$, an element of $\mathit{Sort}$\\
$\inh{B[a]}$ & pattern & its inhabitants, i.e., the positions of $a$\\
$B_M[a]$ & set & the interpretation of $\inh{B[a]}$ in a model $M$\\
$X^{B[a]}$ & pattern & the elements representing functions from $\inh{B[a]}$ to $X$, undefined outside $\inh{B[a]}$; $\initMorph$ if $\inh{B[a]}=\bot$ (Section~\ref{sec:cont-ml})\\
$\top^{B[a]}$ & pattern & $X^{B[a]}[\top/X]$: the elements representing functions from $\inh{B[a]}$ to arbitrary elements\\
$\mathit{Fun}_a(S)$ & set & the interpretation of $X^{B[a]}$ when $X$ is valued to $S$; its elements are the codes\\
$\abs{f}$ & function & the function $B_M[a]\to S$ represented by a code $f\in\mathit{Fun}_a(S)$\\
$\ulcorner g\urcorner_a$ & element & the unique code of $g:B_M[a]\to L$ in $\mathit{Fun}_a(L)$ (Lemma~\ref{lem:codes})\\
$c(a,f)$ & element & the unique element matching $\cons\;\pr{a,f}$, for $f\in\mathit{Fun}_a(M)$\\
\hline
\end{tabular}
\caption{Notation for positions and for elements representing functions.}
\label{tab:notation}
\end{table}

\paragraph{Computing the least fixpoint}
Let $\varphi_F(X)$ denote the pattern $\exists a\cln A\ld  \cons\;\pr{a, X^{B[a]}}$, and let $\Phi_F:\pset{M}\to\pset{M}$ be the function $S\mapsto\ev{\varphi_F(X)}{M, \rho[S/X]}$. By the definitions above,
\[
\Phi_F(S)=\{c(a,f)\mid a\in A_M,\ f\in\mathit{Fun}_a(S)\},
\]
and, by \axname{No Junk}, $\ev{\inh{\mu F}}{M}=\mathbf{lfp}(\Phi_F)$.

\begin{lem}\label{lem:continuity}
If $F$ is finitary in $M$, then $\Phi_F$ preserves unions of increasing chains. Therefore $\ev{\inh{\mu F}}{M}=\mathbf{lfp}(\Phi_F)=\bigcup_{n\ge 0}\Phi^n_F(\emptyset)$.
\end{lem}
\begin{proof}
$\Phi_F$ is monotone, because $\varphi_F$ is positive in $X$. Let $S_0\subseteq S_1\subseteq\cdots$ and $S=\bigcup_n S_n$. If $f\in\mathit{Fun}_a(S)$, then $\abs{f}$ has the finite domain $B_M[a]$, so all its values lie in some $S_N$, and $f\in\mathit{Fun}_a(S_N)$. Hence $\Phi_F(S)=\bigcup_n\Phi_F(S_n)$. The second claim is Kleene's fixpoint theorem.
\end{proof}

Since $\Phi_F(\emptyset)=\ev{\varphi_F(\bot)}{M, \rho}$ and writing $\varphi(\psi)$ for $\varphi[\psi/X]$, we (informally) obtain that $\mathbf{lfp}(\Phi_F)$ is the interpretation of the 
infinite disjunction
\[
\bot \lor \varphi_F(\bot) \lor \varphi^2_F(\bot) \lor \varphi^3_F(\bot) \lor \cdots \tag{\textsc{Lfp}}\label{eq:ini-ml}
\]
according to $M$ and $\rho$. We have $\varphi^n_F(\bot)\limplies \varphi^{n+1}_F(\bot)$, and each $\varphi^n_F(\bot)$ gives an approximation of $\inh{\mu F}$. So,  in order to understand how the elements of the initial algebra look like, we have to investigate these ML patterns.
\begin{description}
\item[ $ \varphi_F(\bot)$.] The pattern $\bot^{B[a]}$ (i.e., $X^{B[a]}[\bot/X]$) is $\initMorph$ if $\inh{B[a]}=\bot$, and $\bot$ otherwise, because $\forall b\cln B[a]\ld \exists y\ld f\;b = y \land y\in \bot$ is $\bot$ when $B[a]$ is non-empty.
We obtain 
\[ 
  \varphi_F(\bot) = \exists a_1\cln A\ld \cons\pr{a_1,\initMorph}\land (\inh{{B[a_1]}}=\bot)
\]
i.e., each $a_1\cln A$, with its corresponding dependent sort $B[a_1]$ empty, defines a constant constructor.
\\
If $\forall a\cln A\ld \inh{B[a]}\not =\bot$, then $\varphi_F(\bot)=\bot$, hence $\varphi^n_F(\bot)=\bot$ for all $n$, and $\inh{\mu F}=\bot$ (see also Remark~\ref{rem:ini-hyp}).
\item[$ \varphi^2_F(\bot)=\exists a_2\cln A\ld  \cons\;\pr{a_2, \varphi_F(\bot)^{B[a_2]}}$.] 
If $\inh{B[a_2]}=\bot$, then $\varphi_F(\bot)^{B[a_2]}=\bot ^{B[a_2]}=\initMorph$, which gives again the constant constructor $\cons\;\pr{a_2,\initMorph}$.
If $\inh{B[a_2]}\not=\bot$, then we have 
\[
\begin{aligned}\varphi_F(\bot)^{B[a_2]}=\exists f\ld f&\land (\forall b\cln B[a_2]\ld \exists a_1\cln A\ld f\;b=\cons\pr{a_1,\initMorph}\land (\inh{{B[a_1]}}=\bot))\\&\land(\forall b'\ld\neg(b'\in\inh{B[a_2]})\limplies f\;b'=\bot)\end{aligned}
\] 
We obtain
\[
\varphi^2_F(\bot) = \varphi_F(\bot) \lor \exists a_2\cln A\ld  \cons\;\pr{a_2, \varphi_F(\bot)^{B[a_2]}} \land (\inh{B[a_2]}\not=\bot)
\]
\item[\ldots]
\end{description}

\paragraph{Axioms for codes}
The axioms above do not say which elements of $M$ represent functions. The theory of the initial algebra also includes the following two axioms, where $(\inh{\mu F})^{B[a]}$ denotes the pattern $X^{B[a]}[\inh{\mu F}/X]$:\\
$\begin{aligned}
&\forall a\cln A\ld \forall f,f'\ld f\in(\inh{\mu F})^{B[a]}\land f'\in(\inh{\mu F})^{B[a]}\land(\forall b\cln B[a]\ld f\;b=f'\;b)\limplies f=f' && \axname{Ext}\\
&\forall a\cln A\ld (\forall b\cln B[a]\ld \exists x\cln\mu F\ld \pr{b,x}\in R\land\forall x'\ld \pr{b,x'}\in R\limplies x'=x)\limplies{}\\
&\qquad\exists f\ld f\in(\inh{\mu F})^{B[a]}\land\forall b\cln B[a]\ld \pr{b,f\;b}\in R && \axname{Tab}
\end{aligned}$\\
In \axname{Tab}, $R$ is a free set variable. Since an axiom holds in $M$ when it holds for all valuations, \axname{Tab} speaks about every subset $R$ of $M$, in particular about the graph of every function from $B_M[a]$ to $\inh{\mu F}$.
Let $L=\ev{\inh{\mu F}}{M}$.

\begin{lem}\label{lem:codes}
Let $M$ be a model of $\Gamma^\mu_F$. Then:
\begin{enumerate}
\item for all $a\in A_M$ and $f,f'\in\mathit{Fun}_a(L)$, if $\abs{f}=\abs{f'}$ then $f=f'$;
\item for all $a\in A_M$ and every function $g:B_M[a]\to L$, there is $f\in\mathit{Fun}_a(L)$ with $\abs{f}=g$.
\end{enumerate}
\end{lem}
\begin{proof}
(1) is the semantics of \axname{Ext}. For (2), value $R$ to $\{\pr{b,g(b)}\mid b\in B_M[a]\}$; these pairs exist, and are pairwise distinct, by the axioms of $\PAIR$ for the sorts $B[a]$ and $\mu F$. The premise of \axname{Tab} holds, so there is $f\in\mathit{Fun}_a(L)$ with $\pr{b,f\apd b}\in R$, hence $f\apd b=\{g(b)\}$, for all $b\in B_M[a]$.
\end{proof}

So every function $g:B_M[a]\to L$ has exactly one code in $\mathit{Fun}_a(L)$, which we denote $\ulcorner g\urcorner_a$.

\paragraph{Adequacy}
Let $F_M$ be the container functor extracted from $M$, and let $\mathit{sup}:F_M\,W_M\to W_M$ be its initial algebra, i.e., the W-type of $a\cln A_M\vdash B_M[a]$~\cite{abbottAG05}, whose elements are the well-founded trees with nodes labelled by shapes $a$ and branching $B_M[a]$. The \emph{algebra extracted from $M$} is
\[
\alpha_\mu(M)=(L,\alpha),\qquad \alpha:F_M\,L\to L,\quad \alpha(a,g)=c(a,\ulcorner g\urcorner_a).
\]
If $F$ is finitary in $M$, every tree $w\in W_M$ has a finite \emph{height}: $\mathit{ht}(\mathit{sup}(a,t))=1+\max_{b\in B_M[a]}\mathit{ht}(t\,b)$, with $\max\emptyset=0$. Let $W_n$ be the set of trees of height at most $n$.

\begin{thm}\label{thm:adequacy-ini}
For every model $M$ of $\Gamma^\mu_F$, $\alpha_\mu(M)$ is an initial $F_M$-algebra: the unique algebra morphism $h:W_M\to L$ is a bijection. If, moreover, $F$ is finitary in $M$, then $h(W_n)=\Phi^n_F(\emptyset)$ for all $n$.
\end{thm}
\begin{proof}
Since $L=\mathbf{lfp}(\Phi_F)$ is a fixpoint of $\Phi_F$, we have $\alpha(a,g)\in\Phi_F(L)=L$, so $(L,\alpha)$ is an $F_M$-algebra. Initiality of $W_M$ gives a unique $h$ with $h(\mathit{sup}(a,t))=c(a,\ulcorner h\circ t\urcorner_a)$.

\emph{$h$ is injective.} The subtree relation on $W_M$ is well-founded. We show by well-founded induction on $(w,w')$ that $h(w)=h(w')$ implies $w=w'$. Let $w=\mathit{sup}(a,t)$ and $w'=\mathit{sup}(a',t')$. From $c(a,\ulcorner h\circ t\urcorner_a)=c(a',\ulcorner h\circ t'\urcorner_{a'})$, \axname{No Confusion} gives $a=a'$ and $\ulcorner h\circ t\urcorner_a=\ulcorner h\circ t'\urcorner_a$, hence $h\circ t=h\circ t'$. By the induction hypothesis, $t\,b=t'\,b$ for all $b$, and $w=w'$.

\emph{$h$ is surjective.} The least fixpoint is the union of the transfinite sequence
\[
\begin{aligned}
\Phi^0_F(\emptyset)&=\emptyset,\\
\Phi^{\xi+1}_F(\emptyset)&=\Phi_F(\Phi^{\xi}_F(\emptyset)),\\
\Phi^{\lambda}_F(\emptyset)&=\textstyle\bigcup_{\xi<\lambda}\Phi^{\xi}_F(\emptyset)\qquad\text{for a limit ordinal }\lambda.
\end{aligned}
\] We show $\Phi^{\xi}_F(\emptyset)\subseteq h(W_M)$ by induction on $\xi$; the limit case is immediate. Let $x\in\Phi^{\xi+1}_F(\emptyset)$, i.e., $x=c(a,f)$ with $f\in\mathit{Fun}_a(\Phi^\xi_F(\emptyset))\subseteq\mathit{Fun}_a(L)$. By the induction hypothesis and injectivity, for each $b\in B_M[a]$ there is a unique $t\,b\in W_M$ with $h(t\,b)=\abs{f}(b)$. Then $\abs{f}=h\circ t$, so $f=\ulcorner h\circ t\urcorner_a$ by Lemma~\ref{lem:codes}, and $x=h(\mathit{sup}(a,t))$.

A bijective algebra morphism is an isomorphism of algebras, so $(L,\alpha)$ is initial.

If $F$ is finitary in $M$, then $h(W_n)=\Phi^n_F(\emptyset)$ by induction on $n$: for $w=\mathit{sup}(a,t)\in W_{n+1}$, every $t\,b$ is in $W_n$, so $\ulcorner h\circ t\urcorner_a\in\mathit{Fun}_a(\Phi^n_F(\emptyset))$ and $h(w)\in\Phi^{n+1}_F(\emptyset)$; the converse inclusion is the successor step above, where the trees $t\,b$ are in $W_n$.
\end{proof}

\begin{cor}
If, moreover, $M$ interprets shapes and positions standardly, i.e., there are a bijection $\iota:A\to A_M$ and bijections $B[a]\cong B_M[\iota(a)]$ for all $a\in A$, then $\alpha_\mu(M)$ is isomorphic to the initial $F$-algebra $\mu F$, the W-type of $a\cln A\vdash B[a]$.
\end{cor}

\begin{prop}\label{prop:soundness-ini}
Let $X\mapsto F\;X=\sum_{a\cln A}X^{B[a]}$ be a container functor, specified as above. There is a model $M$ of $\Gamma_F$ (hence of $\Gamma^\mu_F$) with a bijection $\iota:A\to A_M$ and bijections $B[a]\cong B_M[\iota(a)]$ for all $a\in A$, in which $\inh{\mu F}$ with $\cons$ is isomorphic to the W-type of $a\cln A\vdash B[a]$.
\end{prop}
\begin{proof}
The model is constructed in Appendix~\ref{ap:model} (Theorem~\ref{thm:standard-model}). Its carrier is built in transfinite stages, up to a regular cardinal larger than every $\abs{B[a]}$, by closing under pairs, finite sequences, partial applications, codes of functions on position sets, and fresh terms; there, $\inh{\mu F}$ is the set of well-founded trees of the M-type, and the only elements that represent functions into it are the codes.
\end{proof}

\begin{rem}\label{rem:ini-hyp}
The axioms \axname{Ext} and \axname{Tab} are needed for Theorem~\ref{thm:adequacy-ini}; Proposition~\ref{prop:necessity}(1) and~(2) gives, for each of them, a structure that satisfies all the other axioms of $\Gamma^\mu_F$ and whose extracted algebra is not initial.
Without \axname{Tab}, $\inh{\mu F}$ may miss trees: it contains only the trees whose families of subtrees have codes in $M$.
Without \axname{Ext}, two codes $f\neq f'$ with $\abs{f}=\abs{f'}$ give, by \axname{No Confusion}, two different elements $c(a,f)\neq c(a,f')$ for the same tree.
Finitarity is needed only for Lemma~\ref{lem:continuity} and for the description of $\inh{\mu F}$ by the approximations $\varphi^n_F(\bot)$. For $F\,X=\objOne+X^{\mathbb N}$, the sets $\Phi^n_F(\emptyset)$ contain only trees of finite height, while $\mu F$ also contains, e.g., the tree whose $k$-th subtree has height $k$, which is reached only after $\omega+1$ iterations.
If every $B[a]$ is non-empty, then $\Phi_F(\emptyset)=\emptyset$, so $\emptyset$ is the least fixpoint and $\inh{\mu F}=\bot$. This covers the Moore machines of Section~\ref{sec:Moore}, for any non-empty set of inputs $I$.
\end{rem}

\begin{rem}[Codes and graphs]\label{rem:graphs}
By Lemma~\ref{lem:codes}, codes and functions are in bijection in every model: \axname{Ext} gives injectivity and \axname{Tab} surjectivity. An alternative representation, used for the function sorts of~\cite{ChenCLR26}, represents a function by its \emph{graph}: an element $\alpha$ of a power sort $2^{s_1\otimes s_2}$, whose extension is a subset of $\inh{s_1\otimes s_2}$, with application given by $f\;x=\exists y.\,y\wedge\pr{x,y}\in\mathit{extension}\;f$. The axioms of power sorts make $\mathit{extension}$ injective, and surjective onto the subsets of $\inh{s_1\otimes s_2}$. So, if the function sort is defined as the set of \emph{all} total and single-valued relations (and not only included in it), \axname{Ext} and \axname{Tab} become consequences of the axioms of power sorts, and neither strictness nor the case of $\initMorph$ is needed: a graph determines its domain, and the function on an empty domain is the empty relation.

We do not follow this approach here, for two reasons. First, a graph takes its values in a sort, while $X^{B[a]}$ in \axname{No Junk} and \axname{No Cojunk} maps positions into a set variable $X$. A graph-based version must use a codomain sort, such as $\mu F$ itself, so that the axiom defining $\inh{\mu F}$ refers to $\inh{\mu F}$ on both sides, which complicates the adequacy proofs; a universal codomain sort is excluded by Cantor's theorem, since it would contain its own power sort. Second, the theory of sorts of~\cite{ChenCLR26} requires every sort to be non-empty, while containers need empty sorts of positions, e.g., for the leaves of lists, and may have an empty initial algebra, e.g., for streams. The two representations nevertheless carry the same information: in every model of $\Gamma^\mu_F$, the map $f\mapsto\{(b,\abs f(b))\mid b\in B_M[a]\}$ is a bijection from $\mathit{Fun}_a(L)$ onto the total single-valued relations from $B_M[a]$ to $L$, by Lemma~\ref{lem:codes}, and the same holds for $\nu F$ by Proposition~\ref{prop:local-global}.
\end{rem}

\begin{rem}
If $F$ is finitary in $M$, then $h$ maps $W_n$, i.e., the image of $F^n\,\objZero$ in the colimit~\eqref{eq:ini}, onto $\Phi^n_F(\emptyset)$, the interpretation of $\varphi^n_F(\bot)$. In this sense, the ML pattern~\eqref{eq:ini-ml} is the counterpart of the colimit~\eqref{eq:ini}.
\end{rem}

\paragraph{Deriving Induction Principle}
The following \emph{Induction Principle} is derivable:
\[
\dfrac{\forall a\cln A\ld \cons\;\pr{a,\psi^{B[a]}} \limplies \psi}{\inh{\mu F} \limplies \psi}
\]
where $a$ does not occur free in $\psi$, and $\psi^{B[a]}$ is the pattern $X^{B[a]}[\psi/X]$, i.e., the notation $X^{B[a]}$ from Section~\ref{sec:cont-ml} with $\psi$ in place of $X$.
Indeed, since $a$ is not free in $\psi$, the premise is equivalent to $\varphi_F(\psi)\limplies\psi$.
The Knaster--Tarski rule of matching logic~\cite{CR19}, which derives $\mu X\ld\varphi\limplies\psi$ from $\varphi[\psi/X]\limplies\psi$, then gives $\mu X\ld\varphi_F(X)\limplies\psi$, and \axname{No Junk} rewrites $\mu X\ld\varphi_F(X)$ as $\inh{\mu F}$.

\subsection{Specifying Final Coalgebra}
\label{sec:fin-ml}

Let $X\mapsto F\;X= \sum_{a\cln A}X^{B[a]}$ be a container functor. The final $F$-coalgebra is specified by:
\begin{itemize}
\item the constructor $\cons$ with the axiom \axname{Functional};
\item a sort $\nu F$ with final semantics:\\
$~~~~\begin{aligned}
&\nu F\cln\Sort && \axname{Nu Sort}\\
&\inh{\nu F} = \nu X\ld \exists a\cln A\ld  \cons\;\pr{a, X^{B[a]}} && \axname{No Cojunk}
\end{aligned}$
\item two destructors $\out, \nxt\in \Sigma$ together with the notations:\\
$~~~~\begin{aligned}
& x.\out \equiv \out\;x\\
& x.\nxt \equiv \nxt\;x
\end{aligned}$ 
\\
and the axioms:\\
$~~~~\begin{aligned}
&\forall a\cln A\ld \forall f\ld f\in\top^{B[a]}\limplies (\cons\;\pr{a, f}).\out = a && \axname{No Coconfusion.1}\\
&\forall a\cln A\ld \forall f\ld f\in\top^{B[a]}\limplies (\cons\;\pr{a, f}).\nxt = f && \axname{No Coconfusion.2}
\end{aligned}$
\item the bisimilarity on $\nu F$, defined with the destructors as usual, by the notations\\
$~~~~\begin{aligned}
&\mathit{bis}(R) \equiv \exists x,y\cln \nu F\ld \pr{x,y}\land x.\out = y.\out \land{}\\
&\qquad\qquad\forall b\cln B[x.\out]\ld \exists u,v\ld (x.\nxt)\;b = u\land (y.\nxt)\;b = v\land \pr{u,v}\in R\\
&\mathit{Bisim} \equiv \nu R\ld \mathit{bis}(R)
\end{aligned}$\\
(the set variable $R$ occurs positively in $\mathit{bis}(R)$), and the axiom\\
$~~~~\begin{aligned}
&\forall x,y\cln \nu F\ld \pr{x,y}\in\mathit{Bisim}\limplies x = y && \axname{Bisimilarity}
\end{aligned}$
\end{itemize}

The axiom \axname{No Confusion} of Section~\ref{sec:ini-ml} and the expected inverse property of the destructors are consequences of these axioms.

\begin{lem}\label{lem:derived}
The following patterns are derivable from \axname{Functional}, \axname{No Cojunk}, \axname{No Coconfusion.1}, and \axname{No Coconfusion.2}, and hold in every model of these axioms:
\begin{enumerate}
\item \axname{No Confusion};
\item $\forall x\cln \nu F\ld \cons\;\pr{x.\out, x.\nxt} = x$, which we call \axname{No Coconfusion.3}.
\end{enumerate}
\end{lem}
\begin{proof}
(1) Let $a,a'\cln A$, $f\in\top^{B[a]}$, and $f'\in\top^{B[a']}$ with $\cons\;\pr{a,f}=\cons\;\pr{a',f'}$. By \axname{Functional}, both sides are matched by one element $x$. By \axname{No Coconfusion.1}, $a=x.\out=a'$, and by \axname{No Coconfusion.2}, $f=x.\nxt=f'$.
(2) Let $x\cln\nu F$. By \axname{No Cojunk}, $x\in\exists a\cln A\ld\cons\;\pr{a,(\inh{\nu F})^{B[a]}}$, so $x=\cons\;\pr{a,f}$ for some $a\cln A$ and $f\in(\inh{\nu F})^{B[a]}$, hence $f\in\top^{B[a]}$. By \axname{No Coconfusion.1--2}, $x.\out=a$ and $x.\nxt=f$, so $\cons\;\pr{x.\out,x.\nxt}=\cons\;\pr{a,f}=x$.
Both arguments use only the elimination of existential quantifiers and the substitution of equals in application contexts, which are available in the ML proof system~\cite{CR19,CLR21a}; by soundness, the patterns hold in every model.
\end{proof}

\paragraph{Computing the greatest fixpoint}
We use the notation of Section~\ref{sec:ini-ml}, which does not depend on finitarity, and let $U=\ev{\inh{\nu F}}{M}$. By \axname{No Cojunk}, $U=\mathbf{gfp}(\Phi_F)$.

\begin{lem}\label{lem:cocontinuity}
$\Phi_F$ preserves intersections of decreasing chains. Therefore $U=\mathbf{gfp}(\Phi_F)=\bigcap_{n\ge 0}\Phi^n_F(M)$.
\end{lem}
\begin{proof}
Let $S_0\supseteq S_1\supseteq\cdots$ and $S=\bigcap_n S_n$. By monotonicity, $\Phi_F(S)\subseteq\bigcap_n\Phi_F(S_n)$. Conversely, let $x\in\bigcap_n\Phi_F(S_n)$. For each $n$, $x=c(a_n,f_n)$ with $f_n\in\mathit{Fun}_{a_n}(S_n)$. By \axname{No Confusion}, all pairs $(a_n,f_n)$ are equal to one pair $(a,f)$. Then all values of $\abs{f}$ are in every $S_n$, hence in $S$, so $f\in\mathit{Fun}_a(S)$ and $x\in\Phi_F(S)$. The second claim is the dual of Kleene's fixpoint theorem.
\end{proof}

Hence, informally, $\inh{\nu F}$ is the interpretation of the infinite conjunction
\[
 \top \land \varphi_F(\top) \land \varphi^2_F(\top) \land \varphi^3_F(\top) \land \cdots \tag{\textsc{Gfp}}\label{eq:fin-ml}
\]
The elements that match $\varphi_F(\top)$ are those of the form $x=\cons\;\pr{a,f}$, where $a\cln A$ and $f\in\top^{B[a]}$. By \axname{No Coconfusion.1--2}, these elements satisfy $x.\out\in\inh{A}$ and $x.\nxt\in\top^{B[x.\out]}$. The converse does not hold in general, since the destructors are not constrained outside the image of $\cons$. Further, $\varphi^{n+1}_F(\top)$ is matched by those elements of $\varphi_F(\top)$ whose children $(x.\nxt)\;b$, for $b\cln B[x.\out]$, all match $\varphi^n_F(\top)$. So $\varphi^n_F(\top)$ is matched by the elements that can be unfolded $n$ times by the destructors, and $\inh{\nu F}$ by those that can be unfolded indefinitely.

\begin{rem}
The ML pattern~\eqref{eq:fin-ml} can be seen as an informal ML counterpart of the limit~\eqref{eq:fin}: $\Phi^n_F(M)$ contains the elements that can be unfolded $n$ times, and $F^n\,\objOne$ contains the behaviours truncated at depth $n$.
\end{rem}

\paragraph{Paths and realisation}
\axname{Bisimilarity} ensures that two elements of $\inh{\nu F}$ with the same behaviour are equal. The converse requirement, that every behaviour is the behaviour of some element, needs an axiom about the existence of elements. To state it, we describe a behaviour by the shapes found along the finite paths of a tree.
Let $P\equiv\sum_{a\cln A}B[a]$ be the sort of positions tagged with their shapes (a dependent sum, Section~\ref{sec:fc-ml}), and let $\mathit{List}(P)$ be the sort of finite lists over $P$, specified as a term algebra~\cite{ChenLR26} by symbols $\mathit{List},\mathit{pnil},\pcons\in\Sigma$, the notation $\mathit{List}(P)\equiv\mathit{List}\;P$, and the axioms\\
$~~~~\begin{aligned}
&\mathit{List}(P)\cln\Sort &&\axname{List Sort}\\
&\exists y\cln\mathit{List}(P)\ld \mathit{pnil}=y\qquad
\forall q\cln P\ld\forall p\cln\mathit{List}(P)\ld\exists y\cln\mathit{List}(P)\ld \pcons\;q\;p=y\\
&\forall q\cln P\ld\forall p\cln\mathit{List}(P)\ld \pcons\;q\;p\neq\mathit{pnil}\\
&\forall q,q'\cln P\ld\forall p,p'\cln\mathit{List}(P)\ld \pcons\;q\;p=\pcons\;q'\;p'\limplies q=q'\land p=p'\\
&\inh{\mathit{List}(P)}=\mu Y\ld \mathit{pnil}\lor\pcons\;\inh{P}\;Y
\end{aligned}$\\
A list $\pcons\;\pr{a_n,b_n}\;(\cdots(\pcons\;\pr{a_1,b_1}\;\mathit{pnil})\cdots)$ represents the path that starts at the root, whose shape is $a_1$, goes to its child at position $b_1$, whose shape is $a_2$, and so on (the most recent step comes first).
A set $T$ of pairs $\pr{p,a}$, with $p\cln\mathit{List}(P)$ and $a\cln A$, describes a tree when it satisfies $\mathit{Tree}(T)$, the conjunction of
\[
\begin{aligned}
&\forall p\cln\mathit{List}(P)\ld\forall a,a'\cln A\ld \pr{p,a}\in T\land\pr{p,a'}\in T\limplies a=a' &&\text{(one shape per path)}\\
&\exists a\cln A\ld \pr{\mathit{pnil},a}\in T &&\text{(a root)}\\
&\forall p\cln\mathit{List}(P)\ld\forall a\cln A\ld \pr{p,a}\in T\limplies\forall b\cln B[a]\ld\exists a'\cln A\ld \pr{\pcons\;\pr{a,b}\;p,a'}\in T &&\text{(all children)}
\end{aligned}
\]
The pairs $\pr{p,y}$ such that $y\cln\nu F$ is reached from $x$ along the path $p$ are given by
\[
\begin{aligned}
\mathit{Sub}(x)\equiv\mu S\ld {}&\pr{\mathit{pnil},x}\lor{}\\
&\exists p\cln\mathit{List}(P)\ld\exists y\cln\nu F\ld\exists b\cln B[y.\out]\ld \pr{p,y}\in S\land\pr{\pcons\;\pr{y.\out,b}\;p,(y.\nxt)\;b}
\end{aligned}
\]
and $x$ has the behaviour described by $T$ when
\[
\mathit{Beh}(x,T)\equiv\forall p\cln\mathit{List}(P)\ld\forall y\cln\nu F\ld \pr{p,y}\in\mathit{Sub}(x)\limplies\pr{p,y.\out}\in T.
\]
The theory of the final coalgebra also includes the axiom
\[
\mathit{Tree}(T)\limplies\exists x\cln\nu F\ld\mathit{Beh}(x,T) \tag*{\axname{Realisation}}
\]
where $T$ is a free set variable; as for \axname{Tab}, the axiom speaks about every subset $T$ of $M$.

\begin{rem}
A set $T$ with $\mathit{Tree}(T)$ may contain pairs $\pr{p,a}$ whose path $p$ is not reachable from the root; such pairs are ignored by $\mathit{Beh}$ and play no role. Although $\mathit{Beh}(x,T)$ is stated as an implication, it means that the behaviour of $x$ is exactly the reachable part of $T$ (Lemma~\ref{lem:paths}).
The reachable part is itself definable, by a least fixpoint pattern in the style of $\mathit{Sub}(x)$:
\[
\begin{aligned}
\mathit{Reach}(T)\equiv\mu S\ld {}&(\exists a\cln A\ld \pr{\mathit{pnil},a}\land\pr{\mathit{pnil},a}\in T)\lor{}\\
&(\exists p\cln\mathit{List}(P)\ld\exists a\cln A\ld\exists b\cln B[a]\ld\exists a'\cln A\ld \pr{p,a}\in S\land{}\\
&\qquad\pr{\pcons\;\pr{a,b}\;p,a'}\land\pr{\pcons\;\pr{a,b}\;p,a'}\in T) .
\end{aligned}
\]
Restricting \axname{Realisation} to the sets with $T\subseteq\mathit{Reach}(T)$, all of whose paths are reachable from the root, gives an equivalent axiom: if $\mathit{Tree}(T)$ holds, then $\mathit{Reach}(T)$ satisfies $\mathit{Tree}$ and all its paths are reachable, and an element with the behaviour described by $\mathit{Reach}(T)$ has the behaviour described by $T$, since $\mathit{Reach}(T)\subseteq T$. We keep the unrestricted form, which needs no additional fixpoint pattern.
The axiom \axname{Realisation} is strong: through the free set variable $T$, it requires an element for every subset of the model that describes a tree. It is the ML counterpart of the requirement that the final coalgebra contains all behaviours; without it, $\inh{\nu F}$ may miss behaviours (Remark~\ref{rem:realisation}), and it is sound (Appendix~\ref{ap:model}).
\end{rem}

\paragraph{Adequacy}
By \axname{No Coconfusion.3} and \axname{No Confusion} (Lemma~\ref{lem:derived}), every $x\in U$ is $c(a,f)$ for exactly one $a\in A_M$ and $f\in\mathit{Fun}_a(M)$, namely $a=x.\out$ and $f=x.\nxt$; since $U=\Phi_F(U)$, we have $f\in\mathit{Fun}_a(U)$. The \emph{coalgebra extracted from $M$} is
\[
\alpha_\nu(M)=(U,\gamma),\qquad\gamma:U\to F_M\,U,\quad \gamma(x)=(x.\out,\abs{x.\nxt}).
\]
Let $\zeta:Z_M\to F_M\,Z_M$ be the final $F_M$-coalgebra, i.e., the M-type of $a\cln A_M\vdash B_M[a]$~\cite{abbottAG05}, whose elements are the possibly infinite trees with nodes labelled by shapes $a$ and branching $B_M[a]$.

Let $\beta:(U,\gamma)\to(Z_M,\zeta)$ be the unique coalgebra morphism.
The following lemma shows that finite paths faithfully describe trees, and relates this description to the patterns $\mathit{Tree}$, $\mathit{Sub}$, and $\mathit{Beh}$.
Let $P_M=\{\pr{a,b}\mid a\in A_M,\ b\in B_M[a]\}$. A \emph{path} is a finite sequence $\pi=(\pr{a_1,b_1},\ldots,\pr{a_n,b_n})$ of elements of $P_M$. A tree $z\in Z_M$ \emph{has the path $\pi$ ending in the shape $a$} if, starting at the root, for each $i$ the node reached has shape $a_i$ and one moves to its child at position $b_i$, and the node finally reached has shape $a$. The \emph{path description} of $z$ is $T_z=\{(\pi,a)\mid z\text{ has }\pi\text{ ending in }a\}$.
A set $T$ of pairs $(\pi,a)$, with $\pi$ a path and $a\in A_M$, is \emph{admissible} if (i) $(\pi,a),(\pi,a')\in T$ imply $a=a'$, (ii) $(\varepsilon,a)\in T$ for some $a$, and (iii) $(\pi,a)\in T$ and $b\in B_M[a]$ imply $(\pi\cdot\pr{a,b},a')\in T$ for some $a'$. A path is \emph{reachable} in $T$ if it is $\varepsilon$, or it is $\pi\cdot\pr{a,b}$ with $\pi$ reachable, $(\pi,a)\in T$, and $b\in B_M[a]$; the \emph{reachable part} $R(T)$ consists of the pairs $(\pi,a)\in T$ with $\pi$ reachable.

\begin{lem}\label{lem:paths}
Let $M$ be a model of the axioms of $\Gamma^\nu_F$ other than \axname{Bisimilarity} and \axname{Realisation}.
\begin{enumerate}
\item For every $z\in Z_M$, $T_z$ is admissible and $R(T_z)=T_z$; moreover, $T_z=T_{z'}$ implies $z=z'$.
\item For every admissible $T$, there is exactly one tree $z_T\in Z_M$ with $T_{z_T}=R(T)$.
\item The interpretation of $\mathit{List}(P)$, with $\mathit{pnil}$ and $\pcons$, is in bijection with the set of paths; we write $[\pi]$ for the list that represents $\pi$ (the most recent step first). For $x\in U$, the interpretation of $\mathit{Sub}(x)$ is the set of pairs $\pr{[\pi],x_\pi}$, where $\pi$ is a path of $\beta(x)$ and $x_\pi\in U$ is the element reached from $x$ along $\pi$; $\beta(x)$ has $\pi$ ending in $x_\pi.\out$. Consequently, if a set of pairs $\pr{[\pi],a}$ satisfies $\mathit{Tree}$, then the corresponding set $T$ of pairs $(\pi,a)$ is admissible, and it satisfies $\mathit{Beh}(x,\cdot)$ iff $\beta(x)=z_T$.
\end{enumerate}
\end{lem}
\begin{proof}
(1) Conditions (i)--(iii) hold because every path of $z$ ends in exactly one node, $z$ has a root, and every node of shape $a$ has a child at each $b\in B_M[a]$; every path of $z$ is reachable in $T_z$, by induction on its length. If $T_z=T_{z'}$, then the relation $\{(u,u')\mid T_u=T_{u'}\}$ on $Z_M$ is a bisimulation: related trees have the same root shape $a$, and, for $b\in B_M[a]$, the path descriptions of their children at $b$ are obtained from $T_u=T_{u'}$ by removing the first step $\pr{a,b}$. Since the final coalgebra is strongly extensional, $z=z'$.

(2) For a reachable path $\pi$, let $a_\pi$ be the unique shape with $(\pi,a_\pi)\in T$; it exists by (ii) and (iii), by induction on $\pi$, and it is unique by (i). The reachable paths form an $F_M$-coalgebra, with $\pi\mapsto(a_\pi,\ b\mapsto\pi\cdot\pr{a_\pi,b})$. Let $z_T\in Z_M$ be the image of $\varepsilon$ under the unique morphism into $(Z_M,\zeta)$. By induction on paths, $z_T$ has exactly the reachable paths $\pi$, ending in $a_\pi$, so $T_{z_T}=R(T)$. Uniqueness follows from (1).

(3) The bijection between $\mathit{List}(P)$ and the paths follows from the no-junk and no-confusion axioms of $\mathit{List}(P)$, by the argument of~\cite[Theorem~44]{ChenLR26}. The interpretation of $\mathit{Sub}(x)$ is the least fixpoint of a map that adds, for each pair $\pr{p,y}$ already obtained and each $b\in B_M[y.\out]$, the pair
\[\pr{\pcons\;\pr{y.\out,b}\;p,\ (y.\nxt)\;b};\]
so it is the union of the finite iterations, and its elements are the pairs $\pr{[\pi],x_\pi}$, by induction on the length of $\pi$. Since $\beta$ is a coalgebra morphism, $\gamma(x_\pi)=(x_\pi.\out,\abs{x_\pi.\nxt})$ shows that $\beta(x)$ has $\pi$ ending in $x_\pi.\out$, and that the paths of $\beta(x)$ are exactly these $\pi$. The pattern $\mathit{Tree}$ states (i)--(iii) for the pairs it constrains, so $T$ is admissible, and $\mathit{Beh}(x,T)$ states $T_{\beta(x)}\subseteq T$. If $\beta(x)=z_T$, then $T_{\beta(x)}=R(T)\subseteq T$. Conversely, if $T_{\beta(x)}\subseteq T$, then every path of $\beta(x)$ is reachable in $T$ (by induction, using (i)), and every path reachable in $T$ is a path of $\beta(x)$ (by induction: if $(\pi,a)\in T$ and $\beta(x)$ has $\pi$ ending in $a'$, then $(\pi,a')\in T$, so $a=a'$ by (i), and the children of that node are at the positions $b\in B_M[a]$). So $T_{\beta(x)}=R(T)$, and $\beta(x)=z_T$ by (2).
\end{proof}

\begin{thm}\label{thm:adequacy-fin}
Let $M$ be a model of $\Gamma^\nu_F$. Then:
\begin{enumerate}
\item $U$ is the largest set $S\subseteq M$ with $S\subseteq\Phi_F(S)$;
\item $\alpha_\nu(M)$ is a final $F_M$-coalgebra: the unique coalgebra morphism $\beta:(U,\gamma)\to(Z_M,\zeta)$ is a bijection.
\end{enumerate}
\end{thm}
\begin{proof}
(1) is the Knaster--Tarski characterization of $\mathbf{gfp}(\Phi_F)$.

\emph{$\beta$ is injective.} Let $K=\{(x,y)\in U\times U\mid\beta(x)=\beta(y)\}$. If $(x,y)\in K$, with $\gamma(x)=(a,g)$ and $\gamma(y)=(a',g')$, then $F_M\beta(\gamma(x))=\zeta(\beta(x))=\zeta(\beta(y))=F_M\beta(\gamma(y))$, so $a=a'$ and $\beta\circ g=\beta\circ g'$, i.e., $(g\,b,g'\,b)\in K$ for all $b\in B_M[a]$.
By \axname{No Coconfusion.1--2}, $x.\out=y.\out$ means that $x$ and $y$ have the same shape $a$, and $(x.\nxt)\;b$ is matched by $g\,b$. So, with $\pr{K}=\{\pr{x,y}\mid(x,y)\in K\}$ (pairs of elements of the sort $\nu F$ exist and are injective, by the axioms of $\PAIR$), the previous observation says $\pr{K}\subseteq\ev{\mathit{bis}(R)}{M,\rho[\pr{K}/R]}$. By the Knaster--Tarski theorem, $\pr{K}$ is included in the interpretation of $\mathit{Bisim}$, and \axname{Bisimilarity} gives $x=y$ for all $(x,y)\in K$.

\emph{$\beta$ is surjective.} Let $z\in Z_M$. By Lemma~\ref{lem:paths}(1), $T_z$ is admissible, so the set of pairs $\pr{[\pi],a}$ with $(\pi,a)\in T_z$ satisfies $\mathit{Tree}$. By \axname{Realisation}, there is $x\in U$ with $\mathit{Beh}(x,T_z)$, so $\beta(x)=z_{T_z}=z$ by Lemma~\ref{lem:paths}(1)--(3).

A bijective coalgebra morphism is an isomorphism, so $\alpha_\nu(M)$ is final.
\end{proof}

\begin{cor}
If, moreover, $M$ interprets shapes and positions standardly (a bijection $\iota:A\to A_M$ and bijections $B[a]\cong B_M[\iota(a)]$ for all $a\in A$), then $\alpha_\nu(M)$ is isomorphic to the final $F$-coalgebra $\nu F$, the M-type of $a\cln A\vdash B[a]$.
\end{cor}

\begin{rem}\label{rem:realisation}
The axioms \axname{Bisimilarity} and \axname{Realisation} are needed for Theorem~\ref{thm:adequacy-fin}; Proposition~\ref{prop:necessity}(3)--(6) gives structures that satisfy all the other axioms of $\Gamma_F$ and whose extracted coalgebras are not final.
Without \axname{Bisimilarity}, a model of streams over $E=\{0\}$ may contain two elements $x\neq y$ with $x=\cons\;\pr{0,f_x}$, $y=\cons\;\pr{0,f_y}$, where $f_x$ and $f_y$ map the unique position to $x$ and $y$, respectively; $x$ and $y$ are bisimilar but different.
Without \axname{Realisation}, $\inh{\nu F}$ may even be empty: for the identity functor, there is a model of all the other axioms with $\inh{\nu F}=\bot$ (Section~\ref{sec:boundary}). It may also miss behaviours: for the streams over $E=\{0,1\}$, there is a model of all the other axioms, including \axname{Bisimilarity}, in which $\inh{\nu F}$ consists of the eventually periodic streams; it is constructed in Proposition~\ref{prop:necessity}(5). Closure of this set under $\cons$ and under taking tails gives only that it is a fixpoint of $\Phi_F$, hence included in the greatest one; the converse inclusion holds because the carrier has no other element that can be unfolded indefinitely, by the argument on construction stages of Lemma~\ref{lem:variants}(4). This set is countable, while the final coalgebra $E^\omega$ is not, and \axname{Realisation} fails for the set $T$ describing a stream that is not eventually periodic.
\end{rem}

\begin{prop}\label{prop:soundness-fin}
Let $X\mapsto F\;X=\sum_{a\cln A}X^{B[a]}$ be a container functor, specified as above. There is a model $M$ of $\Gamma_F$ (hence of $\Gamma^\nu_F$) with a bijection $\iota:A\to A_M$ and bijections $B[a]\cong B_M[\iota(a)]$ for all $a\in A$, in which $\inh{\nu F}$ with $\gamma$ is isomorphic to the M-type of $a\cln A\vdash B[a]$.
\end{prop}
\begin{proof}
The model of Appendix~\ref{ap:model} (Theorem~\ref{thm:standard-model}) is a model of the theories of both the initial algebra and the final coalgebra. There, $\inh{\nu F}$ is the set $Z$ of all trees of the M-type, $\cons$ maps a shape and the code of a family of trees to the corresponding tree, fresh terms are excluded from the greatest fixpoint by a well-foundedness argument on construction stages, and \axname{Bisimilarity} and \axname{Realisation} hold because $Z$ is the final coalgebra.
\end{proof}

\paragraph{Deriving Coinduction Principle}
The following \emph{Coinduction Principle} is derivable:
\[
\dfrac{\psi \limplies \exists a\cln A\ld \cons\;\pr{a,\psi^{B[a]}}}{\psi \limplies \inh{\nu F}}
\]
where $\psi^{B[a]}$ is as in the Induction Principle.
The premise is $\psi\limplies\varphi_F(\psi)$.
Since $\nu X\ld\varphi\equiv\neg\mu X\ld\neg\varphi[\neg X/X]$, the Knaster--Tarski rule yields its dual, which derives $\psi\limplies\nu X\ld\varphi$ from $\psi\limplies\varphi[\psi/X]$ (see~\cite{CR19,CLR21a}).
This gives $\psi\limplies\nu X\ld\varphi_F(X)$, and \axname{No Cojunk} rewrites $\nu X\ld\varphi_F(X)$ as $\inh{\nu F}$.
The quantifier on $a$ in the premise is existential, not universal: different elements matching $\psi$ may have different shapes.

\paragraph{Deriving Bisimulation Principle}
The following \emph{Bisimulation Principle}, the usual coinductive proof principle for equality, is derivable:
\[
\dfrac{\psi \limplies \mathit{bis}(\psi)}{\forall x,y\cln\nu F\ld \pr{x,y}\in\psi\limplies x = y}
\]
where $\psi$ is a pattern (of pairs) in which $R$ does not occur free.
Indeed, the dual Knaster--Tarski rule gives $\psi\limplies\nu R\ld\mathit{bis}(R)$, i.e., $\psi\limplies\mathit{Bisim}$, and \axname{Bisimilarity} gives the conclusion.
So, to prove that two elements of $\nu F$ are equal, it is enough to exhibit a bisimulation $\psi$ that relates them.

\subsection{Capturing the Parameters and Comparing the Two Theories}
\label{sec:params}

This subsection concerns both theories: it shows when the shapes and positions, and hence $\mu F$ and $\nu F$ themselves, are captured, and compares the axioms of $\Gamma^\mu_F$ and $\Gamma^\nu_F$.

\paragraph{Capturing the shapes and positions}
Theorems~\ref{thm:adequacy-ini} and~\ref{thm:adequacy-fin} capture $\mu F$ and $\nu F$ relative to the interpretation of the parameter sorts $A$ and $B[a]$. The axioms above do not fix this interpretation: for the constant functor with one shape and no positions, a model with two shapes and no positions satisfies them, and its extracted initial algebra has two elements. The intended data type is obtained in every model when the theory also captures the parameters.

\begin{prop}\label{prop:absolute}
Suppose that the theory also includes axioms such that, in every model $M$, there are a bijection $\iota:A\to A_M$ and bijections $B[a]\cong B_M[\iota(a)]$ for all $a\in A$. Then, in every model $M$, $\alpha_\mu(M)$ is isomorphic to the initial $F$-algebra $\mu F$ and $\alpha_\nu(M)$ is isomorphic to the final $F$-coalgebra $\nu F$.
\end{prop}
\begin{proof}
The bijections form an isomorphism between the containers $a\cln A\vdash B[a]$ and $a\cln A_M\vdash B_M[a]$, hence a natural isomorphism $F\cong F_M$, which maps initial algebras to initial algebras and final coalgebras to final coalgebras. The claim follows from Theorems~\ref{thm:adequacy-ini} and~\ref{thm:adequacy-fin}.
\end{proof}

We now give sufficient conditions for the hypothesis of Proposition~\ref{prop:absolute}.

\begin{defn}\label{def:captured}
A \emph{parameter specification} is a sort $s$ together with a set $\Gamma_s$ of axioms and an intended set $S$. It is \emph{captured} if, in every model $M$ of $\Gamma_0\cup\Gamma_s$, the interpretation $\ev{\inh{s}}{M}$ is in bijection with $S$. A family $a\cln A\vdash B[a]$ over a captured sort $A$, with intended sets $B[a]$, is \emph{captured} if, in every model $M$, there are a bijection $\iota:A\to A_M$ and bijections $B[a]\cong B_M[\iota(a)]$ for all $a\in A$.
The class $\mathcal{C}$ of \emph{standard parameter specifications} is the least class closed under the following rules, where the symbols introduced by different rules are distinct:
\begin{description}
\item[(Fin)] for $n\ge 0$, a sort $s$ with constants $c_1,\ldots,c_n$ and the axioms $s\cln\Sort$, $\exists y\ld c_i=y$, $c_i\neq c_j$ for $i\neq j$, and $\inh{s}=c_1\lor\cdots\lor c_n$ (for $n=0$, $\inh{s}=\bot$), with intended set $\{c_1,\ldots,c_n\}$;
\item[(Prod)] for $s_1,s_2\in\mathcal{C}$, the sort $s_1\otimes s_2$, with intended set $S_1\times S_2$;
\item[(Fam)] for $A\in\mathcal{C}$, a family $a\cln A\vdash B[a]$ that is either \emph{constant}, with the axiom $\forall a\cln A\ld B[a]=s$ for some $s\in\mathcal{C}$, or, if $A$ is given by (Fin) with constants $c_1,\ldots,c_n$, \emph{defined by cases}, with the axioms $B[c_i]=s_i$ for some $s_1,\ldots,s_n\in\mathcal{C}$, or, if $A$ is a dependent sum $\sum_{i\cln I}C[i]$ given by (Sum) with $I$ given by (Fin) with constants $c_1,\ldots,c_n$, \emph{defined by cases on the tag}, with the axioms $\forall x\cln C[c_i]\ld B[\pr{c_i,x}]=s_i$ for some $s_1,\ldots,s_n\in\mathcal{C}$. In all cases, $B[a]=s$ is an equality of sorts, i.e., of elements of $\mathit{Sort}$; it implies $\inh{B[a]}=\inh{s}$, but is stronger;
\item[(Sum)] for $A\in\mathcal{C}$ and a family as in (Fam), the dependent sum $\sum_{a\cln A}B[a]$, with intended set $\sum_{a\in A}B[a]$;
\item[(Ind)] for a container $a\cln A\vdash B[a]$ with $A\in\mathcal{C}$ and a family as in (Fam), the sort $\mu G$ with the axioms of $\Gamma^\mu_G$, and the sort $\nu G$ with the axioms of $\Gamma^\nu_G$, where $G$ is the corresponding container functor, with intended sets the W-type and the M-type of $a\cln A\vdash B[a]$.
\end{description}
The specifications of $\objZero$ and $\objOne$ (Section~\ref{sec:elements}) are the instances $n=0$ and $n=1$ of (Fin), the latter with the constant $\star$, so $\objZero,\objOne\in\mathcal{C}$; similarly, a sum of $n$ sorts of $\mathcal{C}$ is given by (Fin) with $n$ tags, (Fam) by cases, and (Sum).

The class $\mathcal{C}^{\mathit{par}}$ of \emph{parametric specifications} is the least class closed under the rules above, read with $\mathcal{C}^{\mathit{par}}$ in place of $\mathcal{C}$, and under the rule
\begin{description}
\item[(Par)] a sort symbol $s$ with the only axiom $s\cln\Sort$, called a \emph{parameter sort}; it has no intended set.
\end{description}
So $\mathcal{C}\subseteq\mathcal{C}^{\mathit{par}}$. A parameter sort is not captured: its interpretation is arbitrary, possibly empty, and the sorts built from it are captured only relative to this interpretation. An \emph{instance} of a specification in $\mathcal{C}^{\mathit{par}}$ chooses a set $S$ for each parameter sort; the intended sets of the other sorts are then determined by the rules.
\end{defn}

\begin{thm}\label{thm:captured}
Every specification in $\mathcal{C}$ is captured, and so is every family given by (Fam). Consequently, if the shapes $A$ and the family $B$ of a container functor $F$ are given by (Fam) from specifications in $\mathcal{C}$, then, in every model $M$ of $\Gamma_F$ together with these specifications, $\alpha_\mu(M)$ is isomorphic to $\mu F$ and $\alpha_\nu(M)$ is isomorphic to $\nu F$.
\end{thm}
\begin{proof}
By induction on the rules of Definition~\ref{def:captured}.
(Fin): the axioms say that $\inh{s}$ consists of the $n$ pairwise distinct elements denoted by $c_1,\ldots,c_n$.
(Prod): by \prule{Pair Domain} and \prule{Pair Inj}, $\inh{s_1\otimes s_2}$ is in bijection with the product of $\inh{s_1}$ and $\inh{s_2}$, hence, by the induction hypothesis, with $S_1\times S_2$.
(Fam): a constant family has $B_M[a]=\ev{\inh{s}}{M}$ for all $a\in A_M$; a family defined by cases has $B_M[c_i]=\ev{\inh{s_i}}{M}$, and $A_M$ consists of the elements denoted by $c_1,\ldots,c_n$; a family defined by cases on the tag has $B_M[\pr{c_i,x}]=\ev{\inh{s_i}}{M}$, and every element of $A_M$ is such a pair; in all cases the bijections are given by the induction hypothesis.
(Sum): by the axiom of dependent sums and \prule{Pair Inj}, which applies to pairs whose second components belong to different sorts $B[a]$, $\inh{\sum_{a\cln A}B[a]}$ is in bijection with $\sum_{a\in A_M}B_M[a]$, hence with $\sum_{a\in A}B[a]$.
(Ind): by the induction hypothesis the family is captured, so Proposition~\ref{prop:absolute} applies to $G$. By \axname{Mu Sort} and \axname{Nu Sort}, $\mu G$ and $\nu G$ are sorts, so they may occur in (Prod), (Fam), and (Sum), whose axioms quantify over sorts.
The last claim is Proposition~\ref{prop:absolute}, whose hypothesis is the capture of the family.
\end{proof}
The combined theory used in Theorem~\ref{thm:captured} is consistent: it has a model in which all sorts have their intended interpretations, up to canonical bijections (Theorem~\ref{thm:standard-model-ext}). The same theorem covers the parametric specifications: for every instance of a specification in $\mathcal{C}^{\mathit{par}}$, there is a model in which each parameter sort is interpreted by a set in bijection with the chosen one. In particular, the theories of lists over an arbitrary set of elements (Section~\ref{sec:lists}) and of Moore machines over arbitrary sets of outputs and inputs (Section~\ref{sec:Moore}) are consistent, and the results proved for all their models are not vacuous.

\begin{cor}\label{cor:captured-examples}
The following are captured in every model of the corresponding theories, with their intended interpretations: the natural numbers ($\mu$ of the container with shapes $\{z,s\}$ and positions $B[z]=\objZero$, $B[s]=\objOne$); the lists (shapes $\objOne+E$, i.e., $\sum_{i\cln\{l,r\}}C[i]$ with $C[l]=\objOne$ and $C[r]=E$, and positions defined by cases on the tag, $\objZero$ for $l$ and $\objOne$ for $r$) and the streams over a set $E$ given by a specification in $\mathcal{C}$; and the final coalgebra of Moore machines with outputs and inputs given by specifications in $\mathcal{C}$, e.g., deterministic automata with $O=\{0,1\}$ and a finite input alphabet $I$.
\end{cor}

\begin{rem}\label{rem:limits}
Theorem~\ref{thm:captured} has two limits. First, a parameter that is meant to be arbitrary, such as the set of elements of lists in general, is a parameter sort (rule (Par)): it is not fixed by any specification, and the results for it remain relative to its interpretation; the class $\mathcal{C}$ also does not contain sets that are not built by its rules, e.g., the real numbers. Second, (Fam) covers only constant families and families defined by cases over a finite set of shapes or of tags. Families such as $n\cln\mathbb N\vdash[n]$, with $[n]=\{0,\ldots,n-1\}$, are not covered; capturing them would require the family itself to be defined by a captured relation, which we leave for future work.
\end{rem}

\paragraph{Local and global axioms}
The axioms that make the extracted structures isomorphic to the intended ones come in two kinds. \emph{Local} axioms speak about one constructor step: \axname{Ext} and \axname{Tab}. \emph{Global} axioms speak about whole trees: \axname{Bisimilarity} and \axname{Realisation}. The theory of $\mu F$ uses the local ones, and the theory of $\nu F$ the global ones. The following proposition shows that each theory implies the axioms it does not contain, restricted to its own sort.

\begin{prop}\label{prop:local-global}
Each theory implies, on its own sort, the axioms of the other kind:
\begin{enumerate}
\item Let $M$ be a model of $\Gamma^\mu_F$, and $L=\ev{\inh{\mu F}}{M}$. If $K\subseteq L\times L$ is such that, for all $(x,y)\in K$ with $x=c(a,f)$ and $y=c(a',f')$, we have $a=a'$ and $(\abs{f}(b),\abs{f'}(b))\in K$ for all $b\in B_M[a]$, then $K$ is included in the equality. Moreover, every tree of $W_M$ is represented by an element of $L$.
\item Let $M$ be a model of $\Gamma^\nu_F$, and $U=\ev{\inh{\nu F}}{M}$. Then, for all $a\in A_M$, (a) $f,f'\in\mathit{Fun}_a(U)$ and $\abs{f}=\abs{f'}$ imply $f=f'$, and (b) every function $g:B_M[a]\to U$ is $\abs{f}$ for some $f\in\mathit{Fun}_a(U)$.
\end{enumerate}
\end{prop}
\begin{proof}
(1) By Theorem~\ref{thm:adequacy-ini}, every $x\in L$ is $h(w)$ for a unique $w\in W_M$, which gives the second claim. For the first, we show by well-founded induction on $w$ that $(h(w),y)\in K$ implies $h(w)=y$. Let $w=\mathit{sup}(a,t)$, so $h(w)=c(a,\ulcorner h\circ t\urcorner_a)$, and let $y=h(w')$ with $w'=\mathit{sup}(a',t')$. Then $a=a'$ and $(h(t\,b),h(t'\,b))\in K$ for all $b$, so $h(t\,b)=h(t'\,b)$ by the induction hypothesis, hence $h\circ t=h\circ t'$ and $h(w)=h(w')$.

(2a) Since $U=\Phi_F(U)$, the elements $x=c(a,f)$ and $y=c(a,f')$ are in $U$. They have the same shape and the same children, so the set of pairs $\{\pr{x,y}\}\cup\{\pr{z,z}\mid z\in U\}$ is included in the interpretation of $\mathit{bis}(R)$ when $R$ is valued to it. By the Knaster--Tarski theorem it is included in the interpretation of $\mathit{Bisim}$, so $x=y$ by \axname{Bisimilarity}, and $f=f'$ by \axname{No Confusion}.

(2b) Let $z\in Z_M$ with $\zeta(z)=(a,\beta\circ g)$. By Theorem~\ref{thm:adequacy-fin}, $z=\beta(x)$ for some $x\in U$. Let $f=x.\nxt$, so $\gamma(x)=(x.\out,\abs{f})$. Since $\beta$ is a coalgebra morphism, $x.\out=a$ and $\beta\circ\abs{f}=\beta\circ g$, so $\abs{f}=g$ because $\beta$ is injective. Moreover $f\in\mathit{Fun}_a(U)$.
\end{proof}

\begin{rem}
The four kinds of axioms and their roles are summarized below.
\begin{center}
\begin{tabular}{lll}
\hline
Role & $\mu F$ & $\nu F$\\
\hline
no element represents two trees & \axname{Ext}, \axname{No Confusion} & \axname{Bisimilarity}\\
every tree is represented & \axname{Tab} & \axname{Realisation}\\
only the intended elements & \axname{No Junk} (lfp) & \axname{No Cojunk} (gfp)\\
decomposition is unique & \axname{No Confusion} & \axname{No Coconfusion.1--2}\\
\hline
\end{tabular}
\end{center}
The last two rows are dual in the usual sense: least versus greatest fixpoint, and constructor versus destructors. The first two rows are not dual: in both columns they state the injectivity and the surjectivity of the isomorphism with the intended structure. The difference is in which form is primitive. For $\mu F$, the elements are well-founded, so the local axioms suffice and induction yields the global properties (Proposition~\ref{prop:local-global}(1)). For $\nu F$, there is no induction to propagate local properties, and they are too weak: \axname{No Confusion} and \axname{Ext} allow two different bisimilar elements, and \axname{Tab} holds in the model of eventually periodic streams of Remark~\ref{rem:realisation}. So the global axioms are needed, and they imply the local ones (Proposition~\ref{prop:local-global}(2)). This is also why the coalgebra $\gamma$ is well defined without assuming \axname{Ext}.
\end{rem}

\paragraph{Relating the two sorts}
The theories $\Gamma^\mu_F$ and $\Gamma^\nu_F$ use the same constructor, so the two sorts can be compared in the models of $\Gamma_F$. Fix a model $M$ of the axioms of $\Gamma_F$ other than \axname{Realisation}, with $L=\ev{\inh{\mu F}}{M}$ and $U=\ev{\inh{\nu F}}{M}$; the coalgebra $(U,\gamma)$ and the morphism $\beta:(U,\gamma)\to(Z_M,\zeta)$ are defined as in Section~\ref{sec:fin-ml}, and $\beta$ is injective, by the proof of Theorem~\ref{thm:adequacy-fin}. For $x,y\in U$ and $n\ge0$, we say that $x$ and $y$ \emph{agree to depth $n$}, written $x\equiv_n y$, if $n=0$, or $n=m+1$, $x.\out=y.\out$, and $\abs{x.\nxt}(b)\equiv_m\abs{y.\nxt}(b)$ for all $b\in B_M[x.\out]$; agreement to depth $n$ is defined in the same way on $Z_M$, with $\zeta$. Let $d(x,y)=0$ if $x\equiv_n y$ for all $n$, and $d(x,y)=2^{-n}$ otherwise, where $n$ is the greatest number with $x\equiv_n y$.

\begin{prop}\label{prop:mu-nu}
Let $M$ be a model of the axioms of $\Gamma_F$ other than \axname{Realisation}. Then:
\begin{enumerate}
\item \emph{Inclusion.} $L\subseteq U$. Moreover, $\inh{\mu F}\subseteq\inh{\nu F}$, i.e., $\mu F\le\nu F$, is derivable in the ML proof system from \axname{No Junk} and \axname{No Cojunk}.
\item \emph{Well-founded part.} $L$ is closed under the destructors, $\beta(L)$ is the set of the well-founded trees of $Z_M$, and $L=\{x\in U\mid\beta(x)\text{ is well-founded}\}$.
\item \emph{Density.} $d$ is an ultrametric on $U$. If $L\neq\emptyset$, then $L$ is dense in $U$: for every $x\in U$ and every $n$, there is $w\in L$ with $x\equiv_n w$.
\item \emph{Completeness.} If $L\neq\emptyset$, then $M$ satisfies \axname{Realisation} iff the metric space $(U,d)$ is complete. So, in a model of $\Gamma_F$ with $\inh{\mu F}$ non-empty, $U$ is the Cauchy completion of $L$.
\end{enumerate}
\end{prop}
\begin{proof}
(1) $L$ is the least fixpoint of $\Phi_F$ and $U$ is a fixpoint of $\Phi_F$. In the proof system, the unfolding of the greatest fixpoint gives $\varphi_F(\nu X\ld\varphi_F(X))\limplies\nu X\ld\varphi_F(X)$, the Knaster--Tarski rule~\cite{CR19} then gives $\mu X\ld\varphi_F(X)\limplies\nu X\ld\varphi_F(X)$, and \axname{No Junk} and \axname{No Cojunk} rewrite the two sides as $\inh{\mu F}$ and $\inh{\nu F}$.

(2) If $x\in L=\Phi_F(L)$, then $x=c(a,f)$ with $a\in A_M$ and $f\in\mathit{Fun}_a(L)$, so $x.\out=a$ and the values of $\abs{x.\nxt}=\abs f$ are in $L$. Let $h:W_M\to L$ be the isomorphism of Theorem~\ref{thm:adequacy-ini}, and let $\iota:W_M\to Z_M$ be the unique coalgebra morphism from $(W_M,\mathit{sup}^{-1})$, which is injective and whose image is the set of the well-founded trees. Since $h(\mathit{sup}(a,t))=c(a,\ulcorner h\circ t\urcorner_a)$, we have $\gamma(h(\mathit{sup}(a,t)))=(a,h\circ t)$, so $h$ is a coalgebra morphism from $(W_M,\mathit{sup}^{-1})$ to $(U,\gamma)$, and $\beta\circ h=\iota$ by the finality of $Z_M$. Hence $\beta(L)=\iota(W_M)$. If $\beta(x)$ is well-founded, then $\beta(x)=\iota(w)=\beta(h(w))$ for some $w$, and $x=h(w)\in L$, since $\beta$ is injective.

(3) Since $\equiv_{n+1}$ is included in $\equiv_n$, $d(x,z)\le\max(d(x,y),d(y,z))$. If $x\equiv_n y$ for all $n$, then the set of all such pairs $(x,y)$ is a bisimulation on $(U,\gamma)$: the two elements have the same shape, and their children at each position agree to every depth. So $x=y$, as in the proof of Theorem~\ref{thm:adequacy-fin}, by \axname{Bisimilarity}; hence $d$ is an ultrametric. Let $L\neq\emptyset$. We show by induction on $n$ that every $x\in U$ agrees to depth $n$ with some $w\in L$. For $n=0$, take any $w\in L$. For $n+1$, let $x=c(a,f)$ and choose, for each $b\in B_M[a]$, some $w_b\in L$ with $\abs f(b)\equiv_n w_b$. By Lemma~\ref{lem:codes}, the function $g:b\mapsto w_b$ has a code $\ulcorner g\urcorner_a\in\mathit{Fun}_a(L)$, and $w=c(a,\ulcorner g\urcorner_a)\in\Phi_F(L)=L$ has $w.\out=a$ and $\abs{w.\nxt}=g$, so $x\equiv_{n+1}w$.

(4) Since $\beta$ is a coalgebra morphism, $x\equiv_n y$ iff $\beta(x)\equiv_n\beta(y)$, by induction on $n$; so $\beta$ is an isometry. On $Z_M$, agreement to depth $n$ is equality of the images in $F_M^n\,\objOne$, and $Z_M$ is the limit of the terminal sequence $\objOne\leftarrow F_M\,\objOne\leftarrow F_M^2\,\objOne\leftarrow\cdots$ (Appendix~\ref{ap:model}); so $Z_M$ is complete: a Cauchy sequence determines a compatible family of elements of the $F_M^n\,\objOne$, i.e., an element of the limit. If $M$ satisfies \axname{Realisation}, then $\beta$ is a bijection (Theorem~\ref{thm:adequacy-fin}), hence an isometry onto a complete space, and $(U,d)$ is complete. Conversely, let $(U,d)$ be complete and $z\in Z_M$. As in (3), by induction on $n$, there is $w_n\in L$ such that $\beta(w_n)$ agrees with $z$ to depth $n$: for $n+1$ and $\zeta(z)=(a,t)$, take $w_{n+1}=c(a,\ulcorner g\urcorner_a)$, where $g(b)\in L$ is such that $\beta(g(b))$ agrees with $t(b)$ to depth $n$. For $m\ge n$, $\beta(w_m)\equiv_n\beta(w_n)$, so $w_m\equiv_n w_n$, and the sequence $(w_n)$ is Cauchy. Its limit $x\in U$ satisfies $\beta(x)\equiv_n z$ for all $n$, so $\beta(x)=z$. Hence $\beta$ is surjective. For $T$ with $\mathit{Tree}(T)$, an element $x$ with $\beta(x)=z_T$ satisfies $\mathit{Beh}(x,T)$, by Lemma~\ref{lem:paths}(2) and~(3), so \axname{Realisation} holds. The last claim follows from (3).
\end{proof}

\begin{rem}\label{rem:approx}
Proposition~\ref{prop:mu-nu} makes precise the idea that, when $\inh{\mu F}$ is non-empty, its elements approximate those of $\inh{\nu F}$, as the finite lists approximate the infinite ones: an infinite list $e_0e_1e_2\cdots$ is the limit of the finite lists $e_0\cdots e_{n-1}$. Three points should be noted.
First, the hypothesis is sharp: $L\neq\emptyset$ iff some shape has no positions, i.e., $B_M[a]=\emptyset$ for some $a\in A_M$, since $\Phi_F(\emptyset)$ consists of the elements $c(a,\initMorph)$ for these shapes. For the streams and for the Moore machines with a non-empty set of inputs, $\inh{\mu F}$ is empty and nothing approximates the elements of $\inh{\nu F}$.
Second, the approximation is metric, not an information order: the finite list $e_0e_1$ agrees with the infinite list $e_0e_1e_2\cdots$ to depth $2$, but it is not a partial version of it, since the two differ at depth $3$, where one ends and the other continues. An order of approximation would need partial elements, e.g., the functor $F+\objOne$ with an undefined leaf, which the theories do not have.
Third, density does not imply \axname{Realisation}: that every behaviour is the limit of elements of $\inh{\mu F}$ does not give an element of $\inh{\nu F}$ with this behaviour, as the rational approximations of $\sqrt2$ do not give a rational number. Proposition~\ref{prop:necessity}(6) constructs a structure that satisfies all the other axioms, in which $\inh{\mu F}$ is non-empty, hence dense, and \axname{Realisation} fails. By Proposition~\ref{prop:mu-nu}(4), when $\inh{\mu F}$ is non-empty, \axname{Realisation} states exactly that these limits exist. This reading does not replace the axiom: it is not available when $\inh{\mu F}$ is empty, where \axname{Realisation} is still needed, and the inclusion of (1) is derivable, while density and completeness are semantic properties.
\end{rem}

\subsection{Summary of the Guarantees}
\label{sec:summary}

The results of this section, and of Appendix~\ref{ap:model}, are of different kinds. We distinguish the requirements for \emph{defining} the matching logic theories, for their \emph{satisfiability}, and for their \emph{adequacy} with respect to the intended set-theoretic functor, and we separate semantic results from results derivable in the ML proof system.

\paragraph{Defining the theories}
The theories $\Gamma^\mu_F$, $\Gamma^\nu_F$, and $\Gamma_F$ are fixed by Definition~\ref{def:gamma}. They are well-formed ML theories: every fixpoint pattern they use, namely those in \axname{No Junk}, \axname{No Cojunk}, $\mathit{Bisim}$, $\mathit{Sub}$, and the no-junk axiom of $\mathit{List}(P)$, has a body positive in its bound variable, as required by Definition~\ref{def:AML-syntax}, and so do the notations $\mu X\cln s\ld\varphi$ and $\nu X\cln s\ld\varphi$ (Section~\ref{sec:ml}). The notations $X^{B[a]}$ (Section~\ref{sec:cont-ml}) and for partially functional symbols are meaningful for empty and nested sorts, and the table following Definition~\ref{def:gamma} shows on which arguments each constructor and destructor application is defined. The sorts $\mu F$ and $\nu F$ are declared by \axname{Mu Sort} and \axname{Nu Sort}, and \axname{No Confusion} and \axname{No Coconfusion.3} are derived in $\Gamma^\nu_F$ (Lemma~\ref{lem:derived}).

\paragraph{Satisfiability}
Theorem~\ref{thm:standard-model} constructs a model of $\Gamma_F$ in which all sorts have their intended interpretation, up to canonical bijections, and Theorem~\ref{thm:standard-model-ext} extends it to several containers and the parameter specifications of Definition~\ref{def:captured}. So these theories are consistent; Propositions~\ref{prop:soundness-ini} and~\ref{prop:soundness-fin} are consequences. Since Theorem~\ref{thm:standard-model-ext} holds for every instance of a parametric specification, the theories of the case studies have models for arbitrary parameters (Corollary~\ref{cor:case-studies-consistent}).

\paragraph{Necessity of the axioms}
Proposition~\ref{prop:necessity} constructs, for each of \axname{Ext}, \axname{Tab}, \axname{Bisimilarity}, and \axname{Realisation}, a structure that satisfies all the other axioms and whose extracted algebra or coalgebra is not initial or final. So none of these axioms follows from the others, and each is needed for the adequacy theorems.

\paragraph{Adequacy}
The adequacy results are \emph{semantic}: they are statements about all models of a theory, proved in set theory, not derivations in the ML proof system. They are:
Theorems~\ref{thm:adequacy-ini} and~\ref{thm:adequacy-fin} ($\alpha_\mu(M)$ is initial and $\alpha_\nu(M)$ is final for $F_M$, in every model), with Lemmas~\ref{lem:codes}, \ref{lem:cocontinuity}, and~\ref{lem:paths}; Proposition~\ref{prop:local-global}; and Theorem~\ref{thm:moore}. Matching logic with fixpoints has no complete proof system, so these semantic consequences of the axioms, e.g., that bisimilar elements of $\inh{\mu F}$ are equal (Proposition~\ref{prop:local-global}(1)), need not be derivable.

\paragraph{Derivable proof principles}
The Induction, Coinduction, and Bisimulation Principles of Sections~\ref{sec:ini-ml} and~\ref{sec:fin-ml} are \emph{derivable} in the ML proof system, from the Knaster--Tarski rule or its dual and the axioms \axname{No Junk}, \axname{No Cojunk}, and \axname{Bisimilarity}; by soundness, they also hold in every model. The derivations are given at the level of proof rules, not as formal proof objects.

\paragraph{Additional assumptions}
Two kinds of results need assumptions beyond the theories. Finitarity is needed only for Lemma~\ref{lem:continuity}, for the description of $\inh{\mu F}$ by the approximations $\varphi^n_F(\bot)$, and for the correspondence with the colimit~\eqref{eq:ini}. A standard interpretation of the shapes and positions is needed to compare $\alpha_\mu(M)$ and $\alpha_\nu(M)$ with $\mu F$ and $\nu F$ themselves, rather than with $\mu F_M$ and $\nu F_M$ (the corollaries of Theorems~\ref{thm:adequacy-ini} and~\ref{thm:adequacy-fin}, and Proposition~\ref{prop:absolute}); Theorem~\ref{thm:captured} gives specifications that ensure it.

\begin{center}\small
\begin{tabular}{l>{\raggedright\arraybackslash}p{0.74\textwidth}}
\hline
Kind & Results\\
\hline
well-formedness & Definition~\ref{def:gamma}, positivity of the fixpoint bodies, the notations, the typing table\\
satisfiability & Theorems~\ref{thm:standard-model} and~\ref{thm:standard-model-ext}; Corollary~\ref{cor:case-studies-consistent}\\
necessity of the axioms & Proposition~\ref{prop:necessity}, with Lemma~\ref{lem:variants}\\
semantic adequacy & Theorems~\ref{thm:adequacy-ini}, \ref{thm:adequacy-fin}, \ref{thm:moore}; Proposition~\ref{prop:local-global}; Lemmas~\ref{lem:continuity}, \ref{lem:cocontinuity}, \ref{lem:paths}\\
relating $\mu F$ and $\nu F$ & Proposition~\ref{prop:mu-nu} (inclusion, density, completeness)\\
derivable & Induction, Coinduction, and Bisimulation Principles; Lemma~\ref{lem:derived}; Proposition~\ref{prop:mu-nu}(1)\\
extra assumptions & finitarity (Lemma~\ref{lem:continuity}); standard parameters (Proposition~\ref{prop:absolute}, Theorem~\ref{thm:captured})\\
\hline
\end{tabular}
\end{center}

\subsection{Boundary Cases}
\label{sec:boundary}

The following cases test the specifications at their boundaries. In each of them, the theories of Definition~\ref{def:gamma} behave as intended, and the table indicates which axioms and results the case depends on.

\begin{center}\small
\begin{tabular}{>{\raggedright\arraybackslash}p{2.9cm}>{\raggedright\arraybackslash}p{1.9cm}>{\raggedright\arraybackslash}p{9.1cm}}
\hline
Case & $\mu F$, $\nu F$ & Behaviour of the theories\\
\hline
(a) no shapes, $A=\emptyset$ & $\emptyset$, $\emptyset$ & $\varphi_F(X)=\bot$, so both fixpoints are empty; \axname{Ext}, \axname{Tab}, and \axname{Bisimilarity} hold vacuously; \axname{Realisation} is vacuous, since $\mathit{Tree}(T)$ requires a root shape; $P$ is empty and $\mathit{List}(P)$ contains only $\mathit{pnil}$.\\
(b) no positions, all $B[a]=\emptyset$ (constant functor $A$) & $A$, $A$ & $\mathit{Fun}_a(S)=\{\initMorph\}$ for all $S$, so $\varphi_F(X)$ does not depend on $X$ and both fixpoints are the set of constants $c(a,\initMorph)$; codes are not used. Depends on the $\initMorph$-case of $X^{B[a]}$, on \axname{Ini Functional}, and on the fact that $X^{B[a]}$ and the notation for partially functional symbols allow empty domains.\\
(c) empty exponent domain, $(F\,X)^{\emptyset}$ & $\objOne$, $\objOne$ & By Section~\ref{sec:poly-as-cont}, there is exactly one shape, the empty function $\emptyset\to A$, with no positions, even if $A$ is empty; this is case (b). For Moore machines with $I=\emptyset$, $\Mo\;X\approxeq O$.\\
(d) infinite position sets, e.g., $X^{\mathbb N}$ & W-type, M-type & Theorems~\ref{thm:adequacy-ini} and~\ref{thm:adequacy-fin} and Lemma~\ref{lem:cocontinuity} hold; Lemma~\ref{lem:continuity} and the approximation~\eqref{eq:ini-ml} fail (Remark~\ref{rem:ini-hyp}). \axname{Tab} concerns functions on infinite domains, through the free set variable $R$; the standard model uses a regular cardinal larger than every $\abs{B[a]}$; paths remain finite.\\
(e) $A\neq\emptyset$ and all $B[a]\neq\emptyset$ (e.g., $F\,X=X$, streams, Moore with $I\neq\emptyset$) & $\emptyset$, non-empty & $\Phi_F(\emptyset)=\emptyset$, so $\inh{\mu F}=\bot$ without finitarity; \axname{Ext} and \axname{Tab} are vacuous ($\mathit{Fun}_a(L)=\emptyset$, and the premise of \axname{Tab} is false); $\nu F$ is described below.\\
\hline
\end{tabular}
\end{center}

Case (e) shows the role of \axname{Realisation} most clearly. For the identity functor $F\,X=X$ (one shape $\star$ with one position), $\nu F$ has exactly one element, a self-referential $x=\cons\;\pr{\star,f}$, where $f$ maps the unique position to $x$ itself. \axname{Bisimilarity} makes this element unique, but only \axname{Realisation} makes it exist: the construction of Appendix~\ref{ap:model} without the trees of the M-type, in which $\cons$ always yields fresh terms, satisfies all the other axioms of $\Gamma_F$, and, by the argument on construction stages, the greatest fixpoint of $\Phi_F$ is empty there, so $\inh{\nu F}=\bot$ (Proposition~\ref{prop:necessity}(4)).

\section{Case Study: Lists Using Container Functors}
\label{sec:lists}

The set $E$ of elements is a parameter: it is represented by a parameter sort $E$, given by (Par) in Definition~\ref{def:captured}, and the results below hold for the lists over its interpretation $E_M$ (Section~\ref{sec:params}); the sorts $\objZero$ and $\objOne$ are captured by the axioms of Section~\ref{sec:elements}, the instances $n=0$ and $n=1$ of (Fin) in Definition~\ref{def:captured}.
First, we express $\Lst\; X = \objOne + E\times X$ as a unary container functor, using $\objOne \approxeq \sum_{\star\cln \objOne} X^{\objZero[\star]}$,  $E\approxeq\sum_{e\cln E}X^{\objZero[e]}$, and $X \approxeq \sum_{\star\cln \objOne}X^{\objOne[\star]}$:
\begin{align*}
\Lst\; X & = \sum_{\star\cln \objOne} X^{\objZero[\star]} + \sum_{e\cln E}X^{\objZero[e]} \times \sum_{\star\cln \objOne}X^{\objOne[\star]}\\
&= \sum_{\star\cln \objOne} X^{\objZero[\star]} + \sum_{\pr{e,\star}\cln E\times \objOne}X^{\objZero[e] + \objOne[\star]}\\
&= \sum_{a\cln \objOne+E\times \objOne}X^{[\objZero, \langlebar \objZero, \objOne\ranglebar ][a]}
\end{align*}
Using the isomorphisms $E\times \objOne\approxeq E$ and $\langlebar \objZero, \objOne\ranglebar [\pr{e,\star}] = \objZero[e] + \objOne[\star] = \objZero + \objOne \approxeq \objOne$ in the category $\catC$, we obtain the reduced form $\Lst^c\; X=\sum_{a\cln\objOne+E}X^{[\objZero,\objOne][a]}$ of the container functor $\Lst$. From the definition of the sum of container functors we deduce that the elements of $\sum_{a\cln\objOne+E}X^{[\objZero,\objOne][a]}$ are pairs $\{\pr{\star, f} \mid f:\objZero\to X\}\uplus \{\pr{e, f'} \mid e:E, f':\objOne\to X\}$. The only function $\objZero\to X$ is $\initMorph$, and $f':\objOne\to X \approxeq f'\in X$.

\paragraph{The ML specification}
The lists are specified by a literal instance of the theories of Definition~\ref{def:gamma}, with parameter specifications as in Definition~\ref{def:captured}:
\begin{itemize}
\item a sort $\mathit{Tag}$ with constants $l,r$, given by (Fin); and the sorts $\objZero$ and $\objOne$ of Section~\ref{sec:elements}, i.e., (Fin) with $n=0$ and with $n=1$;
\item the shapes $A\equiv\sum_{i\cln\mathit{Tag}}C[i]$, with $C[l]=\objOne$ and $C[r]=E$, so that the shapes are $\pr{l,\star}$ and $\pr{r,e}$ for $e\cln E$;
\item the positions, defined by cases on the tag: $\forall x\cln\objOne\ld B[\pr{l,x}]=\objZero$ and $\forall x\cln E\ld B[\pr{r,x}]=\objOne$.
\end{itemize}
The constructor $\cons$, the sorts $\mu \Lst^c$ and $\nu \Lst^c$, the destructors, and their axioms are those of $\Gamma^\mu_{\Lst^c}$ and $\Gamma^\nu_{\Lst^c}$, without change: nothing specific to lists is postulated. The sort $E$ is a parameter sort, so the specification is in $\mathcal{C}^{\mathit{par}}$ and has a model for every set of elements (Theorem~\ref{thm:standard-model-ext}); if $E$ is given by a specification in $\mathcal{C}$, Theorem~\ref{thm:captured} applies.

\paragraph{Transport along the coding map}
In this instance, the tail of a non-empty list is not the list element $\ell$ itself, but a code of the function from the unique position $\star$ to $\ell$. The set-theoretic isomorphism $X^{\objOne}\approxeq X$ is transported into the theory by the \emph{coding map}, the instance $B[a]=\objOne$ of the codes $\ulcorner g\urcorner_a$ of Section~\ref{sec:ini-ml}, which we define by the notation
\[
\kappa(\psi)\equiv\exists f\ld f\land f\in\top^{\objOne}\land \exists x\ld (f\;\star=x\land x\in\psi) .
\]
The value $f\;\star$ is required to be \emph{in} $\psi$, not equal to it: the equality $f\;\star=\psi$ would compare $\psi$ with the singleton $f\;\star$, so it would fail whenever $\psi$ denotes a set with several elements, and $\psi$ would occur negatively in it. With membership, $\psi$ occurs only positively, and in a model $\kappa(\psi)$ is matched by the codes on $\objOne$ whose value at $\star$ belongs to the interpretation of $\psi$. For an element $\ell$, $x\in\ell$ means $x=\ell$, so $\kappa(\ell)$ is matched by the codes whose value at $\star$ is $\ell$; for a set, $\kappa$ is the union of its values on the elements of the set.
In fact, $\kappa(\psi)$ is the instance $\psi^{\objOne}$ of the notation $X^{B[a]}$ of Section~\ref{sec:cont-ml}: since $\inh{\objOne}=\star$ is non-empty, both are matched, for every interpretation of $\psi$, by the elements $f$ such that $f\;\star$ is a singleton included in it and $f$ is undefined outside $\inh{\objOne}$. The usual constructors and destructors of lists are then notations:
\[
\begin{aligned}
&\nil\equiv\cons\;\pr{\pr{l,\star},\initMorph}, &&
\cons_E\;e\;\ell\equiv\cons\;\pr{\pr{r,e},\kappa(\ell)},\\
&\hd\;x\equiv\exists e\cln E\ld e\land x.\out=\pr{r,e}, &&
\tl\;x\equiv(x.\nxt)\;\star .
\end{aligned}
\]

\begin{prop}\label{prop:lists-transport}
The usual list operations are recovered by transport along $\kappa$. Here $\Gamma^\mu_{\Lst^c}$ and $\Gamma^\nu_{\Lst^c}$ include the parameter specification above, and $E_M$ is the interpretation of $\inh{E}$:
\begin{enumerate}
\item Let $M$ be a model of $\Gamma^\mu_{\Lst^c}$ and $L=\ev{\inh{\mu \Lst^c}}{M}$. If $E_M\neq\emptyset$, then, for every $\ell\in L$, $\kappa(\ell)$ is matched by exactly one element, and $\kappa$ is a bijection between $L$ and the codes on $\objOne$ with values in $L$. Moreover, for every $E_M$, the following hold in $M$:
\[
\begin{aligned}
&\exists y\cln\mu \Lst^c\ld \nil=y \qquad \forall e\cln E\ld\forall \ell\cln\mu \Lst^c\ld\exists y\cln\mu \Lst^c\ld \cons_E\;e\;\ell=y\\
&\forall e\cln E\ld\forall \ell\cln\mu \Lst^c\ld \nil\neq\cons_E\;e\;\ell\qquad
\forall e,e'\cln E\ld\forall \ell,\ell'\cln\mu \Lst^c\ld \cons_E\;e\;\ell=\cons_E\;e'\;\ell'\limplies e=e'\land\ell=\ell'\\
&\inh{\mu \Lst^c}=\mu X\ld \nil\lor\cons_E\;\inh{E}\;X
\end{aligned}
\]
so $(L,\nil,\cons_E)$ is isomorphic to the algebra of finite lists over $E_M$.
\item Let $M$ be a model of $\Gamma^\nu_{\Lst^c}$ and $U=\ev{\inh{\nu \Lst^c}}{M}$. If $E_M\neq\emptyset$, then $\kappa$ is a bijection between $U$ and the codes on $\objOne$ with values in $U$. For every $E_M$, the statements of (1) hold with $\nu \Lst^c$ and $\nu X$ in place of $\mu \Lst^c$ and $\mu X$, and $\hd\;(\cons_E\;e\;\ell)=e$, $\tl\;(\cons_E\;e\;\ell)=\ell$, $\hd\;\nil=\bot$, and $\tl\;\nil=\bot$. Hence $U$, with $\nil$, $\cons_E$, and the partial destructors $\hd$ and $\tl$, is isomorphic to the set $E_M^*\cup E_M^\omega$ of finite and infinite lists (colists) over $E_M$, with its usual structure.
\end{enumerate}
\end{prop}
\begin{proof}
(1) Let $E_M\neq\emptyset$ and $e\in E_M$. The shape $\pr{r,e}$ has the position set $\{\star\}$, so, by Lemma~\ref{lem:codes} for this shape, every function from $\{\star\}$ to $L$ has exactly one code with values in $L$; a code matching $\kappa(\ell)$ has the single value $\ell\in L$, so $\kappa(\ell)$ is matched by exactly one element, and $\kappa$ is a bijection onto these codes. If $E_M=\emptyset$, there is no shape with the position set $\{\star\}$, so \axname{Ext} and \axname{Tab} say nothing about the codes on $\{\star\}$, and $\kappa(\ell)$ may be matched by no element or by several; but then $\cons_E$ is never applied: the statements below that quantify over $e\cln E$ hold vacuously, $\cons_E\;\inh{E}\;X$ is $\bot$, and $L=\{\nil\}$. We continue with the statements that hold for every $E_M$; where $\kappa(\ell)$ is used, $E_M\neq\emptyset$. The element $\nil$ is $c(\pr{l,\star},\initMorph)\in\Phi_F(L)=L$, and $\cons_E\;e\;\ell=c(\pr{r,e},\kappa(\ell))\in\Phi_F(L)=L$ for $\ell\in L$. The two confusion statements follow from \axname{No Confusion}, the injectivity of pairs, $l\neq r$, and the injectivity of $\kappa$. For the last statement, the body $\nil\lor\cons_E\;\inh{E}\;X$ is positive in $X$, since $X$ occurs only in $\kappa(X)$, under $x\in X$. Let $\Psi$ be its interpretation as a function of $X$. For every $S\subseteq M$, $\Psi(S)=\Phi_F(S)$: the shapes are $\pr{l,\star}$, with $B[\pr{l,\star}]=\objZero$ and $S^{\objZero}=\{\initMorph\}$, which gives $\nil$, and $\pr{r,e}$ for $e\in E_M$, with $B[\pr{r,e}]=\objOne$ and $S^{\objOne}=\kappa(S)$, which gives the elements $c(\pr{r,e},f)$ with $f\in\kappa(S)$. Hence $\mu X\ld\nil\lor\cons_E\;\inh{E}\;X$ and $\mu X\ld\varphi_F(X)$ have the same interpretation, which is $L$ by \axname{No Junk}. The isomorphism with finite lists follows from these no-junk and no-confusion properties, as for term algebras~\cite[Theorem~44]{ChenLR26}.

(2) If $E_M\neq\emptyset$, then, by Proposition~\ref{prop:local-global}(2) for a shape $\pr{r,e}$, every function from $\{\star\}$ to $U$ has exactly one code with values in $U$, so $\kappa$ is a bijection as in (1); if $E_M=\emptyset$, then $U=\{\nil\}$ as in (1). In both cases, the first statements follow as in (1), with $\Phi_F(U)=U$. For the fixpoint, $\Psi=\Phi_F$ as in (1), so $\nu X\ld\nil\lor\cons_E\;\inh{E}\;X$ and $\nu X\ld\varphi_F(X)$ have the same interpretation, which is $U$ by \axname{No Cojunk}. Finally, by the coconfusion axioms, $(\cons_E\;e\;\ell).\out=\pr{r,e}$ and $(\cons_E\;e\;\ell).\nxt=\kappa(\ell)$, whose value at $\star$ is $\ell$; and $\nil.\out=\pr{l,\star}\neq\pr{r,e}$ and $\nil.\nxt=\initMorph$, with $\initMorph\;\star=\bot$. For the last claim, $\Lst^c_M\,X\cong\objOne+E_M\times X$ naturally: the shape $\pr{l,\star}$ has no positions, and a shape $\pr{r,e}$ has one position, whose labels are transported along $\kappa$. By Theorem~\ref{thm:adequacy-fin}, $\alpha_\nu(M)$ is a final $\Lst^c_M$-coalgebra, hence isomorphic to the final coalgebra of $X\mapsto\objOne+E_M\times X$, which is $E_M^*\cup E_M^\omega$ with the empty list observed as $\nil$ and a non-empty list observed by its head and its tail; under this isomorphism, $\hd$ and $\tl$ are the usual partial head and tail.
\end{proof}

Thus the usual specifications of lists and of possibly infinite lists, with $\nil$, $\cons_E$, and the partial destructors $\hd$ and $\tl$, as in~\cite{CLR21a}, are recovered as properties of every model of the generic theories, rather than postulated. More generally, any isomorphism of container functors, such as $X^{\objOne}\approxeq X$ here, is transported along the coding bijections in the same way; this also justifies working with the simplified container $\Mo^c$ in Section~\ref{sec:Moore}.

\section{Case Study: Moore Machines}
\label{sec:Moore}

Here we consider an example of a (constant) exponent functor.
Moore machines (automata) have the signature given by the functor $\Mo\;X=  O \times X^I$, where $O$ is the set of outputs and $I$ the set of inputs (see Example~\ref{ex:moore}). Both are parameters, represented by parameter sorts given by (Par) in Definition~\ref{def:captured}, and the family of positions is the constant family of (Fam), so the theory below has a model for all sets $O$ and $I$ (Theorem~\ref{thm:standard-model-ext}); for deterministic automata, $O=\{0,1\}$ can be captured by the axioms for finite sets, rule (Fin) of Definition~\ref{def:captured} (Section~\ref{sec:params}).

\subsubsection*{The Container Form of $\Mo$}

We express $\Mo$ as a unary container functor, using $O\approxeq\sum_{o\cln O}X^{\objZero[o]}$, $X\approxeq\sum_{\star\cln\objOne}X^{\objOne[\star]}$, and the exponent and product constructions of Section~\ref{sec:poly-as-cont}:
\begin{align*}
\Mo\;X &= O\times X^I\\
&\approxeq \sum_{o\cln O}X^{\objZero[o]} \times \Big(\sum_{\star\cln \objOne}X^{\objOne[\star]}\Big)^{I}\\
&\approxeq \sum_{o\cln O}X^{\objZero[o]} \times \sum_{g\cln  I\to \objOne}X^{\sum_{i\cln I}\objOne[g\;i]}\\
&\approxeq  \sum_{\pr{o,g}\cln O\times (I\to \objOne)}X^{\objZero[o] + \sum_{i\cln I}\objOne[g\;i]}\\
&\approxeq \sum_{o\cln O}X^{I}=\Mo^c\;X
\end{align*}
where the last step uses $I\to\objOne\approxeq\objOne$ and $\objZero+\sum_{i\cln I}\objOne\approxeq I$.
So $\Mo^c$ is the container with shapes $O$ and the constant family of positions $o\cln O\vdash I$: the shape of a state is its output, and its children are its successors, one for each input.

\subsubsection*{Capturing the Final $\Mo$-Coalgebra in ML}

The instantiation of the ML specification of Section~\ref{sec:fin-ml}, with $A=O$ and the constant family $B[o]=I$, is a literal instance of Definition~\ref{def:gamma}, and is as follows, where $\top^I$ is the pattern $X^{I}[\top/X]$:
\begin{itemize}
\item the constructor $\cons\in\Sigma$ is specified by the following axioms:\\
$\begin{aligned}
&\forall o\cln O\ld \forall f\ld f\in \top^{I}\limplies \exists x\ld \cons\;\pr{o,f} = x  && \axname{Functional}
\end{aligned}$
\item the sort $\nu \Mo^c$ with final semantics:\\
$~~~~\begin{aligned}
&\nu \Mo^c\cln\Sort && \axname{Nu Sort}\\
&\inh{\nu \Mo^c} = \nu X\ld \exists o\cln O\ld  \cons\;\pr{o, X^{I}} && \axname{No Cojunk}
\end{aligned}$
\item the destructors $\out, \nxt\in \Sigma$, specified by the axioms:\\
$~~~~\begin{aligned}
&\forall o\cln O\ld \forall f\ld f\in \top^{I} \limplies (\cons\;\pr{o, f}).\out = o && \axname{No Coconfusion.1}\\
&\forall o\cln O\ld \forall f\ld f\in \top^{I} \limplies (\cons\;\pr{o, f}).\nxt = f && \axname{No Coconfusion.2}
\end{aligned}$
\item the bisimilarity, which is here the usual bisimilarity of Moore machines: two states are related when they have the same output and their successors under each input are related,\\
$~~~~\begin{aligned}
&\mathit{bis}(R) \equiv \exists x,y\cln \nu \Mo^c\ld \pr{x,y}\land x.\out = y.\out \land
\forall i\cln I\ld \exists u,v\ld (x.\nxt)\;i = u\land (y.\nxt)\;i = v\land \pr{u,v}\in R\\
&\forall x,y\cln \nu \Mo^c\ld \pr{x,y}\in\nu R\ld\mathit{bis}(R)\limplies x = y \qquad\qquad\qquad\axname{Bisimilarity}
\end{aligned}$
\item the axiom \axname{Realisation} of Section~\ref{sec:fin-ml}, with paths given by lists over $P=\sum_{o\cln O}I$, i.e., over the pairs $\pr{o,i}$ of an output and an input.
\end{itemize}
By Lemma~\ref{lem:derived}, \axname{No Confusion} and \axname{No Coconfusion.3}, $\forall x\cln \nu \Mo^c\ld \cons\;\pr{x.\out, x.\nxt} = x$, hold in every model.
For each $n$, the pattern $\varphi^n_{\Mo}(\top)$, where $\varphi_{\Mo}(X)\equiv \exists o\cln O\ld \cons\;\pr{o,X^{I}}$, is matched by the states that can be unfolded $n$ times: they have an output, all their successors have an output, and so on, up to input words of length $n-1$. By Lemma~\ref{lem:cocontinuity}, $\inh{\nu\Mo^c}$ is matched by the states that can be unfolded indefinitely.
If $I$ is non-empty, every shape has positions, so $\mu\Mo^c$ is empty (Remark~\ref{rem:ini-hyp}): there are no finite Moore machine behaviours. If $I$ is empty, $\Mo\;X\approxeq O$, and both $\mu\Mo^c$ and $\nu\Mo^c$ are $O$.

\subsubsection*{The Behaviour of a State}

Let $M$ be a model of the theory above, $U=\ev{\inh{\nu\Mo^c}}{M}$, and let $O_M$ and $I_M$ be the interpretations of the sorts $O$ and $I$. For $x\in U$ and $i\in I_M$, we write $\delta(x,i)$ for the successor $\abs{x.\nxt}(i)\in U$. We extend it to input words $w\in I_M^*$ by $\delta^*(x,\varepsilon)=x$ and $\delta^*(x,i\cdot w)=\delta^*(\delta(x,i),w)$, and define the \emph{behaviour} of $x$ as the function
\[
\mathit{obs}(x):I_M^*\to O_M,\qquad \mathit{obs}(x)(w)=\delta^*(x,w).\out .
\]

\begin{thm}\label{thm:moore}
For every model $M$ of the theory above, $\mathit{obs}:U\to O_M^{I_M^*}$ is a bijection, and it is an isomorphism between the coalgebra $\alpha_\nu(M)=(U,\gamma)$ and the final $\Mo$-coalgebra $(\overline{\mathit{out}},\overline{\mathit{tr}}):O_M^{I_M^*}\to O_M\times(O_M^{I_M^*})^{I_M}$ of Example~\ref{ex:moore}. In particular, for $x\in U$ and $f\in O_M^{I_M^*}$,
\[
\mathit{obs}(x)=f\quad\text{iff}\quad x.\out=f(\varepsilon)\ \text{and}\ \mathit{obs}(\delta(x,i))=\overline{\mathit{tr}}(f)(i)\ \text{for all}\ i\in I_M.
\]
\end{thm}
\begin{proof}
$\mathit{obs}$ is a coalgebra morphism: $\overline{\mathit{out}}(\mathit{obs}(x))=\mathit{obs}(x)(\varepsilon)=x.\out$, and, for $i\in I_M$ and $w\in I_M^*$,
\[
\overline{\mathit{tr}}(\mathit{obs}(x))(i)(w)=\mathit{obs}(x)(i\cdot w)=\delta^*(\delta(x,i),w).\out=\mathit{obs}(\delta(x,i))(w).
\]
Since $\gamma(x)=(x.\out,\abs{x.\nxt})$, these two equations say exactly that $(\overline{\mathit{out}},\overline{\mathit{tr}})\circ\mathit{obs}=\Mo\,\mathit{obs}\circ\gamma$. By Theorem~\ref{thm:adequacy-fin}, $(U,\gamma)$ is a final $\Mo^c$-coalgebra, hence a final $\Mo$-coalgebra up to the isomorphism $\Mo\approxeq\Mo^c$, and $(\overline{\mathit{out}},\overline{\mathit{tr}})$ is a final $\Mo$-coalgebra~\cite{rutten00}. A coalgebra morphism between final coalgebras is the unique isomorphism between them, so $\mathit{obs}$ is a bijection. The last claim is the morphism property, read as a characterization: $\mathit{obs}(x)$ is determined by its value at $\varepsilon$ and by the functions $\overline{\mathit{tr}}(\mathit{obs}(x))(i)$.
\end{proof}

For $O=\{0,1\}$, Moore machines are deterministic automata, $\mathit{obs}(x)$ is (the characteristic function of) the language accepted from the state $x$, and Theorem~\ref{thm:moore} says that every model contains exactly one state for each language over $I_M$. In this case, the Bisimulation Principle of Section~\ref{sec:fin-ml} is the usual coinductive proof method for language equivalence: to show that two states are equal, it is enough to exhibit a relation that relates them, preserves outputs, and is closed under successors.

\section{Conclusion}
The results reported in this paper show that both the initial models and the final models of polynomial functors (in the container sense) can be fully captured in matching logic (ML), using their representation as unary container functors, in the precise sense of Section~\ref{sec:intro}, which comprises semantic adequacy for all models and the derivability of the induction, coinduction, and bisimulation principles, and not of other principles such as iteration: for every model $M$ of the corresponding theory, the algebra $\alpha_\mu(M)$ is an initial algebra (Theorem~\ref{thm:adequacy-ini}) and the coalgebra $\alpha_\nu(M)$ is a final coalgebra (Theorem~\ref{thm:adequacy-fin}) of the container functor extracted from $M$, i.e., the one given by the interpretation in $M$ of the shapes and positions. So the data types are captured relative to the parameter sorts of shapes and positions, and absolutely whenever these sorts are themselves captured (Proposition~\ref{prop:absolute}); Theorem~\ref{thm:captured} shows that this is the case for finite sets, products, dependent sums, and (co)inductive sorts built from them. Moreover, the induction, coinduction, and bisimulation principles are derivable in the ML proof system. Section~\ref{sec:summary} separates the requirements for defining these theories, for their satisfiability, and for their adequacy, and distinguishes the semantic results from the derivable ones. Besides the axioms of the constructor, of the fixpoints, and of the destructors, the theories of Definition~\ref{def:gamma} use only the base theory $\Gamma_0$, i.e., $\DEF$, $\INH$, $\PAIR$ with \prule{Pair Total}, the specification of $\initMorph$, and the indexed family $a\cln A\vdash B[a]$, together with the dependent sum $P$ of positions and the lists $\mathit{List}(P)$ for final coalgebras; functions are represented directly by the notation $X^{B[a]}$, and finite sums by dependent sums over finite sorts of tags. The general specifications of sum sorts and function sorts~\cite{CLR21a} are not used, and dependent products are not specified.

Two features of ML are essential. Fixpoint patterns express \axname{No Junk} and \axname{No Cojunk}, and free set variables allow the existence axioms \axname{Tab} and \axname{Realisation} to speak about every subset of a model; without them, the theories have non-standard models (Remarks~\ref{rem:ini-hyp} and~\ref{rem:realisation}). The comparison of the two theories (Proposition~\ref{prop:local-global}) also clarifies the duality between them: the fixpoint axioms and the (co)confusion axioms are dual, while the remaining axioms play the same role in both theories, in a local form for initial algebras, where induction propagates them, and in a global form for final coalgebras, where it does not.

A classical polynomial functor can be translated into a container functor, because container functors are closed under sum, product, and exponentiation by a constant object (Section~\ref{sec:poly-as-cont}). The result may be cumbersome, because the construction starts from the constant and identity functors, so it is preferable to simplify it using isomorphisms in the category of sets, as we did for lists and Moore machines.

These results can help in defining in ML programming languages that use both inductive and coinductive data types, and in a better understanding of the abstract constructions of category theory. A possible use is as follows:
\begin{itemize}
\item define in a front-end a suitable syntax for data types intended to be defined as initial algebras or final coalgebras;
\item extract the canonical container form of the functor underlying the front-end definition;
\item generate the corresponding ML theory;
\item derive the proof principles needed to soundly handle the defined data type.
\end{itemize}

Future work will focus on the following aspects of the proposed approach:
\begin{itemize}
\item formal proof objects for the derived principles;
\item the iteration and primitive recursion principles, and their duals for final coalgebras~\cite{GeuversP07}, which are not covered by the notion of capture used here; for term algebras, the iteration principle is derived in ML in~\cite{ChenLR26};
\item capturing more general families of position sets, such as $n\cln\mathbb N\vdash[n]$ (Remark~\ref{rem:limits});
\item extending the approach to larger classes of functors admitting initial algebras and final coalgebras, e.g., indexed containers~\cite{altenkirchGHMM15}, the bounded natural functors (BNFs) underlying Isabelle/HOL's datatypes~\cite{TraytelPB12}, or the quotients of polynomial functors, experimentally implemented in Lean~\cite{AvigadCH19}.
\end{itemize}

\paragraph{AI disclosure}
Parts of the manuscript were prepared with the support of Claude Fable. All AI-generated suggestions were carefully reviewed and edited by the author to ensure scientific accuracy and compliance with academic standards. The author takes full responsibility for the content of the publication.

\bibliographystyle{eptcs}
\bibliography{refs}

\appendix

\section{The Pair Sort in ML}
\label{ap:sorts}

We recall from~\cite{CLR21a} the ML specification of pair sorts, with the adaptations described below.
The specification of sum sorts is not recalled: a finite sum of sorts is represented as a dependent sum $\sum_{i\cln\mathit{Tag}}C[i]$ over a finite sort of tags (Definitions~\ref{def:gamma} and~\ref{def:captured}).
The specification $\PAIR$ states \prule{Pair Inj} for all elements, not only for elements of sorts, and adds \prule{Pair Total}, which makes pairing defined on all elements. \prule{Pair Total} is needed because the constructor $\cons$ is applied to pairs $\pr{a,f}$ whose second component represents a function and belongs to no sort; the unrelativized \prule{Pair Inj} is needed for dependent pairs $\pr{a,b}$, whose second components $b\cln B[a]$ need not belong to a common sort.
The specification of function sorts is not recalled either: the theories represent functions by the notation $X^{B[a]}$ of Section~\ref{sec:cont-ml}.
Of the specifications of sort operators of~\cite{CLR21a}, only $\PAIR$ is part of the theories $\Gamma_F$, $\Gamma^\mu_F$, and $\Gamma^\nu_F$ of Definition~\ref{def:gamma}.

\begin{lstlisting}[language=matchinglogic]
spec $\PAIR_{s_1,s_2}$ (*\Impts*) $\INH$                
(*\Smbs*) $\mathit{Pair},\mathit{pair},  \mathit{fst},   \mathit{snd}$
(*\Sugar*) $s_1 \otimes s_2 \equiv \mathit{Pair} \ s_1 \ s_2,
$ $\pr{\varphi_1,\varphi_2} \equiv \mathit{pair}\ \varphi_1 
\ \varphi_2$
(*\Axms*) (*$\sslash$ all axioms are quantified by ``$\forall s_1 , s_2 \cln 
\Sort $''*)
$\begin{array}{ll}
\prule{Pair Sort} & (s_1 \otimes s_2) \cln \Sort \\
\prule{Pair} & 
\forall x_1 \cln s_1 \ld \forall x_2 \cln s_2 \ld 
\pr{x_1,x_2} \cln (s_1 \otimes s_2) \\
\prule{Pair Fst} &
\forall x_1 \cln s_1 \ld \forall x_2 \cln s_2 \ld
\mathit{fst} \ \pr{x_1,x_2} = x_1 \\
\prule{Pair Snd} &
\forall x_1 \cln s_1 \ld \forall x_2 \cln s_2 \ld
\mathit{snd} \ \pr{x_1,x_2} = x_2 \\
\prule{Pair Inj} & 
\forall x_1,x_2,y_1,y_2 \ld \pr{x_1,x_2} {=} \pr{y_1,y_2} \limplies x_1 {=} y_1 \land x_2 {=} y_2 \\&\quad\text{(not relativized to sorts)} \\
\prule{Pair Domain} &  \inh{s_1 \otimes 
s_2} = \pr{\inh{s_1}, \inh{s_2}} \\
\prule{Pair Total} & \forall x\ld\forall y\ld\exists z\ld \pr{x,y}=z \quad\text{(not relativized to sorts)}
\end{array}$
endspec 
\end{lstlisting}

\section{A Standard Model}
\label{ap:model}

In this appendix we construct, for a container functor $X\mapsto F\;X=\sum_{a\cln A}X^{B[a]}$, one model of the theories of Section~\ref{sec:fc-ml} in which $\inh{\mu F}$ is the W-type and $\inh{\nu F}$ is the M-type of $a\cln A\vdash B[a]$. This proves Propositions~\ref{prop:soundness-ini} and~\ref{prop:soundness-fin}.

\paragraph{The theory}
Recall (Definition~\ref{def:gamma}) that $\Gamma_F$ is the union of $\DEF$, $\INH$, $\PAIR$ (Appendix~\ref{ap:sorts}), the specification of $\initMorph$ (Section~\ref{sec:elements}), the indexed family $a\cln A\vdash B[a]$, the theory of the initial algebra (Section~\ref{sec:ini-ml}), and the theory of the final coalgebra, including the dependent sum $P\equiv\sum_{a\cln A}B[a]$ and $\mathit{List}(P)$ (Section~\ref{sec:fin-ml}). The model below also interprets $\objOne$, $\star$, and $\finMorph$, which are used by the parameter specifications and the examples, and satisfies \axname{No Confusion} and \axname{No Coconfusion.3}, which are consequences of $\Gamma^\nu_F$ (Lemma~\ref{lem:derived}).

\paragraph{Disjoint representatives}
The construction needs the shapes and the positions to be distinct from each other and from the objects it builds, which is not the case for arbitrary sets: for the identity functor, $A=B[\star]=\{\star\}$, and a position set may itself contain ordered pairs. We therefore use tagged copies. Let $A^\sharp=\{\mathsf{sh}\}\times A$ and, for $a\in A$, $B^\sharp[(\mathsf{sh},a)]=\{\mathsf{pos}\}\times B[a]$, whose elements are taken as fresh atoms, distinct from all the other objects below. The tagging bijections $\iota_A:A\to A^\sharp$ and $\iota_{B,a}:B[a]\to B^\sharp[\iota_A(a)]$ form an isomorphism between the containers $a\cln A\vdash B[a]$ and $a\cln A^\sharp\vdash B^\sharp[a]$, hence a natural isomorphism between the corresponding functors, and isomorphisms between their W-types and between their M-types. Positions are tagged by their role only, not by their shape, so equality and inclusion between position sets are preserved: shapes with the same position set still share their codes, and nested position sets remain nested. We carry out the construction for $a\cln A^\sharp\vdash B^\sharp[a]$ and, to keep the notation light, we write $A$ and $B[a]$ for $A^\sharp$ and $B^\sharp[a]$ in the rest of this appendix, except in the statement of Theorem~\ref{thm:standard-model}.

\paragraph{Trees}
Let $\zeta:Z\to F\,Z$ be the final $F$-coalgebra, whose elements are the possibly infinite trees with nodes labelled by shapes $a\in A$ and with branching $B[a]$; by Lambek's lemma, $\zeta$ is a bijection. Let $W_Z\subseteq Z$ be the set of well-founded trees. By well-founded recursion, the W-type of $a\cln A\vdash B[a]$ is isomorphic to $W_Z$ with the algebra structure $\zeta^{-1}$ restricted to $F\,W_Z$; equivalently, $W_Z$ is the least set $S\subseteq Z$ with $\zeta^{-1}(a,g)\in S$ for all $a\in A$ and $g:B[a]\to S$.

\paragraph{External and internal finality}
The final coalgebra $(Z,\zeta)$ exists in ZFC, e.g., as the limit of the terminal sequence $1\leftarrow F\,1\leftarrow F^2\,1\leftarrow\cdots$, since container functors preserve connected limits, in particular limits of $\omega^{\mathrm{op}}$-chains. The construction below uses the finality of $(Z,\zeta)$ and the description of $W_Z$ as a least fixpoint only \emph{externally}, as set-theoretic facts used to check that $M$ satisfies the axioms, in particular \axname{No Junk}, \axname{No Cojunk}, and \axname{Bisimilarity}. The \emph{internal} characterization, i.e., that the algebra $\alpha_\mu(M)$ extracted from $M$ is initial and the coalgebra $\alpha_\nu(M)$ is final, is not assumed: once $M$ is shown to be a model, it follows from the axioms by Theorems~\ref{thm:adequacy-ini} and~\ref{thm:adequacy-fin}.

\paragraph{Tagging convention}
All objects of the construction are tagged with their kind, so that objects of different kinds are distinct: a symbol atom is $(\mathsf{sym},\sigma)$, a sort name is $(\mathsf{srt},n)$, a shape and a position are $(\mathsf{sh},a)$ and $(\mathsf{pos},b)$ as above, a tree is $(\mathsf{tree},z)$, a pair is $(\mathsf{pair},x,y)$, a sequence is $(\mathsf{seq},x_1,\ldots,x_n)$, a partial application is $(\mathsf{part},\sigma,x)$, a code is $(\mathsf{code},\mathit{graph}(g))$, and a fresh term is $(\mathsf{sup},a,f)$; the element $\star$ is $(\mathsf{star})$. Objects of the same kind are equal iff their components are equal, and every constructed object differs from its components. These objects are defined set-theoretically, independently of the stages below, which only determine which of them belong to $M$; we keep the usual notation, e.g., $\pr{x,y}$ for $(\mathsf{pair},x,y)$ and $z$ for $(\mathsf{tree},z)$. The tagged copies $A^\sharp$ and $B^\sharp[a]$ are an instance of this convention. In the combined construction below, the trees of $Z_j$ and the fresh terms of $G_j$ are tagged with $\mathsf{tree}_j$ and $\mathsf{sup}_j$, and the constants of (Fin) with $\mathsf{fin}$.

\paragraph{The carrier}
The \emph{atoms} are the following pairwise distinct objects, which are also distinct from all objects constructed below: an atom $\hat\sigma$ for each symbol $\sigma$ of $\Gamma_F$; the elements of $A$ and of $\bar B=\bigcup_{a\in A}B[a]$; the element $\star$; the trees of $Z$; and the \emph{sort names}, i.e., the elements of the least set $N$ that contains $\mathsf{Srt}$, $\mathsf{1}$, $\mathsf{A}$, $\mathsf{B}_a$ for $a\in A$, $\mathsf{P}$, $\mathsf{LP}$, $\mathsf{mu}$, and $\mathsf{nu}$, and contains $s\otimes s'$ for all $s,s'\in N$.
Fix a regular cardinal $\theta>\aleph_0$ with $\abs{B[a]}<\theta$ for all $a\in A$. Let $M_0$ be the set of atoms. For an ordinal $\xi<\theta$, let $M_{\xi+1}$ be $M_\xi$ together with the following new objects:
\begin{description}
\item[(C1)] the pairs $\pr{x,y}$, for $x,y\in M_\xi$;
\item[(C2)] the finite sequences $(x_1,\ldots,x_n)$, $n\ge 0$, of elements of $M_\xi$;
\item[(C3)] the partial applications $\sigma_x$, for $\sigma\in\{\mathit{pair},\pcons,\mathit{Pair},\mathit{depOf},\mathit{DepSum}\}$ and $x\in M_\xi$;
\item[(C4)] the \emph{codes} $\ulcorner g\urcorner$, for $a\in A$ with $B[a]\neq\emptyset$ and $g:B[a]\to M_\xi$; a code is determined by the graph of $g$, so shapes with the same position set share their codes;
\item[(C5)] the \emph{fresh terms} $\mathit{sup}(a,f)$, for $a\in A$ and $f\in M_\xi$.
\end{description}
At limit ordinals we take unions, and we let $M=\bigcup_{\xi<\theta}M_\xi$. The \emph{stage} of $x\in M$ is the least $\xi$ with $x\in M_\xi$. Since $\theta$ is regular and $\abs{B[a]}<\theta$, every function $g:B[a]\to M$ has its values in some $M_\xi$, so $\ulcorner g\urcorner\in M$; similarly, $M$ is closed under (C1)--(C3) and (C5).

\paragraph{The interpretation}
Each symbol $\sigma$ is interpreted as $\{\hat\sigma\}$, except the sort symbols, which are interpreted as their sort names ($\mathit{Sort}$ as $\{\mathsf{Srt}\}$, $\objOne$ as $\{\mathsf 1\}$, $A$ as $\{\mathsf A\}$, $\mu F$ as $\{\mathsf{mu}\}$, $\nu F$ as $\{\mathsf{nu}\}$), $\mathit{star}$, which is interpreted as $\{\star\}$, and $\mathit{pnil}$, which is interpreted as $\{()\}$.
For $a\in A$, let $\mathit{Fun}_a(M)$ be defined as in Section~\ref{sec:ini-ml}: $\{\hat\initMorph\}$ if $B[a]=\emptyset$, and otherwise the set of $f\in M$ such that $f\apd b$ is a singleton for $b\in B[a]$ and $f\apd b=\emptyset$ for $b\notin B[a]$. Application is $\emptyset$ except in the following cases:
\begin{itemize}
\item $\hat{\smbceil}\apd y=M$; $\hat{\mathit{inh}}\apd s=\top(s)$ for $s\in N$, where $\top$ is defined below;
\item $\hat{\mathit{pair}}\apd x=\{\mathit{pair}_x\}$, $\mathit{pair}_x\apd y=\{\pr{x,y}\}$, $\hat{\mathit{fst}}\apd\pr{x,y}=\{x\}$, $\hat{\mathit{snd}}\apd\pr{x,y}=\{y\}$;
\item $\hat{\mathit{Pair}}\apd s=\{\mathit{Pair}_s\}$ and $\mathit{Pair}_s\apd s'=\{s\otimes s'\}$, for $s,s'\in N$;
\item $\hat{\mathit{depOf}}\apd\hat B=\{\mathit{depOf}_{\hat B}\}$ and $\mathit{depOf}_{\hat B}\apd a=\{\mathsf B_a\}$, for $a\in A$; $\hat{\mathit{DepSum}}\apd\mathsf A=\{\mathit{DepSum}_{\mathsf A}\}$ and $\mathit{DepSum}_{\mathsf A}\apd\hat B=\{\mathsf P\}$;
\item $\hat{\finMorphIt}\apd y=\{\star\}$ (and $\hat{\initMorphIt}\apd y=\emptyset$);
\item $\ulcorner g\urcorner\apd b=\{g(b)\}$ for $b$ in the domain of $g$;
\item $\hat{\mathit{List}}\apd\mathsf P=\{\mathsf{LP}\}$; $\hat{\pcons}\apd q=\{\pcons_q\}$ and $\pcons_q\apd(x_1,\ldots,x_n)=\{(q,x_1,\ldots,x_n)\}$;
\item for $a\in A$ and $f\in\mathit{Fun}_a(M)$: $\hat{\cons}\apd\pr{a,f}=\{\zeta^{-1}(a,\abs{f})\}$ if all values of $\abs{f}$ are trees, and $\hat{\cons}\apd\pr{a,f}=\{\mathit{sup}(a,f)\}$ otherwise;
\item for $z\in Z$ with $\zeta(z)=(a,g)$: $\hat{\out}\apd z=\{a\}$, and $\hat{\nxt}\apd z=\{\ulcorner g\urcorner\}$ if $B[a]\neq\emptyset$, $\{\hat\initMorph\}$ otherwise; $\hat{\out}\apd\mathit{sup}(a,f)=\{a\}$ and $\hat{\nxt}\apd\mathit{sup}(a,f)=\{f\}$.
\end{itemize}
The case of $\hat\cons$ is well defined: whether $f\in\mathit{Fun}_a(M)$ depends on the applications $f\apd b$, and $\hat\cons$ itself is not in $\mathit{Fun}_a(M)$, since $\hat\cons\apd b=\emptyset$ for the atoms $b\in B[a]$.
The inhabitants of the sort names are: $\top(\mathsf{Srt})=N$, $\top(\mathsf 1)=\{\star\}$, $\top(\mathsf A)=A$, $\top(\mathsf B_a)=B[a]$, $\top(\mathsf P)=\{\pr{a,b}\mid a\in A, b\in B[a]\}$, $\top(\mathsf{LP})$ the set of finite sequences of elements of $\top(\mathsf P)$, $\top(s\otimes s')=\{\pr{x,y}\mid x\in\top(s),y\in\top(s')\}$, $\top(\mathsf{mu})=W_Z$, and $\top(\mathsf{nu})=Z$.

\begin{lem}\label{lem:only-codes}
Let $a\in A$ with $B[a]\neq\emptyset$, and let $S\subseteq M$ consist of trees and fresh terms. Then $f\in\mathit{Fun}_a(S)$ iff $f$ is a code $\ulcorner g\urcorner$ with $g:B[a]\to S$.
\end{lem}
\begin{proof}
Codes on $B[a]$ with values in $S$ are clearly in $\mathit{Fun}_a(S)$. Conversely, let $f\in\mathit{Fun}_a(S)$ and $b\in B[a]$, so $f\apd b=\{s\}$ with $s$ a tree or a fresh term. By the definition of application, the only elements whose application to an atom of $\bar B$ can return a tree or a fresh term are the codes: $\hat\cons$, $\mathit{fst}$, and $\mathit{snd}$ return non-empty sets only on pairs, $\hat\out$ and $\hat\nxt$ only on trees and fresh terms, $\hat\smbceil$ returns $M$, which is not a singleton, and the other cases return elements that are neither trees nor fresh terms (positions are atoms, distinct from sort names, shapes, pairs, and sequences). So $f=\ulcorner g\urcorner$ for some $g:B[a']\to M$, and strictness forces $B[a']=B[a]$: an element of $B[a]\setminus B[a']$ would have an empty application, and an element of $B[a']\setminus B[a]$ a non-empty one.
\end{proof}

\begin{thm}\label{thm:standard-model}
$M$ is a model of $\Gamma_F$ in which $A_M=A^\sharp$ and $B_M[\iota_A(a)]=B^\sharp[\iota_A(a)]$ for $a\in A$, which are in bijection with $A$ and $B[a]$ via $\iota_A$ and $\iota_{B,a}$; $\inh{\mu F}$ is interpreted as $W_Z$, $\inh{\nu F}$ as $Z$, and $\cons$ acts on them as $\zeta^{-1}$. Hence, by Theorems~\ref{thm:adequacy-ini} and~\ref{thm:adequacy-fin} (or directly from this interpretation), $\alpha_\mu(M)$ and $\alpha_\nu(M)$ are an initial algebra and a final coalgebra, isomorphic to the W-type and to the M-type of $a\cln A\vdash B[a]$.
\end{thm}
\begin{proof}
We check the axioms by groups. Patterns are interpreted pointwise, so it suffices to check them on elements.

\emph{Equality, sorts, pairs, $\objOne$, $\initMorph$, $\finMorph$, the indexed family, and the dependent sum.} \prule{Definedness} holds because $\hat\smbceil\apd x=M$. $\mathit{Sort}$ denotes the element $\mathsf{Srt}\in N=\top(\mathsf{Srt})$. For $s,s'\in N$, $s\otimes s'\in N$, and the axioms of $\PAIR$ hold because pairs are injective and $\top(s\otimes s')$ is the set of pairs of inhabitants; \prule{Pair Total} holds because $\pr{x,y}\in M$ for all $x,y\in M$. The axioms of Section~\ref{sec:fc-ml} for $\star$, $\objOne$, $\initMorph$, and $\finMorph$ hold by definition. $A$ denotes $\mathsf A\in N$, $B[a]$ denotes $\mathsf B_a\in N$ for $a\in A$, and $P$ denotes $\mathsf P\in N$, with $\top(\mathsf P)$ the set of dependent pairs. $\mu F$ and $\nu F$ denote $\mathsf{mu},\mathsf{nu}\in N$, so \axname{Mu Sort} and \axname{Nu Sort} hold.

\emph{Constructor.} By Lemma~\ref{lem:only-codes}, the elements of $\mathit{Fun}_a(M)$ whose values are all trees are the codes on $B[a]$ (or $\hat\initMorph$, if $B[a]=\emptyset$). \axname{Functional} holds by the definition of $\hat\cons$. For \axname{No Confusion}, let $c(a,f)=c(a',f')$. If this element is a tree $z$, then $\zeta(z)=(a,\abs{f})=(a',\abs{f'})$, so $a=a'$, and $f$ and $f'$ are codes on $B[a]$ with the same graph, so $f=f'$. Otherwise it is a fresh term, and fresh terms are injective in $(a,f)$.

\emph{Initial algebra.} For $S\subseteq Z$, Lemma~\ref{lem:only-codes} and the closure of $M$ under codes give
$\Phi_F(S)=\{\zeta^{-1}(a,g)\mid a\in A,\ g:B[a]\to S\}\subseteq Z$.
The least fixpoint of $\Phi_F$ is the union of the transfinite iteration of $\Phi_F$ from $\emptyset$, which stays in $Z$; so it is the least $S\subseteq Z$ closed under $\zeta^{-1}$, i.e., $W_Z$, and \axname{No Junk} holds. By Lemma~\ref{lem:only-codes}, $\mathit{Fun}_a(W_Z)$ is the set of codes on $B[a]$ with values in $W_Z$, and every function $B[a]\to W_Z$ has exactly one code. This gives \axname{Ext}, and \axname{Tab}: if $R$ satisfies the premise of \axname{Tab}, the code of the function that maps $b$ to the unique $x$ with $\pr{b,x}\in R$ satisfies the conclusion.

\emph{Final coalgebra.} Every tree $z$ with $\zeta(z)=(a,g)$ is $c(a,f)$, where $f$ is the code of $g$ (or $\hat\initMorph$), so $Z\subseteq\Phi_F(Z)$ and $Z\subseteq U=\mathbf{gfp}(\Phi_F)$. Conversely, $U\subseteq\Phi_F(U)$, so every element of $U$ is a tree or a fresh term. If $\mathit{sup}(a,f)\in U$, then $f\in\mathit{Fun}_a(U)$, so $f$ is a code by Lemma~\ref{lem:only-codes}, and, since $\hat\cons\apd\pr{a,f}$ is not a tree, some value of $f$ is a fresh term in $U$ of a smaller stage. Iterating, we would get an infinite decreasing sequence of ordinals, which is impossible. Hence $U=Z$, and \axname{No Cojunk} holds. The coconfusion axioms hold by the definitions of $\hat\out$ and $\hat\nxt$, using that codes on $B[a]$ are determined by their graphs.
For \axname{Bisimilarity}, let $R$ be a set of pairs with $R\subseteq\ev{\mathit{bis}(R)}{}$; it relates trees with the same shape whose children at each position are related by $R$, so it is a bisimulation on $(Z,\zeta)$. Since $(Z,\zeta)$ is final, related trees are equal (both projections of a bisimulation are coalgebra morphisms into $Z$). The interpretation of $\mathit{Bisim}$ is the union of all such $R$, so \axname{Bisimilarity} holds.

\emph{Lists and realisation.} $\mathit{List}(P)$ denotes $\mathsf{LP}\in N$, so \axname{List Sort} holds, and $\top(\mathsf{LP})$ is the set of finite sequences over $\top(\mathsf P)$, with $\mathit{pnil}$ the empty sequence and $\pcons$ the prepending of an element; the axioms of $\mathit{List}(P)$ hold, and the least fixpoint in its \axname{No Junk} axiom is exactly this set. Let $T\subseteq M$ satisfy $\mathit{Tree}(T)$. The other axioms of $\Gamma^\nu_F$ hold, as shown above, and $\beta$ is the identity of $Z$; so, by Lemma~\ref{lem:paths}(2) and~(3), the tree $z_T$ satisfies $\mathit{Beh}(z_T,T)$, and \axname{Realisation} holds.

Finally, $A_M=\top(\mathsf A)=A^\sharp$ and $B_M[a]=\top(\mathsf B_a)=B^\sharp[a]$, which are in bijection with the original sets via the tagging bijections, and, on $W_Z$ and $Z$, $\cons$ acts as $\zeta^{-1}$ and $(\out,\nxt)$ as $\zeta$, up to replacing children functions by their codes.
\end{proof}

\subsection*{Variants of the Standard Model}

The following variants of the construction give models in which one axiom fails; they show that \axname{Ext}, \axname{Tab}, \axname{Bisimilarity}, and \axname{Realisation} are needed for the adequacy theorems. A \emph{seed} is a set $Z'$ of atoms, tagged as trees, with an injective map $\zeta':Z'\to F\,Z'$. For a seed and $m\in\{0,1,2\}$, the structure $M(Z',\zeta',m)$ is defined as $M$ above, with the following changes:
\begin{itemize}
\item the tree atoms are the elements of $Z'$ instead of those of $Z$;
\item (C4) adds $m$ copies $\ulcorner g\urcorner^0,\ldots,\ulcorner g\urcorner^{m-1}$ of each code, tagged with their index, with $\ulcorner g\urcorner^i\apd b=\{g(b)\}$ for all copies;
\item for $a\in A$ and $f\in\mathit{Fun}_a(M)$, $\hat\cons\apd\pr{a,f}=\{z\}$ if $\zeta'(z)=(a,\abs f)$ for some $z\in Z'$, and $\{\mathit{sup}(a,f)\}$ otherwise;
\item for $z\in Z'$ with $\zeta'(z)=(a,g)$, $\hat\out\apd z=\{a\}$ and $\hat\nxt\apd z=\{\ulcorner g\urcorner^0\}$ if $B[a]\neq\emptyset$, $\{\hat\initMorph\}$ otherwise (for $m=0$ we only use $Z'=\emptyset$);
\item $\top(\mathsf{mu})=\mathbf{lfp}(\Phi_F)$ and $\top(\mathsf{nu})=\mathbf{gfp}(\Phi_F)$, where $\Phi_F(S)=\{c(a,f)\mid a\in A,\ f\in\mathit{Fun}_a(S)\}$ is determined by the application alone.
\end{itemize}
The standard model is $M(Z,\zeta,1)$.

\begin{lem}\label{lem:variants}
Let $M'=M(Z',\zeta',m)$.
\begin{enumerate}
\item $M'$ satisfies $\Gamma_0$, the axioms of the dependent sum $P$ and of $\mathit{List}(P)$, and the axioms \axname{Mu Sort}, \axname{Nu Sort}, \axname{Functional}, \axname{No Junk}, and \axname{No Cojunk}. For $a\in A$ with $B[a]\neq\emptyset$ and a set $S$ of tree atoms and fresh terms, $\mathit{Fun}_a(S)$ is the set of the codes, of any copy, of the functions $g:B[a]\to S$.
\item If $m\le 1$ or $Z'=\emptyset$, then $M'$ satisfies \axname{No Confusion} and \axname{No Coconfusion.1--2}.
\item If $m=1$, then $M'$ satisfies \axname{Ext} and \axname{Tab}.
\item If $m=1$ and $\zeta'$ is a bijection, then $\mathbf{gfp}(\Phi_F)=Z'$.
\end{enumerate}
\end{lem}
\begin{proof}
(1) The axioms of $\Gamma_0$, of $P$, and of $\mathit{List}(P)$ are checked as in the proof of Theorem~\ref{thm:standard-model}; $\hat\cons\apd\pr{a,f}$ is a singleton for $a\in A$ and $f\in\mathit{Fun}_a(M)$; \axname{No Junk} and \axname{No Cojunk} hold by the definition of $\top$ on $\mathsf{mu}$ and $\mathsf{nu}$. The description of $\mathit{Fun}_a(S)$ is proved as Lemma~\ref{lem:only-codes}, whose proof does not depend on the number of copies of the codes or on the set of tree atoms.
(2) Let $c(a,f)=c(a',f')$. If it is a fresh term, then $(a,f)=(a',f')$. If it is $z\in Z'$, then $\zeta'(z)=(a,\abs f)=(a',\abs{f'})$, so $a=a'$ and $\abs f=\abs{f'}$; by (1), $f$ and $f'$ are codes of the same function (or both are $\hat\initMorph$), and $f=f'$ since $m\le 1$. For $Z'=\emptyset$, only the first case occurs. The coconfusion axioms hold by the definitions of $\hat\out$ and $\hat\nxt$: on a fresh term they return $a$ and $f$, and on $z=c(a,f)\in Z'$ they return $a$ and the code $\ulcorner\abs f\urcorner^0$, which is $f$ by (1), since $m\le1$.
(3) Every element of $\mathbf{lfp}(\Phi_F)$ is a value of $\hat\cons$, i.e., a tree atom or a fresh term; by (1), $\mathit{Fun}_a(\mathbf{lfp}(\Phi_F))$ is the set of the codes of the functions from $B[a]$ to $\mathbf{lfp}(\Phi_F)$, and each function has exactly one code.
(4) If $\zeta'(z)=(a,g)$, then $z=c(a,f)$, where $f$ is the code of $g$ (or $\hat\initMorph$), so $Z'\subseteq\Phi_F(Z')$ and $Z'\subseteq\mathbf{gfp}(\Phi_F)$. Conversely, every element of the greatest fixpoint is a value of $\hat\cons$. If it is a fresh term $\mathit{sup}(a,f)$, then $f\in\mathit{Fun}_a(\mathbf{gfp}(\Phi_F))$, and not all values of $\abs f$ are in $Z'$: otherwise $(a,\abs f)\in F\,Z'=\zeta'(Z')$ and $\hat\cons\apd\pr{a,f}$ would be a tree atom. So $f$ is a code, by (1), one of whose values is a fresh term in the greatest fixpoint, of a smaller stage. Iterating gives an infinite strictly decreasing sequence of ordinals, which is impossible.
\end{proof}

\begin{prop}\label{prop:necessity}
Each of the following structures satisfies all the axioms of the indicated theory except the indicated one, and its extracted algebra or coalgebra is not initial or final.
\begin{enumerate}
\item \emph{Without \axname{Ext}.} For the natural numbers (shapes $\{z,s\}$, $B[z]=\emptyset$, $B[s]=\{\star\}$), $M(\emptyset,\emptyset,2)$ satisfies $\Gamma^\mu_F$ except \axname{Ext}. The element $0=c(z,\hat\initMorph)$ is in $\inh{\mu F}$, and the function $\star\mapsto 0$ has two codes $f^0\neq f^1$, so $c(s,f^0)\neq c(s,f^1)$ represent the same tree. For every choice of a code $\ulcorner g\urcorner$ for each function, the map $\alpha(a,g)=c(a,\ulcorner g\urcorner)$ is not surjective onto $\inh{\mu F}$, so it is not the structure map of an initial algebra, which is a bijection.
\item \emph{Without \axname{Tab}.} For the same container, $M(\emptyset,\emptyset,0)$ satisfies $\Gamma^\mu_F$ except \axname{Tab}. No element represents a function on $\{\star\}$ into the fresh terms, so $\inh{\mu F}$ is interpreted as $\{0\}$, while the W-type is $\mathbb N$. \axname{Tab} fails for $R=\{\pr{\star,0}\}$.
\item \emph{Without \axname{Bisimilarity}.} For the streams over $\{0\}$ (one shape $0$ with $B[0]=\{\star\}$), let $Z'=\{x,y\}$ with $\zeta'(x)=(0,\star\mapsto x)$ and $\zeta'(y)=(0,\star\mapsto y)$, a bijection onto $F\,Z'$. Then $M(Z',\zeta',1)$ satisfies $\Gamma_F$ except \axname{Bisimilarity}: $\inh{\nu F}$ is interpreted as $\{x,y\}$, and $\{\pr{x,y}\}$ is included in its image under $\mathit{bis}$, so $\pr{x,y}\in\mathit{Bisim}$ and $x\neq y$. The final coalgebra has one element.
\item \emph{Without \axname{Realisation}, empty $\nu F$.} For the identity functor (one shape with one position), $M(\emptyset,\emptyset,1)$ satisfies $\Gamma_F$ except \axname{Realisation}, and $\inh{\nu F}$ is interpreted as $\emptyset$, while the final coalgebra has one element.
\item \emph{Without \axname{Realisation}, missing behaviours.} For the streams over $E=\{0,1\}$ (shapes $E$, $B[e]=\{\star\}$), let $\mathit{EP}\subseteq E^\omega$ be the set of eventually periodic streams, with $\zeta'(e\cdot s)=(e,\star\mapsto s)$, a bijection onto $F\,\mathit{EP}$. Then $M(\mathit{EP},\zeta',1)$ satisfies $\Gamma_F$ except \axname{Realisation}, and $\inh{\nu F}$ is interpreted as the countable set $\mathit{EP}$, while the final coalgebra $E^\omega$ is uncountable.
\item \emph{Without \axname{Realisation}, with $\inh{\mu F}$ non-empty and dense.} For the colists over $E=\{0,1\}$ (shapes $\{\mathit{n},\mathit{c}_0,\mathit{c}_1\}$, $B[\mathit{n}]=\emptyset$, $B[\mathit{c}_e]=\{\star\}$), let $Z'=E^*\cup\mathit{EP}$ be the set of the finite lists and of the eventually periodic infinite lists, with $\zeta'(\varepsilon)=(\mathit{n},\emptyset\to Z')$ and $\zeta'(e\cdot s)=(\mathit{c}_e,\star\mapsto s)$, a bijection onto $F\,Z'$. Then $M(Z',\zeta',1)$ satisfies $\Gamma_F$ except \axname{Realisation}; $\inh{\mu F}$ is interpreted as the set $E^*$ of the finite lists, which is non-empty and dense in the interpretation $Z'$ of $\inh{\nu F}$ (Proposition~\ref{prop:mu-nu}(3)), and the infinite lists that are not eventually periodic are not represented.
\end{enumerate}
\end{prop}
\begin{proof}
In all cases, Lemma~\ref{lem:variants}(1) and~(2) give the axioms other than \axname{Ext}, \axname{Tab}, \axname{Bisimilarity}, and \axname{Realisation}.

(1) All the values of $\hat\cons$ are fresh terms. \axname{Tab} holds: every function from $\{\star\}$ to $\mathbf{lfp}(\Phi_F)$ has the code of copy $0$, which is in $\mathit{Fun}_s(\mathbf{lfp}(\Phi_F))$ by Lemma~\ref{lem:variants}(1). \axname{Ext} fails for $f^0$ and $f^1$. Since $\mathbf{lfp}(\Phi_F)=\Phi_F(\mathbf{lfp}(\Phi_F))$ contains $c(s,f^0)$ and $c(s,f^1)$, and $\alpha$ reaches at most one of them, $\alpha$ is not surjective.

(2) By Lemma~\ref{lem:variants}(1), $\mathit{Fun}_s(S)=\emptyset$ for every set $S$ of fresh terms, so $\Phi_F(S)=\{c(z,\hat\initMorph)\}=\{0\}$ for such $S$, and $\mathbf{lfp}(\Phi_F)=\{0\}$. \axname{Ext} holds vacuously. The set $R=\{\pr{\star,0}\}$ satisfies the premise of \axname{Tab} for the shape $s$, and no $f$ satisfies its conclusion.

(3) \axname{Ext} and \axname{Tab} hold by Lemma~\ref{lem:variants}(3), and the greatest fixpoint is $Z'$ by Lemma~\ref{lem:variants}(4). The M-type has a single tree $u$, so $\beta(x)=\beta(y)=u$, and, for every $T$ with $\mathit{Tree}(T)$, $z_T=u$ and $\mathit{Beh}(x,T)$ holds by Lemma~\ref{lem:paths}(3); so \axname{Realisation} holds. Since $x.\out=y.\out=0$ and the children of $x$ and $y$ at $\star$ are $x$ and $y$, the pair $\pr{x,y}$ is in the interpretation of $\mathit{bis}(R)$ for $R=\{\pr{x,y}\}$, hence in $\mathit{Bisim}$ by the Knaster--Tarski theorem.

(4) $F\,\emptyset=\emptyset$, since the shape has a position, so the empty map is a bijection and the greatest fixpoint is empty, by Lemma~\ref{lem:variants}(4). \axname{Ext} and \axname{Tab} hold by Lemma~\ref{lem:variants}(3), \axname{Bisimilarity} vacuously. The set $T$ of all the pairs $\pr{p,a}$, with $p$ in $\top(\mathsf{LP})$ and $a$ the shape, satisfies $\mathit{Tree}(T)$, and there is no element of $\inh{\nu F}$, so \axname{Realisation} fails.

(5) The greatest fixpoint is $\mathit{EP}$, by Lemma~\ref{lem:variants}(4), and \axname{Ext} and \axname{Tab} hold by Lemma~\ref{lem:variants}(3). Here $\gamma(e\cdot s)=(e,\star\mapsto s)$, so $\beta$ is the inclusion of $\mathit{EP}$ into $E^\omega$. If a set $R$ of pairs is included in its image under $\mathit{bis}$, then related streams have equal heads and related tails, so they have the same $n$-th element for every $n$, by induction on $n$, and are equal; hence \axname{Bisimilarity} holds. Let $z\in E^\omega\setminus\mathit{EP}$. The set of pairs $\pr{[\pi],a}$ with $(\pi,a)\in T_z$ satisfies $\mathit{Tree}$, by Lemma~\ref{lem:paths}(1), and an element $x$ with $\mathit{Beh}(x,T_z)$ would have $\beta(x)=z$, by Lemma~\ref{lem:paths}(3), which is impossible. So \axname{Realisation} fails.

(6) $s\in Z'$ iff $e\cdot s\in Z'$, so $\zeta'$ is a bijection onto $F\,Z'$, and the greatest fixpoint is $Z'$, by Lemma~\ref{lem:variants}(4). For $S\subseteq Z'$, $\Phi_F(S)$ consists of $\varepsilon$ and of the lists $e\cdot s$ with $s\in S$, so the least fixpoint is $E^*$. \axname{Ext} and \axname{Tab} hold by Lemma~\ref{lem:variants}(3). As in (5), $\beta$ is the inclusion of $Z'$ into the set $E^*\cup E^\omega$ of all finite and infinite lists, a set of pairs included in its image under $\mathit{bis}$ relates lists that have the same shape at every index, hence equal lists, so \axname{Bisimilarity} holds, and \axname{Realisation} fails for $T_z$ with $z\in E^\omega\setminus\mathit{EP}$. Every prefix of such a $z$ is in $E^*$, so $z$ is approximated to every depth by elements of $\inh{\mu F}$.
\end{proof}

\subsection*{Several Containers and Parameter Specifications}

We now extend the construction to $\Gamma_F$ together with parameter specifications from $\mathcal{C}^{\mathit{par}}$ (Definition~\ref{def:captured}), as needed for Theorem~\ref{thm:captured} and for the case studies. We fix an instance, i.e., a set $S_s$ for each parameter sort $s$ introduced by (Par).

\paragraph{The combined construction}
The specifications involve finitely many containers $G_1,\ldots,G_k$ (those introduced by (Ind)), with $G_k=F$, such that the shapes $A_j$ and the positions $B_j[a]$ of each $G_j$ are built by (Par), (Fin), (Prod), (Fam), and (Sum) from $G_1,\ldots,G_{j-1}$. Each $G_j$ has its own symbols $\cons_j$, $\out_j$, $\nxt_j$, and sort names $\mathsf{mu}_j$, $\mathsf{nu}_j$. The atoms now also include the constants of (Fin), as fresh atoms, the elements of the tagged copies $S_s^\sharp=\{\mathsf{par}_s\}\times S_s$ of the sets chosen for the parameter sorts, also as fresh atoms, and the trees of the M-types $Z_1,\ldots,Z_k$, where $Z_j$ is built over the interpretations of $A_j$ and $B_j$ described next. The sorts of the specifications are interpreted by their intended sets: a parameter sort $s$ by $S_s^\sharp$, the constants of (Fin) by themselves, products by pairs, dependent sums by dependent pairs, and $\mu G_j$ and $\nu G_j$ by the well-founded trees $W_{Z_j}$ and all trees $Z_j$. Here the shapes and positions are not tagged atoms, but these objects; for instance, the positions of lists are the element $\star$ of $\objOne$, and positions in a dependent sum are pairs.
The stages are as before, with (C4) adding the codes on all the position sets $B_j[a]$, (C5) adding fresh terms $\mathit{sup}_j(a,f)$ for every $j$, and a regular cardinal $\theta$ larger than all the position sets. Application is defined as before for each $j$: $\hat{\cons_j}\apd\pr{a,f}$, for $a\in A_j$ and $f\in\mathit{Fun}^j_a(M)$, is $\{\zeta_j^{-1}(a,\abs f)\}$ if all values of $\abs f$ are trees of $Z_j$, and $\{\mathit{sup}_j(a,f)\}$ otherwise; $\hat{\out_j}$ and $\hat{\nxt_j}$ are defined on the trees of $Z_j$ and on the fresh terms of $G_j$ as before. Here $\mathit{Fun}^j_a(S)$ is $\{\hat\initMorph\}$ if $B_j[a]=\emptyset$, and otherwise the set of $f$ with $f\apd b$ a singleton included in $S$ for $b\in B_j[a]$ and $f\apd x=\emptyset$ for $x\notin B_j[a]$. Let $\mathit{Fresh}_j$ be the set of fresh terms of $G_j$.

\paragraph{Canonical rather than literal interpretations}
``Intended'' is meant up to canonical bijections: the pairs of the model are the objects of (C1), not set-theoretic pairs, and $\mu G_j$ and $\nu G_j$ are interpreted by the sets of trees $W_{Z_j}$ and $Z_j$, which are isomorphic to the W-type and the M-type of $G_j$. This is all that Theorem~\ref{thm:captured} requires, since it states bijections with the intended sets.
Tagging is needed only for the parameter sorts of (Par). For a single container, tagging was needed because $A$ and $B[a]$ are arbitrary external sets, whose elements might coincide with objects of the construction, such as codes, symbol atoms, or pairs; the same holds for the sets $S_s$, and the tagged copies $S_s^\sharp$ consist of fresh atoms. All the other parameter sets are built inside the construction, from fresh atoms, $\star$, pairs, and trees, so no such coincidence can occur, and the disjointness provided by tagging is replaced by the facts (F1) and (F2) below and by Lemma~\ref{lem:only-codes-ext}. Coincidences between the parameter sets themselves are harmless: for instance, for the identity functor, $\star$ is both the shape and the position, and Lemma~\ref{lem:only-codes-ext} does not use the disjointness of shapes and positions.

\paragraph{Sort names}
The axioms of (Fam) are equations between sort names, not only between sets of inhabitants, so the sort names must be chosen accordingly. The set $N$ of sort names contains $\mathsf{Srt}$, $\mathsf 1$, one name for each sort introduced by (Par), by (Fin), and by (Sum), the names $\mathsf{mu}_j$ and $\mathsf{nu}_j$ of $\mu G_j$ and $\nu G_j$, and the names $\mathsf{P}_j$ and $\mathsf{LP}_j$ of the sorts of positions and of paths of each $G_j$, and it is closed under $\otimes$. The function $\top$ gives each name its intended set: $S_s^\sharp$ for a parameter sort $s$, the constants for (Fin), pairs for $\otimes$, dependent pairs for (Sum), and $W_{Z_j}$ and $Z_j$ for (Ind). A family $B$ introduced by (Fam) creates \emph{no} new sort names. Its symbol is interpreted as the atom $\hat B$, and $\mathit{depOf}_{\hat B}$ returns an existing name: $\mathit{depOf}_{\hat B}\apd a$ is the name of $s$ for every $a$ in the interpretation of $A$, for a constant family; the name of $s_i$ for $a=c_i$, for a family defined by cases; and the name of $s_i$ for $a=\pr{c_i,x}$ with $x\in\top(C[c_i])$, for a family defined by cases on the tag; it is empty otherwise. A compound sort, such as $s_1\otimes s_2$, is returned as its compound name. Similarly, the shapes $A_j$ of $G_j$ denote the name of the sort of $\mathcal{C}^{\mathit{par}}$ that gives them. For each $j$, $\hat{\mathit{DepSum}}\apd n_j=\{\mathit{DepSum}_{n_j}\}$ and $\mathit{DepSum}_{n_j}\apd\hat{B_j}=\{\mathsf P_j\}$, where $n_j$ is the name of $A_j$, and $\hat{\mathit{List}}\apd\mathsf P_j=\{\mathsf{LP}_j\}$, with $\top(\mathsf P_j)=\{\pr{a,b}\mid a\in A_j,\ b\in B_j[a]\}$ and $\top(\mathsf{LP}_j)$ the set of finite sequences over $\top(\mathsf P_j)$; the dependent sums of (Sum) are named in the same way. The fresh names $\mathsf{B}_a$ of the single-container model are used only there, where the family is not constrained by equations.

We use two facts about the sorts of the parameter specifications, which we call \emph{specified sorts}.
\begin{description}
\item[(F1)] Every element of a sort of the parameter specifications has an empty application on every argument: it is an atom of some $S_s^\sharp$, a constant of (Fin), $\star$, a pair, a sequence, or a tree. Specified sorts contain no sort names, symbol atoms, codes, $\hat\initMorph$, partial applications, or fresh terms.
\item[(F2)] The positions of $G_j$ are built only from (Par), (Fin), (Prod), (Sum), and $G_1,\ldots,G_{j-1}$; so neither they nor their components are trees of $Z_j$ or fresh terms of any $G_i$.
\end{description}

\begin{lem}\label{lem:only-codes-ext}
Let $1\le j\le k$, $a\in A_j$ with $B_j[a]\neq\emptyset$, and $S\subseteq Z_j\cup\mathit{Fresh}_j$. Then $f\in\mathit{Fun}^j_a(S)$ iff $f$ is a code $\ulcorner g\urcorner$ with $g:B_j[a]\to S$.
\end{lem}
\begin{proof}
Codes on $B_j[a]$ with values in $S$ are in $\mathit{Fun}^j_a(S)$. Conversely, let $f\in\mathit{Fun}^j_a(S)$ and $b\in B_j[a]$, so that $f\apd b=\{s\}$ with $s\in Z_j\cup\mathit{Fresh}_j$. We examine all the elements of $M$ by their kind.
\begin{enumerate}
\item $\hat\smbceil\apd b=M$, which is not a singleton.
\item $\hat{\mathit{inh}}$ has non-empty applications only on sort names, and positions are not sort names, by (F1).
\item $\hat{\mathit{pair}}$, $\hat{\mathit{Pair}}$, $\hat{\mathit{depOf}}$, $\hat{\mathit{DepSum}}$, $\hat{\mathit{List}}$, $\hat\pcons$, and their partial applications yield partial applications, pairs, sort names, or sequences, which are not in $Z_j\cup\mathit{Fresh}_j$.
\item $\hat{\mathit{fst}}$ and $\hat{\mathit{snd}}$ yield, on a pair-shaped position $\pr{x,y}$, its components $x$ or $y$, which are not in $Z_j\cup\mathit{Fresh}_j$ by (F2). Moreover, they are defined on every pair of $M$, e.g., on pairs of sort names, which are not positions, so they are not strict on $B_j[a]$.
\item $\hat\initMorph\apd b=\emptyset$, and $\hat{\finMorphIt}\apd b=\{\star\}$, with $\star\notin Z_j\cup\mathit{Fresh}_j$.
\item $\hat{\cons_i}$, for any $i$, has non-empty applications only on pairs $\pr{a',f'}$ with $f'\in\mathit{Fun}^i_{a'}(M)$. If $b=\pr{x,y}$ is a pair-shaped position, then $y$ is an element of a specified sort; by (F1) it has an empty application everywhere and is not $\hat\initMorph$, so $y\notin\mathit{Fun}^i_{a'}(M)$, and $\hat{\cons_i}\apd b=\emptyset$.
\item $\hat{\out_i}$ and $\hat{\nxt_i}$, for any $i$, have non-empty applications only on the trees of $Z_i$ and on the fresh terms of $G_i$. By (F2), a position can be such a tree only if $i<j$, and is never a fresh term. On a tree of $Z_i$, $\hat{\out_i}$ yields a shape of $G_i$, an element of a specified sort built from $G_1,\ldots,G_{i-1}$, hence not in $Z_j\cup\mathit{Fresh}_j$; and $\hat{\nxt_i}$ yields a code or $\hat\initMorph$. Moreover, independently of the values, $\hat{\out_i}$ and $\hat{\nxt_i}$ are defined on the fresh terms of $G_i$, which are not positions, so they are not strict on $B_j[a]$.
\item A code $\ulcorner g\urcorner$, of any container, yields $\{g(b)\}$ for $b$ in the domain of $g$. Strictness forces the domain of $g$ to be $B_j[a]$: an element of $B_j[a]$ outside the domain would have an empty application, and an element of the domain outside $B_j[a]$ a non-empty one.
\item All other elements, i.e., atoms, trees, fresh terms, pairs, sequences, sort names, and the remaining symbol atoms, have empty applications on positions.
\end{enumerate}
So $f$ is a code $\ulcorner g\urcorner$ with domain $B_j[a]$, and its values are in $S$.
\end{proof}
In other words, neither the auxiliary operations nor the operations of the other containers yield elements of $\mathit{Fun}^j_a(S)$ other than the intended codes. A code created for another container counts only when its domain is exactly $B_j[a]$, and then it is an intended code, since codes are determined by their graphs.

\begin{thm}\label{thm:standard-model-ext}
For every instance of a parametric specification, the combined theory, i.e., $\Gamma_{G_1}\cup\cdots\cup\Gamma_{G_k}$ together with the parameter specifications, has a model in which every parameter sort $s$ is interpreted by a set in bijection with the chosen set $S_s$, and every other sort by a set in canonical bijection with its intended set; in particular, it is consistent.
\end{thm}
\begin{proof}
Let $M$ be the structure of the combined construction. We check the axioms by groups.

\emph{Base theory.} \prule{Definedness}, the axioms of $\INH$ and $\PAIR$, and the axioms of $\initMorph$, $\star$, and $\objOne$ hold as in the proof of Theorem~\ref{thm:standard-model}: $N$ is closed under $\otimes$, pairs are injective and total, and $\top(s\otimes s')$ is the set of pairs of inhabitants.

\emph{Parameter specifications.} (a) \emph{Equations between sort names.} The axioms of (Fam), $\forall a\cln A\ld B[a]=s$, $B[c_i]=s_i$, and $\forall x\cln C[c_i]\ld B[\pr{c_i,x}]=s_i$, hold because $\mathit{depOf}_{\hat B}$ returns exactly the name of the required sort; in particular, the family axiom $\forall a\cln A\ld B[a]\cln\mathit{Sort}$ holds. The sorting axioms, i.e., $s\cln\mathit{Sort}$ for (Par) and (Fin), $A\cln\mathit{Sort}$, $(s_1\otimes s_2)\cln\mathit{Sort}$, the sort axioms of the dependent sums, and \axname{Mu Sort}, \axname{Nu Sort}, and \axname{List Sort} for each $G_j$, hold because these names belong to $N=\top(\mathsf{Srt})$. (b) \emph{Equations between inhabitants.} The axioms of (Fin), (Prod), and (Sum), and the equalities $\inh{B[a]}=\inh{s}$, follow from the definition of $\top$ on $N$; the constants of (Fin) are pairwise distinct atoms. A parameter sort $s$ has no further axioms, and $\top$ maps its name to $S_s^\sharp$, in bijection with $S_s$.

\emph{Containers.} Fix $j$. The interpretations of $A_j$ and of $B_j[a]$ are the sets $A_j$ and $B_j[a]$ of the construction, by (a) and (b). We repeat the proof of Theorem~\ref{thm:standard-model} for $G_j$, with Lemma~\ref{lem:only-codes-ext} in place of Lemma~\ref{lem:only-codes}.
\begin{itemize}
\item \emph{Constructor.} \axname{Functional} holds by the definition of $\hat{\cons_j}$. For \axname{No Confusion}, if $c_j(a,f)=c_j(a',f')$ is a tree $z\in Z_j$, then $\zeta_j(z)=(a,\abs f)=(a',\abs{f'})$, so $a=a'$, and, by Lemma~\ref{lem:only-codes-ext} with $S=Z_j$, $f$ and $f'$ are codes on $B_j[a]$ with the same graph (or both are $\hat\initMorph$), so $f=f'$; otherwise it is a fresh term $\mathit{sup}_j(a,f)$, and fresh terms are injective.
\item \emph{Initial algebra.} For $S\subseteq Z_j$, Lemma~\ref{lem:only-codes-ext} and the closure of $M$ under codes give $\Phi_{G_j}(S)=\{\zeta_j^{-1}(a,g)\mid a\in A_j,\ g:B_j[a]\to S\}$, so the least fixpoint is $W_{Z_j}=\top(\mathsf{mu}_j)$, and \axname{No Junk} holds; \axname{Ext} and \axname{Tab} hold because every function $B_j[a]\to W_{Z_j}$ has exactly one code, and $\mathit{Fun}^j_a(W_{Z_j})$ consists of these codes.
\item \emph{Final coalgebra.} $Z_j\subseteq\Phi_{G_j}(Z_j)$, and every element of the greatest fixpoint is a tree of $Z_j$ or a fresh term of $G_j$. A fresh term $\mathit{sup}_j(a,f)$ in the greatest fixpoint has $f\in\mathit{Fun}^j_a(Z_j\cup\mathit{Fresh}_j)$, so $f$ is a code by Lemma~\ref{lem:only-codes-ext}, one of whose values is a fresh term of $G_j$ in the greatest fixpoint, of a smaller stage; this is impossible, as in the proof of Theorem~\ref{thm:standard-model}. So the greatest fixpoint is $Z_j=\top(\mathsf{nu}_j)$, and \axname{No Cojunk} holds. \axname{No Coconfusion.1--2} hold by the definitions of $\hat{\out_j}$ and $\hat{\nxt_j}$, and \axname{Bisimilarity} because a set of pairs included in its image under $\mathit{bis}$ is a bisimulation on the final coalgebra $(Z_j,\zeta_j)$.
\item \emph{Paths and realisation.} $P_j$ denotes $\mathsf P_j$ and $\mathit{List}(P_j)$ denotes $\mathsf{LP}_j$, by the definition of the applications of $\hat{\mathit{DepSum}}$ and $\hat{\mathit{List}}$, and $\top(\mathsf P_j)$ is the set of dependent pairs, so the axiom of the dependent sum $P_j$ holds. $\top(\mathsf{LP}_j)$ is the set of finite sequences over $\top(\mathsf P_j)$, with $\mathit{pnil}$ the empty sequence and $\pcons$ the prepending of an element, so the axioms of $\mathit{List}(P_j)$ hold; sequences over different sets $\top(\mathsf P_j)$ are handled by the same symbols $\mathit{pnil}$ and $\pcons$, which are defined on all sequences. The other axioms of $\Gamma^\nu_{G_j}$ hold, and the unique coalgebra morphism from the extracted coalgebra to $(Z_j,\zeta_j)$ is the identity of $Z_j$; so, for $T$ with $\mathit{Tree}(T)$, the tree $z_T$ of Lemma~\ref{lem:paths}(2) satisfies $\mathit{Beh}(z_T,T)$ by Lemma~\ref{lem:paths}(3), and \axname{Realisation} holds.
\end{itemize}
The symbols of $G_j$ do not occur in the axioms of the other containers, except through the sorts $\mu G_j$ and $\nu G_j$, whose interpretations are $W_{Z_j}$ and $Z_j$. Finally, every sort is interpreted by $\top$ of its name, which is $S_s^\sharp$ for a parameter sort and the intended set, up to the canonical bijections, for the others.
\end{proof}

\begin{cor}\label{cor:case-studies-consistent}
For every set $E$, the theory of lists over $E$ of Section~\ref{sec:lists} has a model in which $E$ is interpreted by a set in bijection with $E$, and, for all sets $O$ and $I$, the theory of Moore machines of Section~\ref{sec:Moore} has a model in which $O$ and $I$ are interpreted by sets in bijection with $O$ and $I$.
\end{cor}
\begin{proof}
The parameter specification of lists consists of the parameter sort $E$, the sorts $\mathit{Tag}$, $\objZero$, and $\objOne$ given by (Fin), the family $C$ defined by cases, the dependent sum $\sum_{i\cln\mathit{Tag}}C[i]$, and the family of positions defined by cases on the tag; that of Moore machines consists of the parameter sorts $O$ and $I$ and the constant family $o\cln O\vdash I$. Both are in $\mathcal{C}^{\mathit{par}}$, so Theorem~\ref{thm:standard-model-ext} applies.
\end{proof}

\section{Changes with Respect to the FROM 2023 Version}
\label{ap:changes}

The main changes with respect to~\cite{Lucanu23} are:
\begin{itemize}
\item \emph{Terminology.} ``Polynomial functor'' now means polynomial in the container sense. The classical polynomial functors are defined with exponentiation of a functor by a constant object, and are shown to be container functors (Section~\ref{sec:pf}).
\item \emph{Corrected axioms.} Several axioms were corrected: the notation for partial functions (which now also allows empty target sorts), the representation $X^{B[a]}$ of functions, which is made strict, the injectivity of pairs (now stated for all elements, as needed for dependent pairs), the generic axiom \axname{Functional}, \axname{No Confusion} (now restricted to elements representing functions), the coconfusion axioms, and the axioms of the list and Moore machine case studies. The initial morphism $\initMorph$ is now required to be functional.
\item \emph{Proof principles.} The induction and coinduction principles are corrected and derived from the Knaster--Tarski rule; a bisimulation principle for equality is added.
\item \emph{Initial algebra.} Two axioms, \axname{Ext} and \axname{Tab}, fix which elements represent functions. Theorem~\ref{thm:adequacy-ini} proves that, in every model, the extracted algebra is initial (a W-type), and Proposition~\ref{prop:soundness-ini} gives a model. The approximation of the least fixpoint by $\varphi^n_F(\bot)$ is proved for finitary functors (Lemma~\ref{lem:continuity}).
\item \emph{Final coalgebra.} Two axioms, \axname{Bisimilarity} and \axname{Realisation}, are added. Theorem~\ref{thm:adequacy-fin} proves that, in every model, the extracted coalgebra is final (an M-type), and Proposition~\ref{prop:soundness-fin} gives a model. Models in which one axiom fails show that the new axioms are needed (Proposition~\ref{prop:necessity}), and Proposition~\ref{prop:local-global} relates the axioms of the two theories.
\item \emph{Parameters.} The notion of capturing a data type is stated for parametric data types: the theories capture $\mu F$ and $\nu F$ relative to the interpretation of the sorts of shapes and positions, and absolutely when these sorts are themselves captured (Proposition~\ref{prop:absolute}); Theorem~\ref{thm:captured} gives sufficient conditions, and its limits are stated in Remark~\ref{rem:limits}.
\item \emph{Theories.} The theories used by the adequacy results are fixed formally (Definition~\ref{def:gamma}), together with the typing of their symbols; Appendix~\ref{ap:sorts} states which recalled specifications are adapted and which are not used, and $\PAIR$ gets an axiom making pairing total. The sorts $\mu F$ and $\nu F$ are declared as sorts, the derivable axioms \axname{No Confusion} and \axname{No Coconfusion.3} are not axioms of $\Gamma^\nu_F$ (Lemma~\ref{lem:derived}), and finite sums are represented as dependent sums over finite sorts of tags, so the specification of sum sorts is no longer recalled; functions are represented by the notation $X^{B[a]}$, so the specification of function sorts is no longer recalled either; the ML specification of dependent products, which no result used, is removed.
\item \emph{Paths.} A lemma (Lemma~\ref{lem:paths}) proves that finite paths faithfully describe the trees of the M-type and relates this description to the patterns used by \axname{Realisation}; the surjectivity proof of Theorem~\ref{thm:adequacy-fin} and the check of \axname{Realisation} in Appendix~\ref{ap:model} rely on it.
\item \emph{Summary.} Section~\ref{sec:summary} classifies the results into well-formedness, satisfiability, semantic adequacy, derivable proof principles, and results that need additional assumptions.
\item \emph{Boundary cases.} Section~\ref{sec:boundary} analyses empty shape and position sets, empty exponent domains, infinite position sets, and functors with empty initial algebra and non-empty final coalgebra.
\item \emph{Standard model.} Appendix~\ref{ap:model} gives a full construction of a model of the theories of the initial algebra and of the final coalgebra, using tagged representatives of the shape and position sets and including several containers used as parameters (Lemma~\ref{lem:only-codes-ext} and Theorem~\ref{thm:standard-model-ext}), which proves that they have models with the intended interpretation, up to canonical bijections (the tagging bijections, for a single container). It also separates the external use of the final coalgebra in the construction from the internal characterization given by the axioms.
\item \emph{Lists.} The list case study is now a literal instance of the generic theories; the usual list constructors are recovered by transport along the coding map $\kappa$ (Proposition~\ref{prop:lists-transport}).
\item \emph{Moore machines.} The container form is corrected, the specification is instantiated completely, and the characterization of the final coalgebra by $O^{I^*}$ is proved (Theorem~\ref{thm:moore}).
\item The notations $\mu X\cln s\ld\varphi$ and $\nu X\cln s\ld\varphi$ for fixpoints within a sort are corrected so that their bodies are positive in $X$ (Section~\ref{sec:ml}).
\item The informal computations of the fixpoint approximations and the definition of coalgebra morphisms were corrected, and typos fixed throughout.
\item \emph{Audit of the revised version.} The class of parametric specifications (rule (Par) of Definition~\ref{def:captured}) and the complete proof of Theorem~\ref{thm:standard-model-ext} give models for the case studies over arbitrary parameters (Corollary~\ref{cor:case-studies-consistent}); $\mathit{List}(P)$ is declared as a sort and interpreted in the standard model; structures in which one axiom fails are constructed (Proposition~\ref{prop:necessity}); the two sorts are related by an inclusion, a density property, and a characterisation of \axname{Realisation} as completeness (Proposition~\ref{prop:mu-nu}); and all statements are numbered with one counter.
\end{itemize}

\end{document}